\documentclass{aa}

\usepackage{graphicx}

\usepackage{txfonts}

\usepackage{xcolor}
\usepackage[colorlinks=true,allcolors=blue]{hyperref}
\bibpunct{(}{)}{;}{a}{}{,} 
\usepackage{paralist}
\usepackage{epstopdf}
\usepackage{epsfig}
\usepackage{amsmath}
\usepackage{tablefootnote}
\usepackage{amssymb}
\usepackage{bm}
\usepackage{tabularx}
\usepackage{lmodern}

\newcommand\msol{{\cal $M_{\odot}$}}
\def\kmsec{\mbox{km~s$^{\rm -1}$}}
\def\hfilt{\mbox{\textit{H}}}
\def\kfilt{\mbox{\textit{K}}}
\def\teff{$\textit{T}_{\rm{eff}}$}
\def\teffldr{$T_{\rm{eff}}$(LDR)}
\def\teffspec{$T_{\rm{eff}}$(spec)}
\def\teffgaia{$T_{\rm{eff}}$(\emph{Gaia})}
\def\logg{\mbox{log~{\it g}}}
\def\vmicro{\mbox{$\xi_{\rm t}$}}
\def\carbiso{\mbox{$^{12}$C/$^{13}$C}}

\begin{document} 

   \title{High-Resolution Infrared Spectroscopy of the Dust-Obscured Metal-Poor Open Cluster Trumpler 5}

   \author{S.~\"Ozdemir
          \inst{1,2}
          \and
          M.~Af\c{s}ar\inst{2}
          \and
          C.~Sneden\inst{3}
          \and
          D.~A.~VandenBerg\inst{4}
          \and
          P.~A.~Denissenkov\inst{4}
          \and
          A.~P.~Milone\inst{5}
          \and
          Z.~Bozkurt\inst{2}
          \and
          H.~Oh\inst{6}
          \and
          K.~Sokal\inst{3}
          \and
          G.~N.~Mace\inst{3}
          \and
          D.~T.~Jaffe\inst{3}
          }

   \institute{Nicolaus Copernicus Astronomical Center, 
              Polish Academy of Sciences, ul. Bartycka 18, 00-716, Warsaw, Poland\\
              \email{sergen@camk.edu.pl, sergenozdemir58@gmail.com}
         \and
              Department of Astronomy and Space Sciences, 
              Ege University, 35100 Bornova, \.{I}zmir, Turkey \\
             \email{melike.afsar@ege.edu.tr}
         \and
              Department of Astronomy and McDonald Observatory, 
		   The University of Texas, Austin, TX 78712\\
         \and
              Department of Physics and Astronomy, University of Victoria, Victoria, BC, V8W 2Y2, Canada \\
         \and
              Dipartimento di Fisica e Astronomia "Galileo Galilei", Univ. di Padova, Vicolo dell'Osservatorio 3, Padova, IT-35122 \\
         \and
              Korea Astronomy and Space Science Institute, 776 Daedeok-daero, Yuseong-gu, Daejeon 34055, Republic of Korea \\
             }

   \date{Received --; accepted --}

  \abstract
   {Open clusters are important tools to investigate the chemistry of the Galactic disk. Trumpler 5 is a moderately old, dust-obscured metal-poor open cluster. In this study, high-resolution near-infrared spectroscopic data of seven giant stars from the Trumpler 5 cluster were analyzed to derive chemical abundances for 20 elements and \carbiso\ ratios. Color-magnitude diagram (CMD) analysis of BV and Gaia photometry has also been performed for a comprehensive study of the cluster.}
   {This work uses high-resolution near-infrared spectroscopy exclusively to derive atmospheric parameters and chemical abundances in the obscured open cluster Trumpler 5. Thanks to the methodology employed, some targets are studied for the first time. Additionally, it provides a detailed color-magnitude diagram analysis using photometric and spectroscopic data.}
   {We gathered high-resolution spectra for seven Trumpler~5 red giants in the near-infrared \hfilt\ and \kfilt\ wavelength domains, using the Immersion Grating INfrared Spectrometer (IGRINS).  Five out of seven targets have been studied for the first time here with high-resolution spectroscopy. We introduced a method to initially estimate the stellar surface gravity (\logg) by using calibrated equivalent widths of the \ion{Ti}{II}\ line at 15873 \AA\ from a large sample. We performed standard spectroscopic analyses to refine the model atmospheric parameters of our targets and determined the chemical abundances primarily through spectrum synthesis. We also performed color-magnitude diagram analyses to extract differential reddening correction to compare cluster parameters both with and without corrections.}
   {We derived stellar parameters for seven members of Trumpler 5 with our method and the results are consistent with both the literature and other methods. We also inferred elemental abundances for more than 20 species, along with the \carbiso\ ratios. The elemental abundances are in good agreement with the literature values for similar targets. Through CMD analysis, we found the reddening value, E(B-V)$\simeq$0.76 and estimated the age of the cluster to be approximately 2.50 Gyr.}
   {}


    \titlerunning{Analysis of Trumpler 5}
    \authorrunning{S. \"Ozdemir et al.}
    \maketitle
\nolinenumbers
\section{Introduction}\label{intro}

Stellar clusters are very important probes to understand the evolution of our Galaxy. Their chemical compositions are crucial for understanding Galactic chemical evolution and their kinematics yield constraints on model predictions of Galactic structure. Notably, open clusters are the main tools to investigate the Galactic disk evolution. There are more than 2000 known open clusters (OCs) in our Galaxy \citep{Donati15} and many of them are hidden behind the interstellar dust along the Galactic disk, towards to center or anti-center of the Galaxy. Therefore, studying many dust-obscured OCs through observations in the infrared (NIR) spectral region is essential to unveil the complete structure of our Galactic disk. 

OCs have a crucial role in understanding many astrophysical processes such as stellar evolution, galactic structure and evolution, and star formation(e.g., \citealt{Friel95, Bragaglia04, Bertelli17}). It is essential to obtain high-resolution spectra of red giant OC members, as they are bright enough to allow us to work at greater distances. Only a small percentage ($15\%$) of open clusters is old enough to have giant members \citep{Donati15}. 

Among open clusters in our Galaxy, Trumpler 5 (hereafter Tr5) has unique characteristics. The basic parameters of Tr5 were collected from the WEBDA database \footnote{https://webda.physics.muni.cz/} and are given in Table \ref{tab:basic}. It is a moderately old ($\sim 3$ Gyr), highly reddened, slightly metal-poor thin disk OC (\citealt{Dow70, Kim03, Piatti04, Donati15}). Tr5 is located towards the Galactic anticenter about 0.04 kpc above the Galactic plane \citep{Piatti04}. It has differential reddening across the cluster field (\citealt{kaluzny98, Kim09, Donati15}) due to an obscuring cloud of interstellar matter. \cite{Piatti04} showed that the differential reddening of Tr5 is about $\Delta \text{E(B-V)} = 0.11 - 0.22$ mag. Later, \cite{Kim09} estimated a high total reddening value of E(B-V)=0.62 $\pm$ 0.08 mag by applying theoretical isochrone fitting to the color-magnitude diagram (CMD) obtained from 2MASS data. Being obscured by a cloud makes Tr5 difficult to observe in the visible region. This also makes it more challenging to determine the cluster age, especially via isochrone fitting techniques to the CMD of the cluster. Establishing the age of Tr5 is very important since only five percent of known open clusters are older than 2 billion years old \citep{Donati15}. The minimum age for Tr5 published so far is 2.4 $\pm$ 0.2 Gyr \citep{Kim03}. Recently, \cite{cantat20} used an artificial neural network to derive a much larger age for the cluster: 4.27 Gyr. Clearly Tr5 is a prime candidate for a more detailed color-magnitude diagram (CMD) analysis.

Dynamical classification via blue stragglers is one method to classify globular clusters. \cite{Vaidya20} showed that this kind of analysis can be also performed on open clusters. Recently, \cite{Rain21} performed this analysis on Tr5 along with two other open clusters and searched for blue stragglers. Among the three OCs, Rain et al. found that Tr5 has a large populations of blue stragglers: 40 candidates. In the dynamical analysis of Tr5 cluster, they found a flat distribution of blue stragglers up to 15 arcmin diameter. Because of the flat distribution, Tr5 can be classified as a member of the Family I type of cluster \citep{Rain21}. The family I corresponds to dynamically young clusters which show flat distribution of blue stragglers \citep[see][for details]{Ferraro2012}. However, the calculated $N_{relax}$ parameter shows that Tr5 is dynamically evolved but still, its distribution is flat \citep{Rain21}. Tr5 is also a good candidate for studies on cluster dynamics.

To find the exact position of a cluster in the Galaxy, accurate distance and proper motion determination of the cluster is essential. Recent studies of Tr5 give reasonably consistent values for the cluster distance as about 3 kpc (\citealt{Kim09, Donati15, cantat18}). However, due to differential reddening, distance determinations still contain some discrepancies, and more precise distance information is still needed.

Most of the previous spectroscopic studies suggest that Tr5 has sub-solar metallicity with an average of about [Fe/H]=$-$0.40 (e.g. \citealt{Carrera07,Kim09}). To the best of our knowledge, only \cite{kaluzny98} assumed solar metallicity and derived lower limits for the reddening of the cluster from the $BVI$ photometry. High-resolution optical spectroscopy has been a challenge for Tr5 members due to their faint apparent magnitudes. Hence, there are only a few studies that focus on the detailed chemical abundances of Tr5 members. \cite{Monaco14} provided information on the abundance analysis of four giant stars in the optical region. They also identified a super lithium-rich red clump star in the cluster. \cite{Donati15} determined the elemental abundances of three giant members of Tr5 using high-resolution spectra obtained with UVES. \cite{Spina2021} compiled a catalog of abundances for open clusters from both GALAH and APOGEE surveys along with their membership analysis, and in their work, there are nine confirmed members from Tr5 with an average metallicity of [Fe/H] = $-$0.44 ($\sigma$=0.02). \cite{Viscasillas22} reported the metallicity obtained from the Gaia-ESO Survey (GES) as [Fe/H]=$-$0.35. Monaco et al. and \cite{Donati15} agreed that Tr5 is a metal-poor open cluster and they obtained elemental abundances of some cluster members. GES results \citep{Viscasillas22} also support that Tr5 is a metal-poor open cluster.

High-resolution infrared spectroscopy plays a critical role in terms of accessing highly reddened clusters because interstellar extinction is substantially smaller in the infrared and target cool red giant stars are much brighter in the infrared. In this study, we use the power of the infrared region to unveil the members of the Tr5 open cluster. We analyzed the detailed chemical compositions of seven selected giant members, for the first time, using high-resolution spectra obtained with the Immersion Grating INfrared Spectrometer \citep[IGRINS;][]{Yuk10, Park14, Mace2016, Mace2018} in the NIR region. We also revisit the kinematics and CMD of the cluster. This work is part of a series of comprehensive chemical abundance studies of selected ref giant (RG) members of OCs using high-resolution spectra from both optical \citep{752vis, 6940vis} and NIR spectral regions \citep{6940IR, 752IR}.

\begin{table}
\caption{Basic parameters of Tr5 open cluster taken from WEBDA$^{\rm a}$ database.}  
\label{tab:basic}      
\centering                          
\begin{tabular}{l l}      
\hline\hline                
   Right Ascension (2000) & 06 36 42  \\ 
   Declination (2000)     & +09 26 00 \\  
   Galactic Longitude     & 202.865  \\ 
   Galactic Latitude      & 1.050  \\  
   Distance [pc]          & 3000 \\  
   Reddening [mag]        & 0.58 \\ 
   Distance Modulus [mag] & 14.18\\  
   log Age                & 9.61 \\  
\hline                                 
\multicolumn{2}{l}{$^{\rm a}$  http://www.univie.ac.at/webda/webda.html}\\
\end{tabular}
\end{table}

In section \ref{observ}, observations and data reduction are summarized. The kinematics of Tr5 are given in Section \ref{kinematics}. In Section \ref{models}, we explain the method applied to obtain the model atmosphere parameters, and the abundance analysis of our targets is summarized in Section \ref{abund}. We give the CMD analysis of Tr5 in Section \ref{sec:CMD}. Our discussion and our conclusion are given in Sections \ref{disc} and \ref{summ}, respectively.

\begin{figure}
\includegraphics[width=\columnwidth]{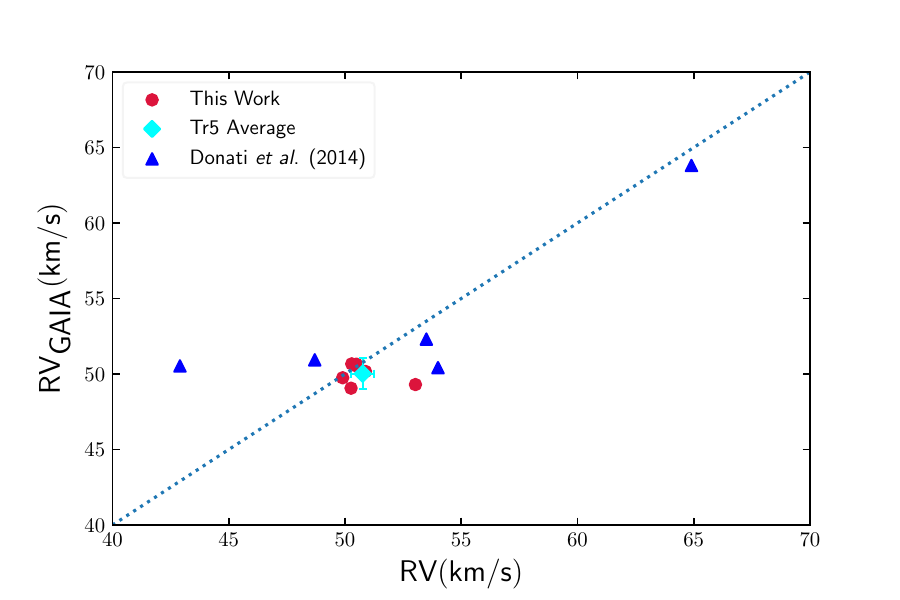}
      \caption{Comparison between the radial velocities from \emph{Gaia} DR3 ($\mbox{RV}_{\rm{Gaia}}$) and the radial velocities measured from IGRINS spectra ($\mbox{RV}_{\rm{IGRINS}}$) and the velocities taken from the literature. Blue diamond shows the average velocity of our target stars and the red dotted line shows the equality of $\mbox{RV}_{\rm{Gaia}}$ and $\mbox{RV}_{\rm{IGRINS}}$ velocities.}
      \label{fig:RV}
\end{figure}

\section{Observations and data reduction}\label{observ}

\begin{table*}
 \caption{Basic parameters of the program stars and summary of IGRINS observations.} 
 \label{tab:observ}       
\centering
\begin{tabular}{c c c c c c c c c}
      \hline  \hline
      Star & RA & Dec & Date & Exposure &V$^{\rm c}$&(B-V)$^{\rm c}$ & G$^{\rm b}$ & (G-G$_{\rm RP}$)$^{\rm b}$  \\ 
      \hline 
      $2MASS^{\rm a}$ & $Gaia^{\rm b}$ & $Gaia^{\rm b}$ &  & $seconds$ &$mag$&$mag$& $mag$ & $mag$ \\  
      \hline
      J06361880+0934066 & 06 36 18.80 & +09 34 06.60 & 21 12 2018 & 240 &12.331&2.040 & 11.186 & 1.138 \\  
      J06361925+0925587 & 06 36 19.25 & +09 25 58.80 & 25 11 2018 & 300 &13.135&1.920 & 12.354 & 1.087 \\  
      J06362606+0931584 & 06 36 26.06 & +09 31 54.84 & 21 12 2018 & 180 &-&-& 12.998 & 1.005 \\  
      J06363256+0925083 & 06 36 32.56 & +09 25 08.40 & 19 03 2019 & 180 &13.400&1.948 & 12.452 & 1.080 \\ 
      J06364731+0932596 & 06 36 47.31 & +09 32 59.64 & 27 11 2018 & 300 &13.721&1.724 & 12.957 & 0.998 \\  
      J06365277+0932270 & 06 36 52.77 & +09 32 26.88 & 21 12 2018 & 240 &-&-& 13.437 & 0.995 \\ 
      J06363185+0940450 & 06 36 31.85 & +09 40 45.12 & 25 03 2016 & 600 &13.600&1.729 & 12.858 & 0.971 \\ 
      \hline      
\multicolumn{7}{l}{$^{\rm a}$ \cite{2MASS}}\\
\multicolumn{7}{l}{$^{\rm b}$  \cite{gaia16}, \cite{gaiadr3}}\\
\multicolumn{7}{l}{$^{\rm c}$ SIMBAD Database}\\
\end{tabular}
\end{table*}

We have selected seven RG members from the Tr5 open cluster according to membership probabilities. The membership probabilities are taken from \cite{Kharchenko13} for 2MASS J06363185+0940450 \footnote{We will call it as J06363185 for the rest of the text and same pattern will be used for other stars.} and from \cite{cantat18} for the rest of the sample. High-resolution (R $\simeq$ 45,000), high S/N ($\sim$100) spectra were taken using the IGRINS spectrograph \citep{Yuk10, Park14, Mace2016, Mace2018}. IGRINS has the advantage of covering $\hfilt$ and $\kfilt$ bands (1.48 \textmu m $-$ 2.48 \textmu m), simultaneously, with only a small gap of about 0.01~$\mu$m (100~\AA) between bands. It has a compact and practical design so that it provides high throughput as a visitor instrument at observing facilities (e.g. the Gemini South telescope). Most of our targets were observed with the 4.3-meter Lowell Discovery Telescope (LDT, formerly known as Discovery Channel Telescope, DCT) at the Lowell Observatory (Arizona). Only the spectrum of J06363185 was obtained with the 2.7-meter Harlan J. Smith Telescope (HJST) at the McDonald Observatory. The observation log is summarized in Table \ref{tab:observ}.

Observations were made using an ABBA nod sequence to effectively remove the sky background, as well as the dark current and telescope background \citep{Kyung14}. The data were processed using the IGRINS Pipeline Package PLP2 \citep{plp2}. More information about the data reduction process can be found in \cite{Afsar16}. For each target, a telluric divisor (rapidly rotating hot stars from B to A type) was observed to remove the water vapor, the main contaminant in the \hfilt-band, and CO$_2$, the main contaminant in the \kfilt-band. We used the ``telluric'' routine of IRAF to divide out the telluric lines from stellar spectra. We then detected some well-known lines to calculate the total velocity shift from the rest in each spectrum. Then, we shifted our spectra by using ``dopcor'' routine of IRAF to a rest laboratory wavelength system. After this reduction process, we checked each spectrum in terms of quality. During this examination, we excluded one of the stars (2MASS J06364685+0938098) from our sample due to broadened lines, which made it unsuitable for our analysis techniques. \cite{Rain21} reported some fast-rotators in Tr5. Our broad-lined star is probably another rapid rotator, and deserves more careful analysis in the future.

We also used \emph{Gaia}'s renormalized unit weight error values (RUWE, an astrometric goodness-of-fit statistic to examine the existence of a possible unresolved companion. We found that the Gaia DR3 RUWE values for our sample range from 0.87 to 1.26, all below the threshold of 1.4, which suggests a lower probability of being binaries \citep{Stassun21}. Therefore, our stars are likely to be single.

\section{Kinematics}\label{kinematics}

To re-confirm the membership and investigate the kinematics of Tr5 we first measured the radial velocities (RVs) of our targets. We used the ``fxcor'' task of IRAF for the RV measurement and Arcturus spectrum (obtained with IGRINS) as the radial velocity standard (as described in \cite{Lee2004} for the infrared region). Our results are given in Table \ref{tab:kinematic} along with the ones from Gaia DR3 (\citealt{GAIA18b,gaiadr3}). Tr5 has been a subject for several studies (\citealt{Donati15, Soubiran18,Carrera19}), which helped us to collect RV values of 21 other members of the cluster. We compare all the RVs in Figure \ref{fig:RV}. As seen in the figure, all values agree well with each other within 5 $\kmsec$ (except for the one star in \cite{Donati15}).  We summarize the average RVs from \emph{Gaia} and our analysis for the cluster in Table \ref{tab:kinematic}. As seen in the Table, our cluster average of 50.76 $\kmsec (\sigma =0.49 \kmsec)$ agrees very well with the \emph{Gaia} DR3 and other values reported in the literature \citep[e.g. average RV values of 49.7, 51.3 and 50.0 km/s from ][respectively]{Donati15, Soubiran18, Carrera19}.

\begin{table*}
 \caption{Kinematic properties of our target stars.} 
 \label{tab:kinematic}   
\centering
\begin{tabular}{@{} c c c c c c c c @{}}
      \hline \hline
      Star & RV$_{\text{IGRINS}}$ & e$_{\text{RV}}$ & RV$_{\text{Gaia}} ^{\rm b}$ & e$_{\text{RV}}$ & Distance$^{\rm b}$ & Distance$^{\rm c}$ & P$_{\text{Memb.}}$ $^{\rm d}$ \\ 
      \hline 
      2MASS$^{\rm a}$   & \kmsec & \kmsec & \kmsec  & \kmsec & pc & pc &  \\ 
      \hline 
      J06363185+0940450 & 50.88 & 1.12 & 50.17 & 0.69 & 3653 (272) & 3423 (214) & 0.90 \\  
      J06361880+0934066 & 50.49 & 0.31 & 50.65 & 0.52 & 3392 (231) & 3104 (156) & 1.00 \\ 
      J06361925+0925587 & 50.29 & 0.22 & 50.67 & 0.50 & 3236 (225) & 2780 (147) & 0.90 \\  
      J06362606+0931584 & 50.45 & 0.28 & 50.56 & 1.23 & 2969 (145) & 2714 (111) & 1.00 \\  
      J06363256+0925083 & 53.03 & 0.55 & 49.30 & 1.85 & 2718 (111) & 2655 (104) & 1.00 \\
      J06365277+0932270 & 50.26 & 0.46 & 49.06 & 1.26 & 3145 (138) & 2755 (112) & 1.00 \\  
      J06364731+0932596 & 49.90 & 0.52 & 49.75 & 1.25 & 3088 (174) & 2816 (126) & 1.00 \\  
      Tr5               & 50.76 & 0.49 & 50.02 & 1.04 & 3172 (300) & 2892 (275) &      \\  
      \hline     
\multicolumn{7}{l}{$^{\rm a}$ \cite{2MASS}}\\
\multicolumn{7}{l}{$^{\rm b}$ \cite{gaia16}, \cite{gaiadr3}}\\
\multicolumn{7}{l}{$^{\rm c}$ \cite{BailerJ21}}\\
\multicolumn{7}{l}{$^{\rm d}$ \cite{cantat18}}\\
\end{tabular}
\tablefoot{Tr5 indicates the average value of our target stars. Errors for distances are given as a standard deviation in parentheses.}
\end{table*}

Distances of our targets have been adopted from \emph{Gaia} DR3 (\citealt{gaia16,gaiadr3}). Tr5 is quite far and the trigonometric parallax method comes with its handicaps for such distances. \cite{BailerJ21} used their algorithm to revise the distances for 1.47 billion stars in \emph{Gaia} eDR3 by using statistical methods and \emph{Gaia} parameters together. They published both geometric and photogeometric distances for those targets. Photogeometric distances have more accuracy and precision \citep{BailerJ21}. We adopt the photogeometric distances given by \cite{BailerJ21} for Tr5 and the values from both \emph{Gaia} DR3 and \cite{BailerJ21} agree that Tr5 has an average distance of $\sim$3$\pm$0.3 kpc with $\sim$11$\pm$0.3 kpc galactocentric distance. The distance information for each star is given in Table \ref{tab:kinematic}. 

\cite{cantat18} focused on the membership probabilities of open clusters by using \emph{Gaia} DR2 data. They also studied the Tr5 members and listed the membership probabilities for more than 3000 stars, including our targets (Table \ref{tab:kinematic}). According to \cite{cantat18}, five of our targets have 100\% of membership probability while J06363185 and J06361925 have 90\%. Taking into account both the RV agreement and high membership probability, we can conclude that all of our targets are kinematic members of Tr5.

Open clusters are mostly located on the Galactic disk. Tr5 has been reported as a thin-disk cluster by some studies \citep{Kim03, Piatti04, Donati15}. The thin disk of our Galaxy has a few hundred-parsec height \citep{Rix2013} and Galactic positions and velocities of stars help us to find the difference between the thin and thick disk stars. \cite{Soubiran18} investigated the kinematics of many OCs including Tr5. They determined the Galactic velocities for Tr5 as \textit{U} = -50.42 \kmsec, \textit{V} = -12.78 \kmsec, \textit{W} = -5.12 \kmsec, respectively. Considering these values, one can also estimate the kinematic position of Tr5 as a thin disk member of the Galaxy by using the well-known Toomre diagram \citep[See Figure 4 of][]{Yan19}.

\section{Stellar model atmospheres from infrared spectra}\label{models}

To the best of our knowledge, this is the first high-resolution spectroscopic study for five of our targets in the literature, except J06362606 and J06364731 have been studied by Gaia-ESO Survey \citep[GES,][]{GESfinal}. Due to the lack of information from the optical region, and the substantial reddening of Tr5, we have developed a new method to determine the atmospheric parameters of our targets from the NIR spectral data. 

There are different approaches to determine model atmospheric parameters. One way is to compare observed spectra to a library of synthetic spectra \citep[APOGEE is a very good example for this approach in the near infrared,][]{Apogee2022}. The traditional way to determine the model atmospheric parameters usually requires the investigation of high-resolution optical spectral data. The most common method is based on the excitation equilibrium and ionization balance of mostly Fe lines present in the optical region. Although our NIR data harbor many \ion{Fe}{i} lines, \ion{Fe}{ii} lines are absent. Therefore, finding the stellar atmospheric parameters only from the high-resolution NIR spectral data requires a different methodological approach. The subsequent subsections provide a comprehensive explanation of the method, particularly the initial estimate of stellar surface gravity, which is essential for starting the model atmosphere solution.

\begin{figure*}
\includegraphics[scale=0.9]{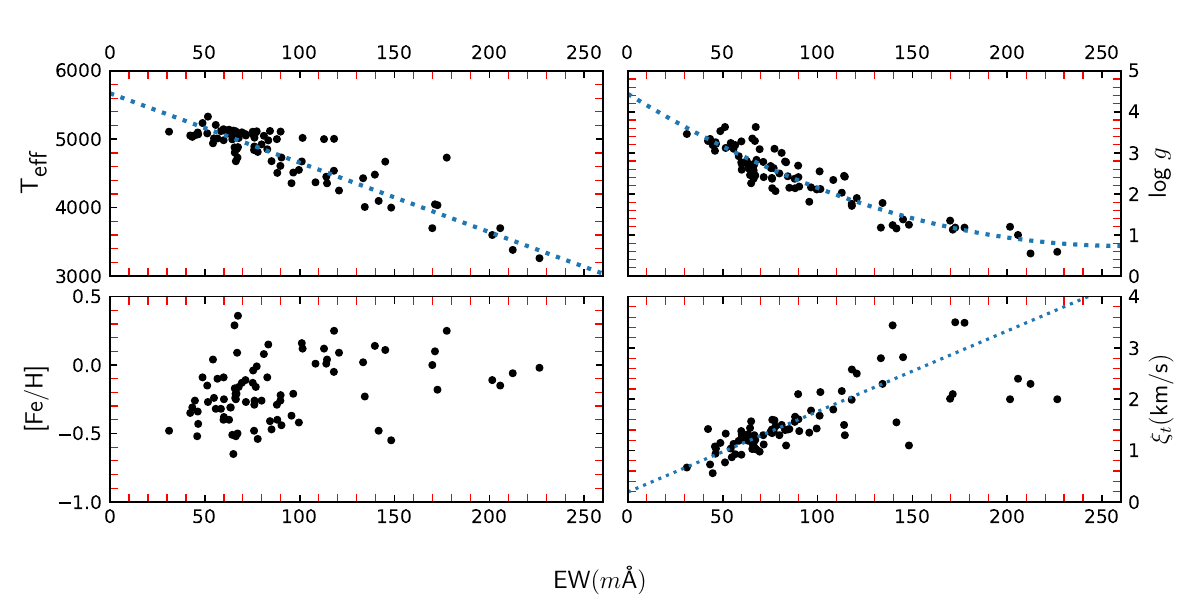}
      \epsfxsize=16cm
      \caption{Equivalent widths (EWs) of Ti II line at H band versus stellar atmospheric parameters. Ti II line EWs show relations for atmospheric parameters except metallicity. Empirical relations for \logg\ and \vmicro\ are used to find initial parameters in our model search.}
      \label{fig:titan}
\end{figure*}

\subsection{Initial \teff\ determination}\label{teff}

Several iterations with reasonable initial parameters are needed to derive the model atmospheric parameters. A useful approach to have an initial value for the \teff\ is to apply the line depth ratio (LDR) method (e.g. \citealt{gray91,kovt06,biazzo07a,biazzo07b}). The LDR method has been frequently used in the optical region. \cite{fukue15} and \cite{Jian19} applied the LDR method in the NIR and reported LDR$-$\teff\ relations for the \hfilt-band region. The LDR method has been also applied by \cite{Taniguchi18}, \cite{Matsunaga21}, and \cite{Taniguchi21} to the YJ-band spectral region. Most recently, \cite{AfsarLDR} has published LDR$-$\teff\ relations for both \hfilt- and \kfilt-band and added new LDRs to the Jian et al. relations in the \hfilt-band and presented the first-ever LDR$-$\teff\ relations in the \kfilt-band. We calculated the LDR temperatures of our targets using the recent LDR$-$\teff\ relations listed in Table 3 of \cite{AfsarLDR} and Table 3 (``metallicity-dependent'' LDR relations) of \cite{Jian19} . Below, we list these values in Table \ref{tab:mods} along with the \emph{Gaia} temperatures. Among these LDR temperatures, we 
set the initial \teff\ estimates of our targets using the metallicity-dependent LDR relations from \cite{Jian19}.

\begin{table}
 \caption{Basic properties of Ti II line and the statistical results of the relations between EW of Ti II line and stellar parameters.} 
 \label{tab:titan}     
 \centering
\begin{tabular}{@{}l c c c @{}}
      \hline \hline
      \textbf{Line}                        & Wavelength(\AA)   & log \textit{gf} & E.P.(eV)               \\ 
      \hline 
      Ti II$^{\rm a}$             &     15873.840      &      -1.90      & 3.123                   \\ 
      \hline 
      \textbf{Relation}                    &  $R^2$             & Std.Error  & Significance \textit{F} \\ 
      \hline 
      \teff                       & 0.79               & 205             & 2.29$*10^{-29}$         \\
      \logg                       & 0.85               & 0.29            & 1.63$*10^{-33}$         \\
      \vmicro                     & 0.64               & 0.23            & 2.43$*10^{-16}$         \\ 
      \hline
\multicolumn{4}{l}{$^{\rm a}$ \cite{Wood14}} \\
\end{tabular}
\end{table}

\begin{figure}
\centering
\includegraphics[width=\columnwidth]{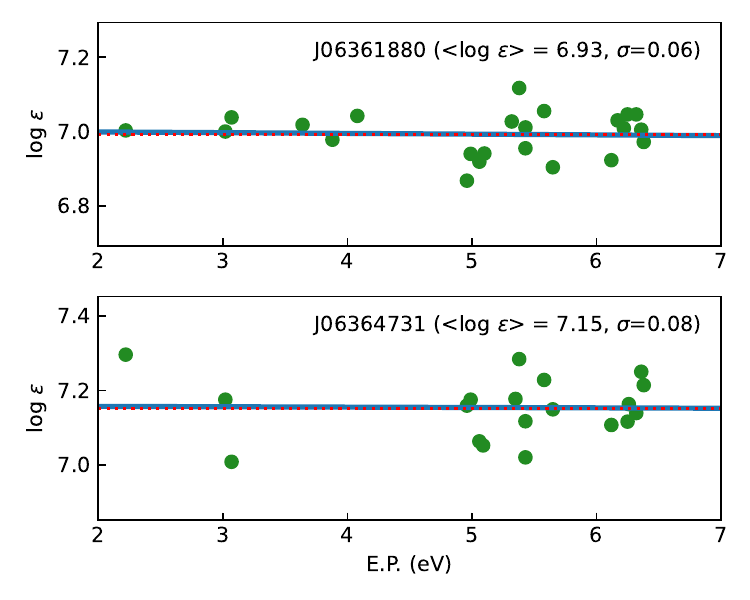}
      \caption{Abundances of Fe I as a function of excitation potential for the coolest (J06361880) and the hottest (J06364731) stars.}
      \label{fig:exc_pot}
\end{figure}

\subsection{Initial \logg\ determination}\label{logg}

It is difficult to obtain an initial spectroscopic value for the \logg\ due to the absence of \ion{Fe}{ii} lines in the NIR spectral region. To overcome this issue we have constructed a different method to determine the initial \logg\ values of our Tr5 targets. Here we have made use of a large sample that contains the IGRINS spectra of about 120 stars. This sample consists of about 70 red horizontal branch (RHB) and red clump (RC) stars with a temperature range between $\sim$4000 $-$ 5500 K. The rest of the spectral data were taken from the IGRINS spectral library \citep{Park18}, to extend the sample toward cooler temperatures down to $\sim$3200 K. RHB and RC stars have been also observed in the optical region and all have their atmospheric parameters determined \citep{afsar18a}. The atmospheric parameters for the rest of the sample were adopted from \cite{Park18} and from the PASTEL catalog when needed \citep{Soubiran16}. Our overall sample contains stars with surface gravities ranging from 0.55 to 3.63.

\begin{figure}
\includegraphics[width=\columnwidth]{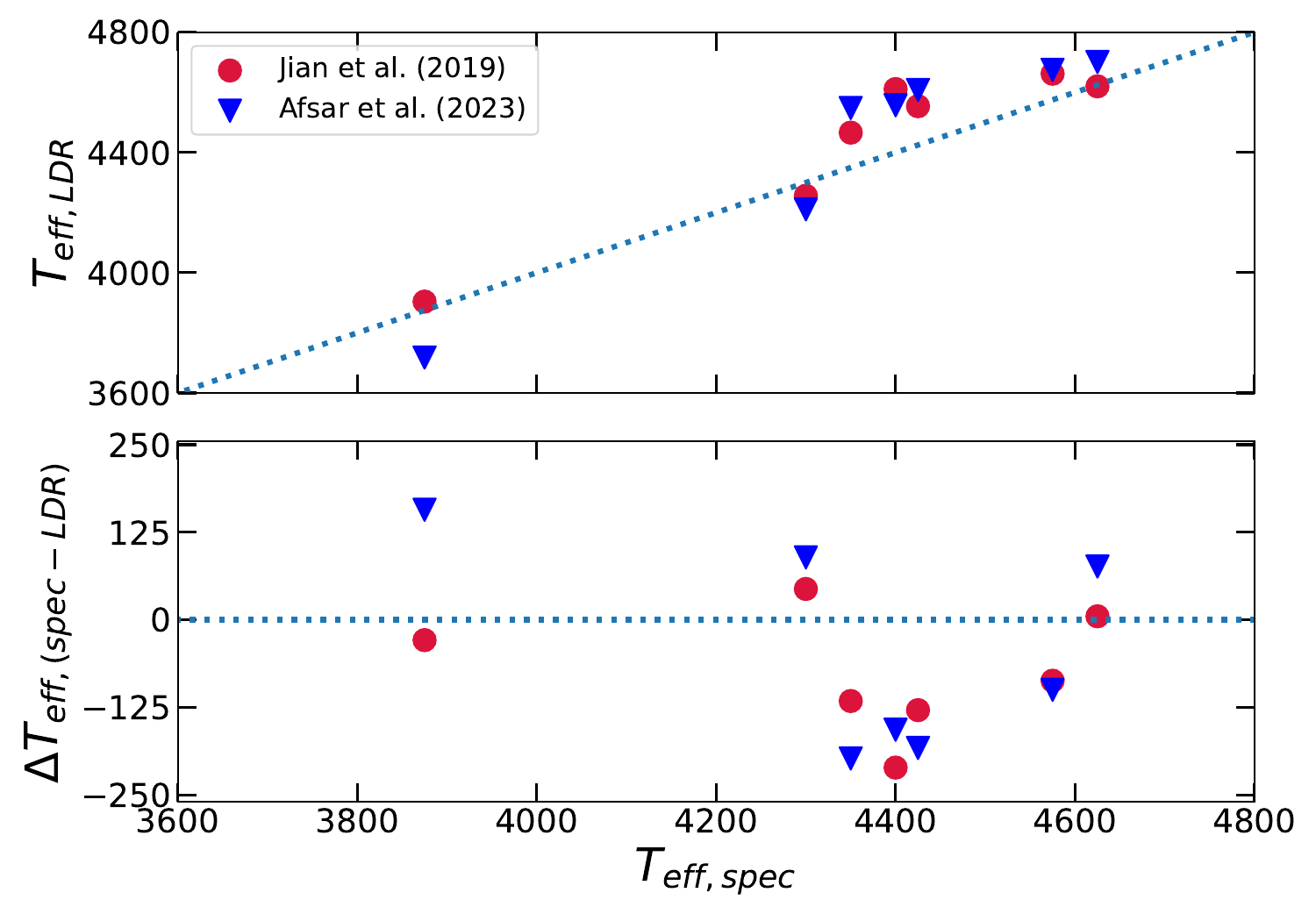}
      \caption{Comparison of LDR temperatures (\teffldr) calculated from the relations (see text for details) and our spectroscopic temperatures (\teffspec). On the bottom panel residuals of temperatures are given.}
      \label{fig:LDR}
\end{figure}
 
We used this sample to investigate the relationship between EW of Ti II 15873 \AA\ and the stellar surface gravity. This line appears to be the only iron-group ionized transition that is strong and has a well-determined laboratory transition probability \citep{Wood14}. The relations obtained through this effort are given in Figure \ref{fig:titan} and Equations \ref{eq:grav}, \ref{eq:teff}, and \ref{eq:turb}. Our main focus was to obtain a reasonable correlation between EW$_{(\ion{Ti}{ii})}$$-$\logg\ for disk giants (Figure \ref{fig:titan}, upper right panel). The equation that represents this correlation is given in Eq.1. Since \teff\ and \logg\ values are related in giant stars, there is also \teff\ dependence on EW measurements. However, we used LDR temperatures as initial independent guesses as described previously. EW$_{(\ion{Ti}{ii})}$ has no dependency on [Fe/H] for stars with [Fe/H]$\gtrsim-0.80$. The stars having lower metallicity values show a trend with decreasing metallicity. We excluded stars with metallicities lower than this value and used 83 stars from our sample for the relations as represented by the equations below. Statistical analysis of final relations and the basic properties of \ion{Ti}{ii} line can be found in Table \ref{tab:titan}.

\begin{equation}
\label{eq:grav}
\logg = 4.53 \times 10^{-5} \times\ \rm {EW^2} - 0.03 \times EW  + 4.36
\end{equation} 

\begin{equation}
    \label{eq:teff}
    \textit{T}_{\rm{eff}} = - 9.35 \times\ \rm {EW} + 5618.06
\end{equation}

\begin{equation}
\label{eq:turb}
\vmicro = 0.01 \times \rm {EW} + 0.56
\end{equation} 

Various studies have established similar empirical relations to determine atmospheric parameters \citep{Park18, Chun2020}. Specifically, \cite{Park18} presented empirical relations using the IGRINS spectral library, with a focus on Ti I and CO lines.

One issue with the \ion{Ti}{ii} line is that it comes with CO contamination, and this contamination becomes stronger for stars with temperatures cooler than $\sim$4500 K and with \logg\ $\lesssim$ 1.5 dex. Therefore, this method should be applied with caution for cooler stars and lower surface gravities. We will describe how we dealt with this issue in the following section.

\subsection{Final model atmosphere parameters}\label{final} 

After determining the initial parameters for the model atmosphere iterations, we first applied our method to two different stars with different temperatures and well-known atmospheric parameters: HIP 54048, the RHB star \citep{afsar18a}, and Arcturus \citep{ramirez11}. Without taking into account their known atmospheric parameters we first made use of the metallicity-dependent LDR relations from \cite{Jian19} and estimated an initial temperature value, then we applied the EW$_{(\ion{Ti}{ii})}$$-$\logg\ correlation for the initial guess on the \logg\  parameter. We started our iterations with assumption of solar metallicity ([Fe/H]) approximation and the initial \vmicro\ value was obtained from the EW$_{(\ion{Ti}{ii})}$$-$\vmicro\ relation given in Eq.\ref{eq:turb}. Then, we measured the equivalent widths (EWs) of \ion{Fe}{i} lines reported in \cite{afsar18b} and used the ``abfind'' driver of current version of MOOG\footnote{https://www.as.utexas.edu/~chris/moog.html} \citep{sneden73}, the local thermodynamic equilibrium (LTE) line analysis and synthetic spectrum code developed by \cite{sneden73}, to determine the \ion{Fe}{i} abundances from their EW measurements. Iterations continued until the \ion{Fe}{i} abundances from the low-and high-excitation \ion{Fe}{i} lines show no trend with the excitation potential ($\chi$) of each line (Figure \ref{fig:exc_pot}), in other words until the final value for \teff\ was achieved. This effort also leads to the estimation of final [Fe/H] and the \vmicro\ values. The spectral \teff\ of our stars are listed in Table \ref{tab:mods}.

Our study is based specifically on Fe I lines that are suitable for both EW measurements and spectrum synthesis analyses. This approach requires rigorous selection and detailed evaluation of iron lines, as described in \cite{afsar18b}. Following the determination of the final atmospheric model parameters, we performed a spectrum synthesis analysis of the Fe I lines. This step was taken to validate that the Fe I abundances derived from the EW method are consistent with those obtained via spectrum synthesis. The results of this cross-validation indicate a high degree of agreement between the two methods. This comprehensive comparison was carried out for all targets in our sample and consistently yielded a similar level of agreement. The mean of absolute differences between the EW and spectrum synthesis analysis results across all targets is 0.010, with a standard deviation of $\sigma$=0.007. A similar analysis was also performed for the Ti lines, yielding a difference of 0.037 ($\sigma$=0.025).

As mentioned earlier, the \ion{Ti}{ii} is blended with a  $^{12}$CO transition. Therefore, before deriving the Ti abundance from the \ion{Ti}{ii} line, an estimated C abundance value was assigned by investigating the $^{12}$CO molecular lines present at and around 15873 \AA. Then, we applied both EW and synthetic spectrum fitting to the \ion{Ti}{ii} for the Ti abundance determination. Both methods are also applied to \ion{Ti}{i} lines for the same purpose. Thereafter, a better value for the \logg\ was reached by demanding the \ion{Ti}{i} and \ion{Ti}{ii} lines producing the same mean abundance. As discussed in detail by \citep{afsar18b} Ti lines have significant isotopic substructure, therefore we used the ``blend'' driver of MOOG to determine the Ti abundance when applying the EW method to the \ion{Ti}{i} lines. We also took into account the isotopic structure during the synthetic spectrum fitting calculations. 

The following steps outline the process we pursued during the iterations to the model atmospheric parameter solution: 

   \begin{enumerate}
      \item Deriving initial \teff\ value from LDR method 
      \item Using EW$_{(\ion{Ti}{ii})}$ $-$\logg\ relation for an initial \logg\ value and synthesizing $^{12}$CO lines at and around 15873 \AA\ region to determine an initial C abundance to set the CO contamination
      \item After setting initial atmospheric parameters, deriving the spectral \teff\ by iterating the models until \ion{Fe}{i} abundances show no trend with the $\chi$ of each line
      \item Changing \logg\ until the mean \ion{Ti}{i} and \ion{Ti}{ii} abundances merge in an agreement.
   \end{enumerate}

The iterations were continued until the difference between the values obtained in the previous and next iterations of all atmospheric parameters became insignificant. The resultant model atmospheric parameters of the HIP 54048 and Arcturus were found to be in quite good agreement with the ones reported by \cite{afsar18a} and \cite{ramirez11}, respectively. Later, we applied this method to our targets and found their model atmospheric parameters as listed in Table \ref{tab:mods}. In Figure \ref{fig:LDR}, we also compare our final spectroscopic \teff\ values with the ones obtained from the LDR relations of \cite{Jian19} (metallicity-dependent) and \cite{AfsarLDR}. The mean of differences between final spectroscopic \teff\ values and initial \teffldr\ values is 77 K($\sigma=97$ K), demonstrating general consistency between the temperatures.

\subsection{Model atmosphere comparison}\label{model_comp}

The majority of our targets have been investigated for the first time in this work. Two of our targets (J06362606 and J06364731) have been studied by the Gaia-ESO Survey \citep[GES,][]{GES}. GES is a spectroscopic survey that uses high-resolution spectroscopy and the final data release of GES has been published \citep{GESfinal}. The method that GES has used for high-resolution spectroscopy of FGK-type stars has been introduced in \cite{Smiljanic14}.

\emph{Gaia} also has published atmospheric parameters in its DR3 for millions of stars (\citealt{gaia16,gaiadr3}). Six of our targets have spectroscopic atmospheric parameters from the \emph{Gaia} RVS spectra, while one of them has photometric atmospheric parameters. Only J06363185 has both of those parameters. Unfortunately, J06365277 has no parameters from either source. To calibrate \logg\ and [Fe/H] values, we used calibrations published by \cite{RecioBlancoRVS}. For [Fe/H] calibration, we used [M/H]$_{OC}$ values from Table 3 in that work. Comparisons of \emph{Gaia} parameters and our final model parameters can be found in Table \ref{tab:mods}. 

The average absolute differences between our final parameters and calibrated \emph{Gaia} spectroscopic parameters (for five stars only) are as follows: 86 K for \teff, 0.27 and 0.22 dex for \logg\ and [M/H], respectively. J06363185 has the biggest difference from our models and if we exclude this star, the average of differences are 41 K for \teff, 0.19 and 0.25 dex for \logg\ and [M/H], respectively. Our parameters are in the acceptable range with spectroscopically derived \emph{Gaia} parameters. The one with photometric parameters has significant differences from our model parameters. This mainly comes from the ISM reddening. One should keep in mind that differential reddening in the direction of Tr5 can cause these significant differences. The GES parameters for stars J06362606 and J06364731 are in very good agreement with our atmospheric parameter solutions. For J06362606, the GES parameters are \teff\ = 4390 K, \logg\ = 1.93, and [Fe/H] = $-$0.47. For J06364731, the GES parameters are \teff\ = 4473 K, \logg\ = 1.91, and [Fe/H] = $-$0.53. Especially, the surface gravity is in perfect agreement which is less than 0.1 dex in both cases.

\begin{table*}
\caption{Model atmosphere parameters of our target stars.}
\label{tab:mods}
\begin{tabularx}{\textwidth}{l*{12}{>{\raggedright\arraybackslash}X}}
\hline \hline
Star & $T_{\mathrm{LDR}}$ & $T_{\mathrm{LDR}}$ & \teff & \logg & [M/H] & \teff & \logg & \vmicro & \teff & \logg & [Fe/H] & \vmicro \\
\hline
2MASS$^{\rm a}$ & Jian$^{\rm b}$ & Af\c{s}ar$^{\rm c}$ & \multicolumn{3}{c}{Gaia$^{\rm d}$ Parameters} & \multicolumn{3}{c}{Ti II Relations} & \multicolumn{4}{c}{Final Parameters} \\
\hline
J06361880+0934066 & 3904 & 3718 & 3849 & 0.84 & $-0.94$ & 4105 & 1.26 & 2.65 & 3875 & 0.70 & -0.50 & 2.20 \\
J06361925+0925587 & 4233 & 4211 & 4270 & 1.86 & $-0.16$ & 4447 & 1.75 & 2.11 & 4300 & 1.90 & -0.38 & 1.85 \\
J06362606+0931584 & 4503 & 4548 & 4931$^{\rm e}$ & 2.41$^{\rm e}$ & $-0.07^{\rm e}$ & 4646 & 2.09 & 1.79 & 4350 & 2.00 & -0.47 & 1.95 \\
J06363256+0925083 & 4554 & 4608 & 4411 & 1.76 & $-0.34$ & 4577 & 1.97 & 1.90 & 4425 & 1.85 & -0.45 & 2.05 \\
J06364731+0932596 & 4620 & 4701 & 4530 & 1.48 & $-0.58$ & 4577 & 1.97 & 1.90 & 4625 & 1.95 & -0.35 & 1.95 \\
J06365277+0932270 & 4662 & 4675 & - & - & - & 4584 & 1.98 & 1.89 & 4575 & 1.90 & -0.40 & 1.90 \\
J06363185+0940450 & 4611 & 4557 & 4666 & 2.63 & $-0.30$ & 4704 & 2.20 & 1.70 & 4400 & 2.05 & -0.39 & 1.80 \\
\hline 
\end{tabularx}
\smallskip 
\parbox{180mm}{
\footnotesize
$^{\rm a}$ \cite{2MASS}\\
$^{\rm b}$ \cite{Jian19}\\
$^{\rm c}$ \cite{AfsarLDR}\\
$^{\rm d}$ Parameters from \cite{gaia16, gaiadr3} calibrated according to \cite{RecioBlancoRVS}\\
$^{\rm e}$ Photometric values by \emph{Gaia} \citep{gaia16, gaiadr3}}

\end{table*}

\subsection{Uncertainty analysis for model atmosphere parameters}\label{error1}

To quantify the internal and external uncertainty levels in the model atmosphere parameters, we have followed a similar method described in, for example, \cite{afsar12}.

For the internal uncertainty level estimations, we ran a series of analysis using the spectral data of two stars from our sample: J06364731 and J06363185. The temperature uncertainties were determined by varying the \teff\ values in $\pm$50 K steps until the mean abundance value of the individual \ion{Fe}{i} lines exceeded $\pm$1$\sigma$ level of scattering. This method produced an average uncertainty level of $\sim$150 K for the \teff. Since our line list does not include ionized Fe lines, the \logg\ uncertainty level was estimated by changing the \logg\ value until the $\pm$1$\sigma$ distance between the \ion{Ti}{i} and \ion{Ti}{ii} abundances was achieved, which resulted in a \logg\ uncertainty level of $\sim$0.25 dex. The uncertainty level of the \vmicro\ was found by applying $\pm$0.1 \kmsec\ variations to the \vmicro\ value while keeping other atmospheric parameters fixed. Similarly the \teff\ uncertainty estimation, the behavior of the low and high-excitation \ion{Fe}{i} line abundances were observed during the trials. The level of uncertainty in \vmicro\ was reached as $\sim$0.25 \kmsec\ when the reduced width slope for low and high-excitation \ion{Fe}{i} abundances almost exceeded 1$\sigma$ level of scatter of the mean Fe abundance. 

The level of external uncertainty could only be estimated for the \teff\ by using the LDR and \emph{Gaia} DR3 \citep{gaiadr3} temperatures of our targets (Table \ref{tab:mods}). We compare our spectroscopic temperatures with both LDR temperatures in Figure \ref{fig:LDR}. We estimated the average external uncertainty level as about 170 K by taking the mean of the standard deviations of $\Delta T_{\rm{spec-LDR}} =$ \teffspec$-$\teffldr\ , and $\Delta T_{\rm{spec}-\emph{Gaia}} =$\teffspec $-$ \teffgaia\ differences. Taking into account both internal and external uncertainties, we adopt 160 K as the uncertainty level of the \teff.

\section{Abundance analysis}\label{abund}

\begin{figure*}
\centering
\includegraphics[width=\linewidth]{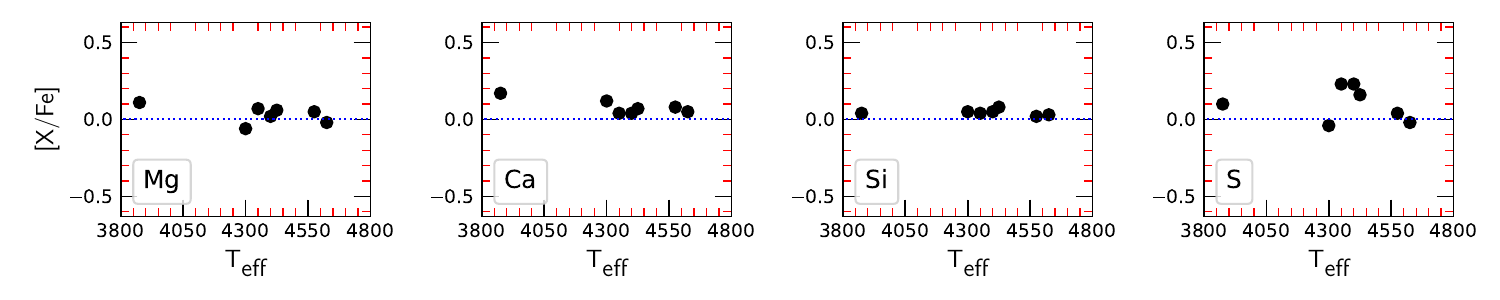}
      \caption{$\alpha$-elements abundances of stars presented in this work. There is no clear evidence of $\alpha$-enrichment in Tr5.}
      \label{fig:alpha}
\end{figure*}

\begin{figure*}
\centering
\includegraphics[scale=0.6]{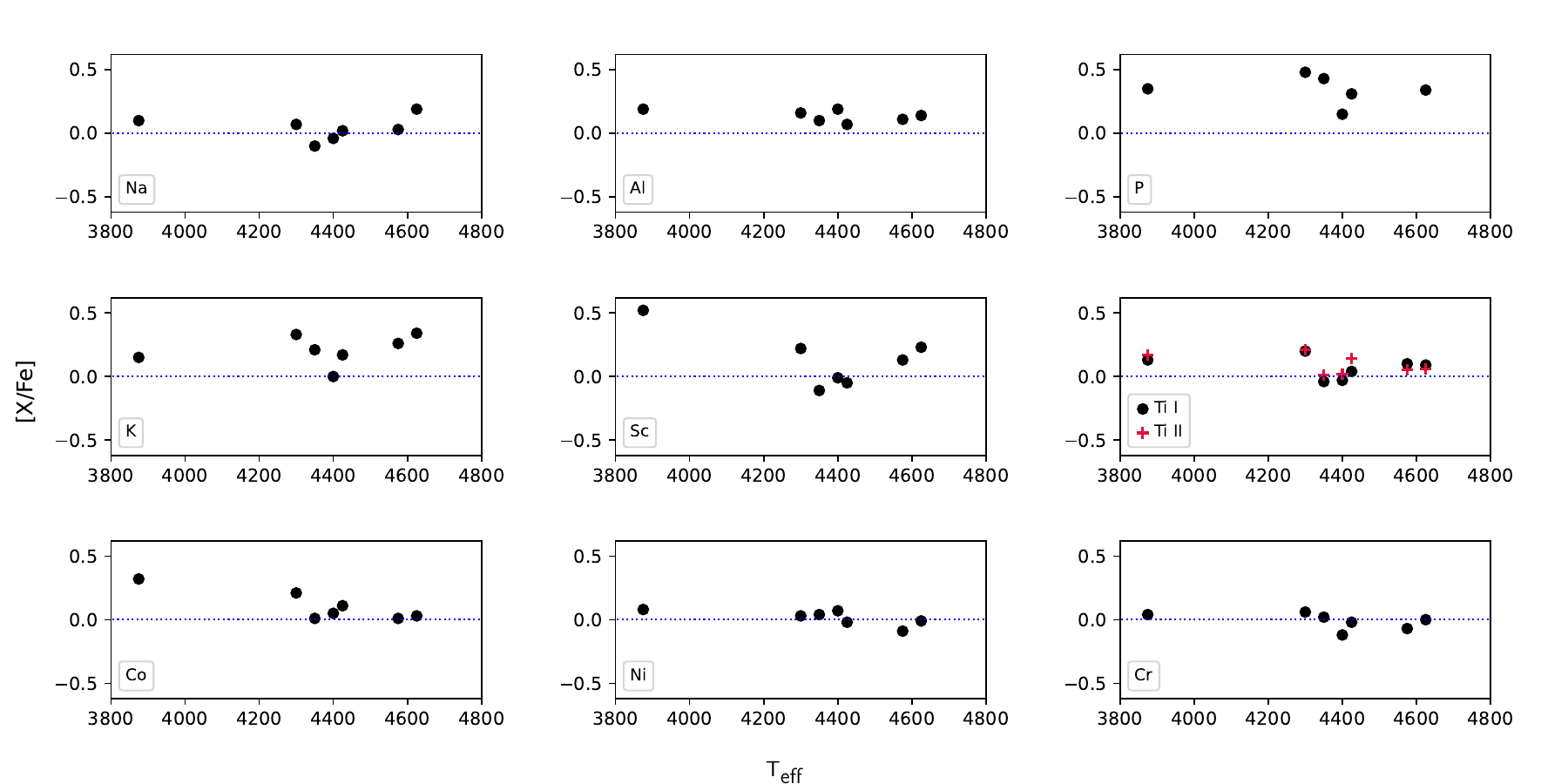}
      \caption{Abundances of different species of stars presented in this work.}
      \label{fig:abund}
\end{figure*}

\begin{figure}
\includegraphics[width=\columnwidth]{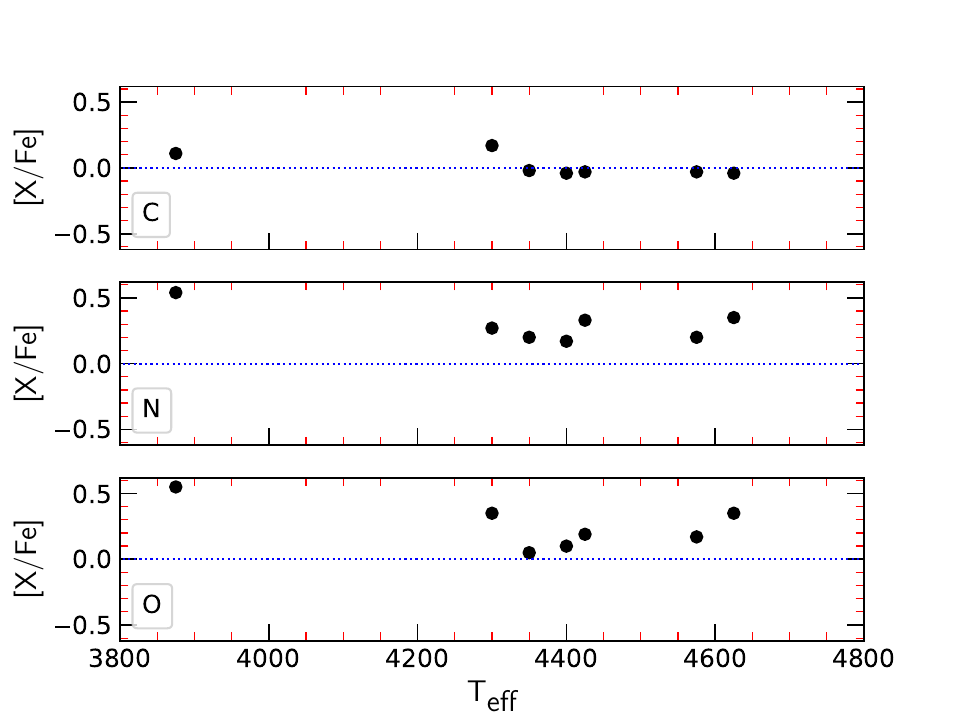}
      \caption{CNO abundances of stars presented in this work.}
      \label{fig:cno}
\end{figure}

High-resolution IGRINS spectra of cool stars are rich in many atomic and molecular features. For the abundance analysis of our stars, we adopted the atomic and molecular line list from \cite{afsar18b}. We determined the elemental abundances of 22 species of 21 elements by applying the spectrum syntheses method. We measured the abundances of $\alpha$-(Mg, Si, S, Ca), odd-Z light (F, Na, Al, P, K), Fe-group (Sc, Ti, Cr, Fe, Co, Ni), n-capture (Ce, Yb), and CNO-group elements. We also used the first overtone ($\nu=2$) band heads of $^{13}$CO (2$-$0) near 23440 \AA\ and $^{13}$CO (3$-$1) near 23730 \AA\ to derive the \carbiso\ ratios for our stars. The differential abundances (relative to the Sun), [X/Fe], \footnote{For elements $A$ and $B$ [$A/B$] = log ($N_{A}$/$N_{B}$)$_{*}$ $-$ log ($N_{A}$/$N_{B}$)$_{\odot}$ and log $\epsilon$($A$) = log ($N_{A}$/$N_{H}$) +12.0,  \citep{Wallerstein59}.} for all species are listed in Table \ref{tab:abnd} along with the \carbiso\ ratios. For the differential abundances, we adopted the solar abundances from \cite{asplund09}. Our results are plotted in Figure \ref{fig:alpha}, \ref{fig:abund} and  \ref{fig:cno}. 

\subsection{$\alpha$ and odd-Z elements}\label{alpha}

$\alpha$-elements are even-Z elements of which their most abundant isotopes have masses as multiples of the mass of He-nucleus, so-called the $\alpha$-particle. Here, we determined the abundances of Mg, Si, S, and Ca of our targets (Table \ref{tab:abnd}). Inspection of Figure 4 immediately suggests that these alpha elements have solar abundance ratios in Tr 5.  We found <$[\alpha/\text{Fe}]$>  = 0.06 $\pm$ 0.02 ($\sigma=$ 0.03) dex.

For all of our targets, we found Mg and Si abundance ratios to be essentially solar. For Mg the cluster average is $\langle$[Mg/Fe]$\rangle$ = 0.03 (Table \ref{tab:abnd}). Mg lines in the \hfilt-band are quite strong, especially in our lower temperature targets. Special attention needs to be given to the NIR region for these lines. As noted in \cite{Osorio20}, the Mg lines in this region may be affected by NLTE conditions. They examined Arcturus, which is a close analog to the stars in our sample, and showed that the difference between LTE and NLTE can reach up to 0.15 dex. Among our targets, J06361880 has the highest Mg abundance ([Mg/Fe] = 0.11). For Si abundance, we found the cluster average is about 0.05 dex, which is relatively solar and consistent in the cluster ($\sigma=$ 0.02).

It is difficult to investigate sulfur abundances in the optical range, due to lack of strong \ion{S}{I} features. But sulfur is critical as a tracer for Galactic nucleosynthesis. Fortunately there are useful \ion{S}{I} lines in the infrared. We found that S has an average value of 0.1 dex. NLTE effects on \ion{S}{I} can be important. \cite{Lucertini2023} studied S abundances of six stars from Tr5 and found 0.05 dex after the NLTE correction. Several works (\citealt{Korotin2009, Taked2016, Lucertini2023}) investigated NLTE effects in the optical region and show that it has an important effect on the derived abundances. S lines in the IR are mostly weak and may be less affected by the NLTE conditions. \cite{afsar18b} showed that the self-consistency in their work is encouraging to use S lines in the near-IR region. Our NIR LTE results are in good agreement with the NLTE results of \citep{Lucertini2023} in the optical region.

Calcium abundances show a similar trend to sulfur abundances around the solar value. We have Ca lines from both \hfilt\ and \kfilt -bands. NLTE analysis for \hfilt-band Ca lines shows that NLTE can create about 10\% better fit than LTE \citep{Osorio20}. <[S/Fe]> and <[Ca/Fe]> are found as 0.10 ($\sigma=$ 0.11) dex and 0.08 ($\sigma=$ 0.05) dex, respectively. In general, J06361880 is the richest star in both Mg and Ca.

We determined the abundances of four odd-Z light elements; Na, Al, P, and K. Most of our targets lie around the solar value with one exception, J06364731 is richer than the Sun but it is still within the uncertainty. J06364731 has the largest Na abundance as 0.19 dex and J06362606 has the lowest Na abundance as $-$0.10 dex. Our line list includes five \ion{Na}{i} lines in the \kfilt -band. Except for J06364731, all lines have been used for the Na abundance determination with relatively low ($<$ 0.05 dex) line-to-line scatter. There is no NLTE analysis done for \kfilt\ -band lines for comparison. However, \cite{Zhou23} showed that NLTE corrections for Na lines in the \hfilt\ -band are small, within 0.05 dex.

There are six \ion{Al}{i} lines present in both \hfilt\ and \kfilt-band and all of them were used to measure Al abundances of our stars. Our analysis shows that the cluster mean for Al is $\langle$[Al/Fe]$\rangle$ = 0.14 ($\sigma$ = 0.05). Star-to-star variation for Al abundance is about 0.05 dex and, unlike Na, is consistent among targets. 

\begin{figure}
\includegraphics[width=\columnwidth]{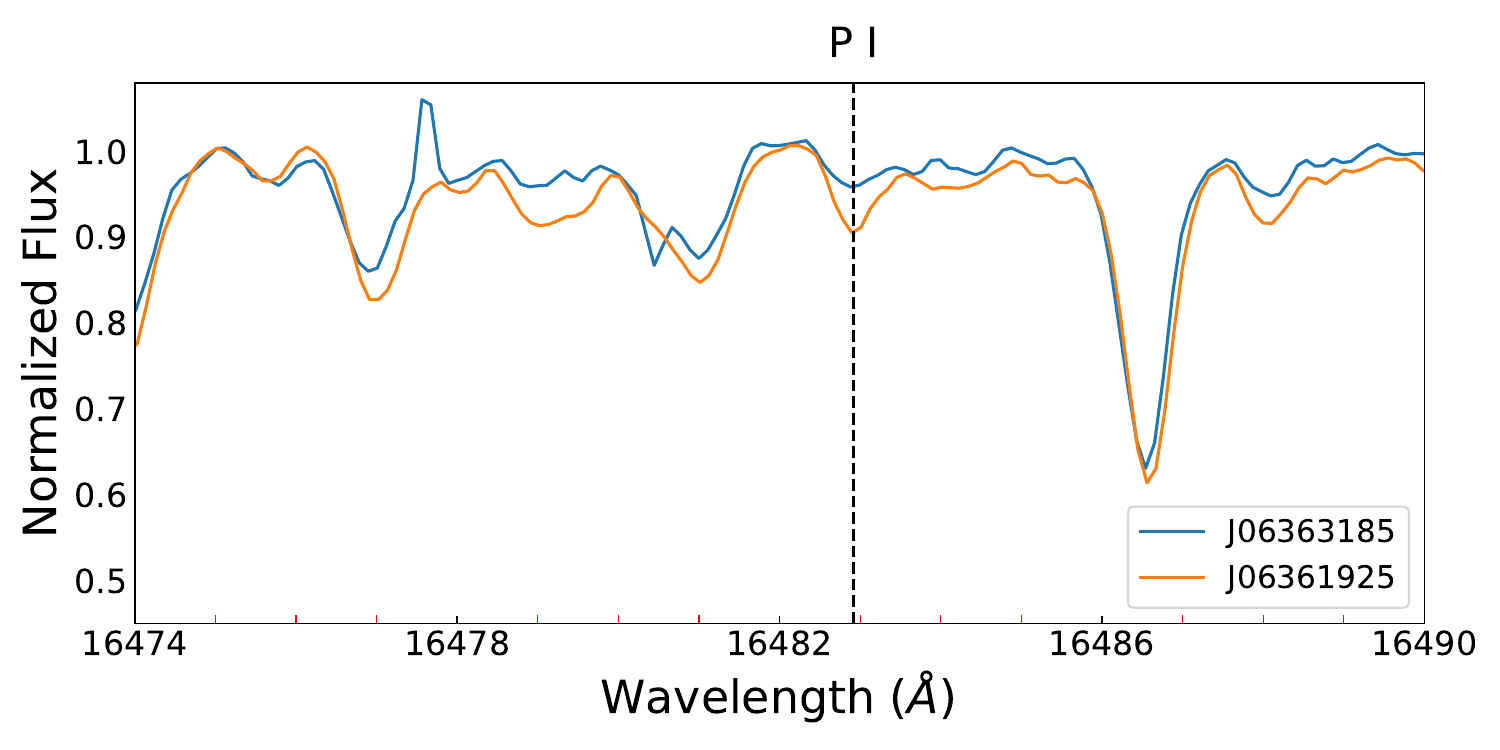}
      \caption{Comparison of P I line in the \hfilt-band for the richest (J06361925) and the poorest (J06363185) stars.}
      \label{fig:P_strong}
\end{figure}

We only have two lines for both \ion{P}{i} (15711.52 and 16482.92  \AA) and \ion{K}{i} (15163.09 and 15168.40 \AA).  We could not measure the abundance of P for J06365277 due to the low S/N around the particular lines. We managed to determine the abundance for only three stars by utilizing the spectral line at 15711.52 \AA. A high internal scatter was observed in the measurements of the two lines. It was noted that this spectral line is significantly affected by telluric line contamination, a finding previously reported by \cite{Nandakumar2022}. Consequently, we excluded this line from our abundance analysis and instead relied on the spectral line at 16482.92 \AA. However, this line also suffers from contamination, specifically from the CO molecular line, as highlighted by \citealt{Nandakumar2022}. Therefore, caution is advised when interpreting the abundance data derived from this line due to the potential for errors in carbon and oxygen abundances to influence the measured phosphorus abundance. We have illustrated the stars with the highest and lowest phosphorus content in Figure \ref{fig:P_strong}. The mean phosphorus abundance for the cluster, denoted as ⟨[P/Fe]⟩ = 0.34, indicates that the phosphorus trends in our sample are consistent with those observed in thin disk stars, as detailed in Figure 4 of \citealt{Nandakumar2022}.

The K abundances of our targets exhibit a similar enrichment pattern as P, averaging at 0.21 dex for the cluster. With the exception of J06363185, we successfully used both \ion{K}{i} lines, and their consistency was confirmed by the low line-to-line scatter. However, for J06363185, we could only derive a solar value based on a single line at 15168.40 \AA. It is crucial to exercise caution when determining abundances since both \ion{K}{i} lines are blended. There is no significant NLTE affects on the K abundances derived from these lines \citep{Osorio20} in the case of Arcturus which is a similar star to those of our sample. The average abundance of odd-Z light elements can be expressed as $\frac{1}{4}$({[Na/Fe]+[Al/Fe]+[P/Fe]+[K/Fe]}) = 0.18 $\pm$ 0.13.

\subsection{Fe-group and Neutron-capture elements}\label{Fe-group}

The abundances of Fe-group elements Sc, Ti, Cr, Fe, Co, and Ni and the abundances of Neutron-capture elements Ce, and Yb were investigated.

We have a total of 27 \ion{Fe}{i} lines in our list, present in both H and K band spectral regions. Fe lines were adopted from \cite{afsar18b}. For our targets, more than 20 Fe lines were used for the analysis. We found <[Fe/H]> = $-$0.43 $\pm$ 0.05 dex for the cluster, which very consistent with literature values \citep[such as $-$0.40 and $-$0.35 found by][respectively]{Carrera07, Viscasillas22}. J06361880 is the most metal poor star in our analysis ([Fe/H]=$-$0.50) This star also has high abundances in some of the heavy elements (e.g. Ti I, Ti II, Sc and Co).

The Fe-group species show values consistent with solar levels, except J06361925 which exhibits distinct variations from the other targets, particularly in the elements Sc, Ti, and Co. Overall, Sc is notably abundant in our targets, with two Sc I lines in the \kfilt-band and a line-to-line scatter below 0.1 dex for our targets. In the \hfilt-band, our line list includes three Cr I lines and six Ni I lines. These lines have been measured for our target stars, and they display high consistency in terms of abundance. The abundances of Cr I and Ni I show a linear trend at the solar value.

We also measured the abundances of two \textit{n}-capture elements: Ce and Yb. All of our targets are rich in neutron-capture elements. Among our targets, J06361925 has the most distinct value of <[\textit{n}-capture/Fe]> = 0.69 dex. Overall, the average \textit{n}-capture abundance for the cluster is about <[\textit{n}-capture/Fe]> = 0.32 dex.

\subsection{The CNO group}\label{cno}
Light elements carbon, nitrogen, and oxygen are important tracers of stellar and Galactic chemical evolution \citep{Salaris2005, Williams2013}. Throughout stellar evolution, those abundances change over time via nucleosynthesis processes, especially the CNO cycles.

We used OH, CO, and CN molecules to determine the abundances of C, N, and O elements \citep[for details see,][]{6940IR, 752IR}. Since C, N and O are bound together through these molecules, we applied an iterative approach to determine their abundance. We started our calculations by using the OH molecules in the \hfilt-band to determine the O abundance. Since the O/C ratio in stars is typically greater than one, the oxygen abundance derived from OH lines is relatively insensitive to the initial carbon abundance. This simplifies the iterative process.

In the \hfilt-band, there are five spectral regions that include useful vibration-rotation lines of OH: 15277-15282 \AA; 15390-15392 \AA; 15504-15507 \AA;15568-15580 \AA\ and 16189-16193 \AA. We investigated and defined a set of clear lines of OH. After O abundances were measured we determined the C abundances from the CO molecular lines. For most of our targets, especially low-temperature ones, we were able to derive the C abundances using CO lines present both in the \hfilt- and \kfilt-bands. As the temperature increases, CO structures in the H-band get weaken, and it becomes more difficult to determine the C abundance from these lines in the H-band. C abundances derived from both bands were in agreement; differences were lower than 0.05 dex. Once the C and O abundances were obtained, we determined the N abundance from the CN bands in the H-band, specifically between 15000 and 15500 \AA. These iteration steps were repeated until the change in abundance of the same species between iterations became negligible.

As can be seen in Figure \ref{fig:cno}, all of our targets have positive [O/Fe] and [N/Fe] abundances. In particular, J06361880 has the largest N and O abundance ratios. For C abundances, J06361880 and J06361925 have unique characteristics. They both have positive values while the others have values around solar (see Table \ref{tab:abnd}). Both stars will be discussed later in detail.

CNO abundances and carbon isotopic ratios are crucial parameters for determining the evolutionary status of stars. \carbiso\ ratios are especially used to explain the mixing process in giant stars \citep[e.g.][]{Lambert1974, Charbonnel94, Gratton2000, Aguilera23, Molaro23}. In the IR region, the \carbiso\ ratios have been derived by previous studies, such as those by \cite{afsar18b} and \cite{6940IR}, using IGRINS spectral data. In this paper, we applied the same spectral synthesis method described by \cite{afsar18b} to determine the \carbiso\ ratios for our targets. In the \kfilt\ band there are two main band heads of the first overtone ($\nu=2$) $^{13}$CO (2$-$0) near 23440 \AA\ and $^{13}$CO (3$-$1) near 23730 \AA. These band heads are easy to determine the \carbiso ratios and our analysis confirms that our target stars from Tr5 are evolved stars. 

We found the mean C, N, and O abundances ([X/Fe]) as 0.02, 0.29, and 0.25 dex, respectively, for Tr5. Also, we found 21 as a mean \carbiso\ ratio. J06364731 and J06365277 have distinct values among our stars. We summarized these abundances in Table \ref{tab:abnd}.

\begin{table*}
\begin{minipage}{180mm} 
 \caption{Relative abundances and \carbiso\ ratios of our target stars from IGRINS spectra.} 
 \label{tab:abnd}           
\begin{tabular}{@{} c c c c c c c c c c @{}}
\hline \hline

Species& J06361880  & J06361925  & J06362606  & J06363256 & J06364731 & J06365277 & J06363185 & Tr5& $\sigma$\\
$[\text{X/Fe}]$ & +0934066 & +0925587   & +0931584  & +0925083  & +0932596  & +0932270  & +0940450 & &  \\
\hline
C (from CO)  & 0.11   & 0.17    & -0.02     & -0.03     & -0.04     & -0.03     & -0.04  & 0.02  & 0.08\\
N (from CN)  & 0.54   & 0.27  & 0.20 & 0.33      & 0.35      & 0.20      & 0.17      & 0.29  & 0.13\\
O (from OH)  & 0.55   & 0.35  & 0.05 & 0.19      & 0.35      & 0.17      & 0.10      & 0.25  & 0.17\\
F (from HF) & 0.19  & 0.21 & -0.14 & 0.13     &   -   &     -      &    -       & 0.10  & 0.16\\
Na I   & 0.10    & 0.07   & -0.10    & 0.02      & 0.19      & 0.03      & -0.04     & 0.04  & 0.10\\
Mg I   & 0.11    & -0.06     & 0.07    & 0.06      & -0.02     & 0.05      & 0.02      & 0.03  & 0.06\\
Al I   & 0.19    & 0.16     & 0.10    & 0.07      & 0.14      & 0.11      & 0.19      & 0.14  & 0.05\\
Si I  & 0.04  & 0.05    & 0.04  & 0.08      & 0.03      & 0.02      & 0.05      & 0.05  & 0.02\\
P I   & 0.35  & 0.48 & 0.43  & 0.31      & 0.34      & -   & 0.15      & 0.34  & 0.11\\
S I  & 0.10 & -0.04   & 0.23  & 0.16      & -0.02     & 0.04      & 0.23      & 0.10  & 0.11 \\
K I   & 0.15   & 0.33  & 0.21   & 0.17      & 0.34      & 0.26      & 0.00      & 0.21  & 0.12\\
Ca I   & 0.17  & 0.12 & 0.04 & 0.07      & 0.05      & 0.08      & 0.04      & 0.08  & 0.05\\
Sc I  & 0.52   & 0.22  & -0.11 & -0.05     & 0.23      & 0.13      & -0.01     & 0.13  & 0.22 \\
Ti I    & 0.13  & 0.20   & -0.04     & 0.04      & 0.09      & 0.10      & -0.03     & 0.07  & 0.09\\
Ti II   & 0.17     & 0.21   & 0.01       & 0.14      & 0.06      & 0.05      & 0.02      & 0.09  & 0.08\\
Cr I   & 0.04    & 0.06   & 0.02 & -0.02     & 0.00      & -0.07     & -0.12     & -0.01 & 0.06\\
Co I  & 0.32  & 0.21  & 0.01  & 0.11      & 0.03      & 0.01      & 0.05      & 0.11  & 0.12\\
Ni I  & 0.08  & 0.03  & 0.04  & -0.02     & -0.01     & -0.09     & 0.07      & 0.01  & 0.06\\
Ce II & 0.24  & 0.69  & 0.31    & 0.27      & 0.31      & 0.16      & 0.28      & 0.32  & 0.17\\
Yb II  & 0.10   & 0.63    & 0.31   & 0.19     & 0.36      & 0.16      & 0.42      & 0.31  & 0.18\\
\carbiso & 21  & 16  & 21  & 16        & 12        & 27        & 20        &    -    &   -  \\
\hline
\end{tabular}
\tablefoot{Tr5 represents the average for our seven targets.}
\end{minipage}
\end{table*}

\subsection{Hydrogen fluoride}\label{HF}
Fluorine synthesis happens in the late stages of stellar evolution, but the details are still poorly understood. \cite{Woosley90} suggest that SN II is the main responsible source for F production. $\alpha$ Boo ([Fe/H]=-0.6, [F/Fe]=0, [F/O]=-0.5) weakens this hypothesis as it shows a solar [F/Fe] abundance ratio in a metal-deficient star \citep{Jorissen92}. He-burning shell can be the main site of F synthesis and AGB stars are good candidates for F synthesis \citep{Li2013, Johnsson2014, Johnsson2017}. Fluorine abundances have been investigated in numerous studies across different components of the Galaxy; such as Galactic disk \citep{Guerco19}, local stellar populations \citep{Brady24}, Galactic nuclear star cluster \citep{Guerco22}, low-metallicity giants \citep{Lucatello11, Li2013}, evolved stars \citep{Abia10, Abia19}, and open clusters \citep{6940IR, Bijavara2024}.

In the visual region, there is a lack of useful F lines but in \kfilt -band there are some rotation-vibration lines of HF. Even though some of them are blended with CN and CO lines, several studies have been focused on HF lines in the NIR region \citep{Pilachowski15, 6940IR, Guerco19, Ryde20, Guerco22, Nandakumar23, Brady24}. The HF at 23358 \AA\ is a relatively unblended transition that has been used as the main F abundance indicator in several previous studies (\citealt{Jorissen92, 6940IR, Guerco19, Guerco22, Nandakumar23, Brady24}). We adopted the line properties from \cite{6940IR} for this line (labeled R9 line in Nandakumar et al.). We also adopted three more lines suggested to use for abundance analysis by \citet{Nandakumar23}. These lines are called R16 (22778  \AA), R18 (22714  \AA) and R19 (22699  \AA) in their work.

\begin{figure}
\includegraphics[width=\columnwidth]{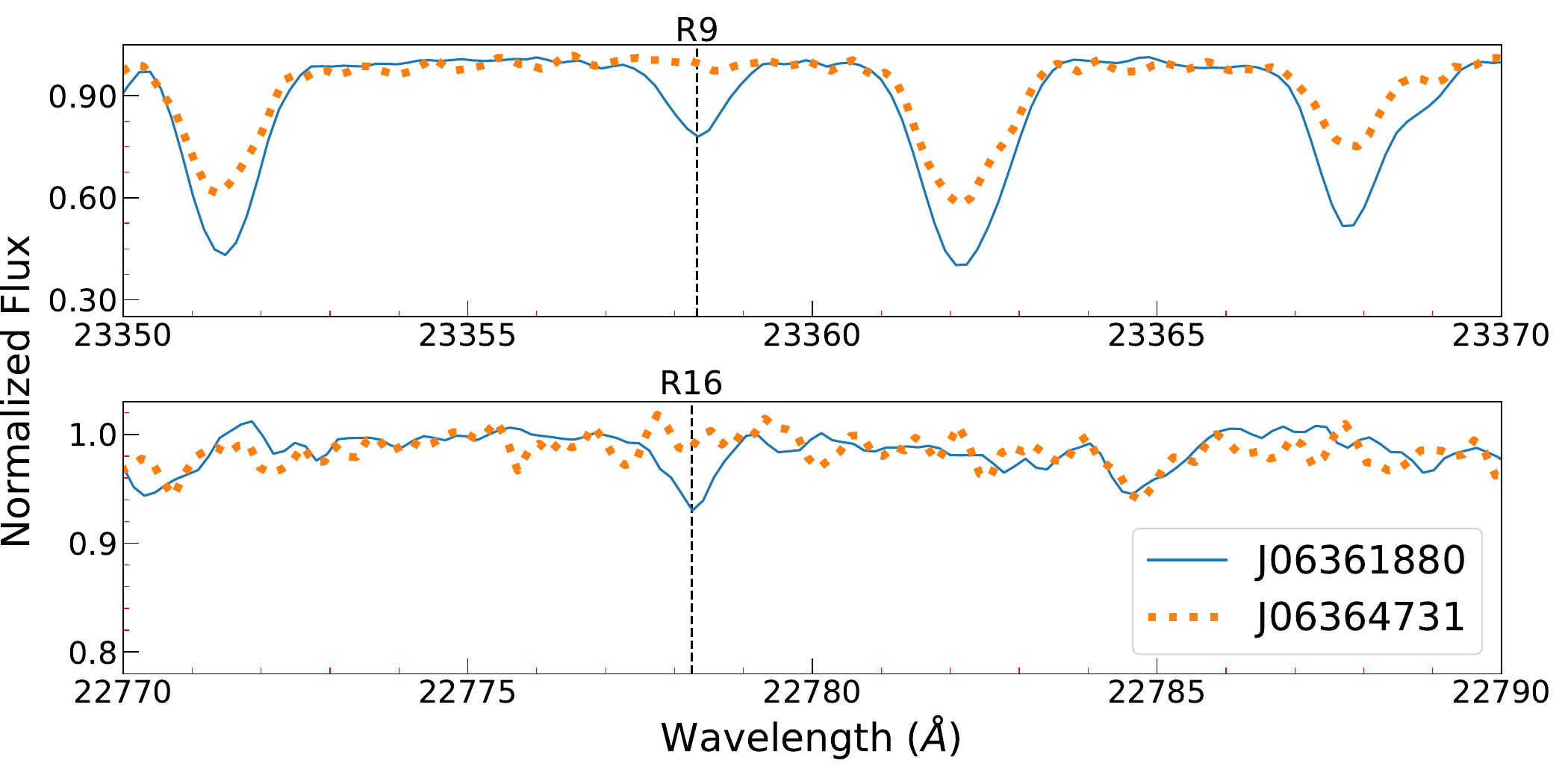}
      \caption{Comparison of HF lines in the \kfilt-band for two of our target stars.}
      \label{fig:HF}
\end{figure}

The F abundance of Trumpler 5 has been studied by \cite{Bijavara2024} along with other open clusters. They used only one star from Tr5 and found [F/Fe]=0.01$\pm$0.09. We were able to derive F abundances for four targets.  We could not measure abundances from R19 and R18 lines since they are blend or weak. We used R9 and R16 for J06361880, J06361925 and J06363256 while we only used R9 for J06362606. One should be careful with HF abundances obtained for J06363256 since its standard deviation is large ($\sigma$= 0.25 dex) while it is consistent for other stars. J06362606 has a negative value of [F/Fe] as $-$0.14 dex. On the other hand, F abundances were found as 0.19, 0.21 and 0.13 dex for J06361880, J06361925 and J06363256, respectively. For our three other targets, we could not measure F abundances because of the weakness of the HF line. The difference between J06361880 and J06364731 can be seen in Figure \ref{fig:HF}.  

\subsection{Sensitivity analysis for line abundances}\label{error2}
\begin{table}[]
    \caption{Sensititivy of our elemental abundances for the changes in the atmospheric parameters.} 
    \label{tab:error}   
    $$
    \begin{array}{cccc}
    \hline \hline
        Species & \Delta \textit{T}_{\rm{eff}} (K) & \Delta \logg & \Delta \vmicro (km/s) \\
                &    +160/-160       &  +0.25/-0.25   &  +0.25/-0.25  \\
                \hline
        \ion{C}{}     & -0.08 / +0.16& -0.01 / +0.07&+0.04 / -0.03 \\
        \ion{N}{}     & -0.15 / -0.07& -0.04 / -0.06&-0.05 / +0.01 \\
        \ion{O}{}     & -0.23 / +0.16& +0.02 / -0.01&-0.01 / -0.01 \\
        \ion{HF}{}    & -0.45 / +0.32& -0.02 / -0.05&-0.02 / -0.03 \\
        \ion{Na}{i}   & -0.14 / +0.14& +0.01 / -0.02&+0.02 / -0.03 \\
        \ion{Mg}{i}   & -0.07 / +0.07& -0.12 / -0.09&-0.10 / -0.15 \\
        \ion{Al}{i}   & -0.16 / +0.16& +0.09 / +0.03&+0.04 / -0.03 \\
        \ion{Si}{i}   & -0.01 / -0.04& -0.12 / -0.01&+0.01 / -0.08 \\
        \ion{P}{i}    & -0.09 / +0.06& -0.19 / +0.16&-0.09 / -0.09 \\
        \ion{S}{i}    & +0.09 / -0.16& -0.08 / +0.08&0.00  / +0.02 \\
        \ion{K}{i}    & -0.06 / +0.15& -0.05 / +0.05&+0.05 / +0.05 \\
        \ion{Ca}{i}   & -0.12 / +0.14& -0.01 / -0.02&+0.02 / -0.03 \\
        \ion{Sc}{i}   & -0.25 / +0.30& -0.01 /  0.00&+0.02 / -0.01 \\
        \ion{Ti}{i}   & -0.24 / +0.26& +0.04 / +0.04&+0.05 / +0.02 \\
        \ion{Ti}{ii}  & -0.02 / -0.07& -0.12 / +0.08&+0.01 / -0.02 \\
        \ion{Cr}{i}   & -0.10 / +0.15& +0.01 / +0.01&+0.03 /  0.00 \\
        \ion{Fe}{i}   & -0.04 / +0.06& -0.02 / +0.05&+0.06 / -0.02 \\
        \ion{Co}{i}   & -0.12 / +0.02& -0.07 / +0.03&-0.02 / -0.02 \\
        \ion{Ni}{i}   & -0.04 / +0.02& -0.01 / +0.06&+0.02 / -0.03 \\
        \ion{Ce}{ii}  & -0.06 / +0.10& -0.07 / +0.16&+0.08 / +0.04 \\
        \ion{Yb}{ii}  & +0.05 / +0.13& -0.10 / +0.18&+0.05 / +0.05 \\
        \carbiso      & 2 / -7       & 3 / -1       & 0 / -2       \\
        \hline
    \end{array}
    $$
\end{table}

We used our error ranges as indicators of the sensitivity for given species. J06362606 has been chosen as the representative of our targets to determine the line uncertainties since it has a higher S/N in the entire spectrum. We changed our model atmosphere parameters in our error ranges described in Section \ref{error1} to follow the response of elemental abundances. In this way, we found how our elemental abundances respond to changes in the stellar parameters. We summarized our results in Table \ref{tab:error}. 

Various atmospheric parameters lead to substantial changes in some of our abundances. Compared to other parameters and typical uncertainty levels, the impact of \vmicro\ is negligible. Notably, \teff\ induces significant abundance variations (exceeding 0.1 dex) in the CNO group, Na I, Al I, Ti I, Sc I, and HF lines, while \logg\ notably influences P I abundances. The most affected abundance is F, with \teff\ exerting a particularly high influence (greater than 0.3 dex) on its levels.

\section{CMD analysis of Trumpler 5}\label{sec:CMD}

\subsection{General considerations}

The usual procedure that is followed when analyzing the CMD of a star cluster that is subject to a relatively low foreground reddening is to deredden the observed colors, transform the relevant apparent magnitudes to the absolute magnitudes that are plotted along the $y$-axis (on the assumption of the best estimate of the apparent distance modulus), and then superimpose isochrones for the observed chemical abundances onto the resultant CMD in order to derive the turnoff (TO) age.  To achieve this superposition, it is frequently necessary to apply a small offset to the predicted colors (e.g., \citealt{Van13}) to compensate for, among other things, errors in the model temperatures and/or the adopted color--\teff\  relations, as well as minor differences in the zero points of the observed and synthetic photometry.  However, when the reddening exceeds $E(B-V) \sim 0.15$ mag or so, the dependence of its effects on the spectral type of a star becomes large enough to significantly affect the morphology of an observed CMD.  As is well known (see, e.g., \citealt{Bessell98}), the reddening caused by a given amount of dust is larger for stars of earlier spectral types; consequently, dust affects the photometry of cluster TO stars more than its giants.

\begin{figure*}
\begin{center}
\includegraphics[width=\textwidth]{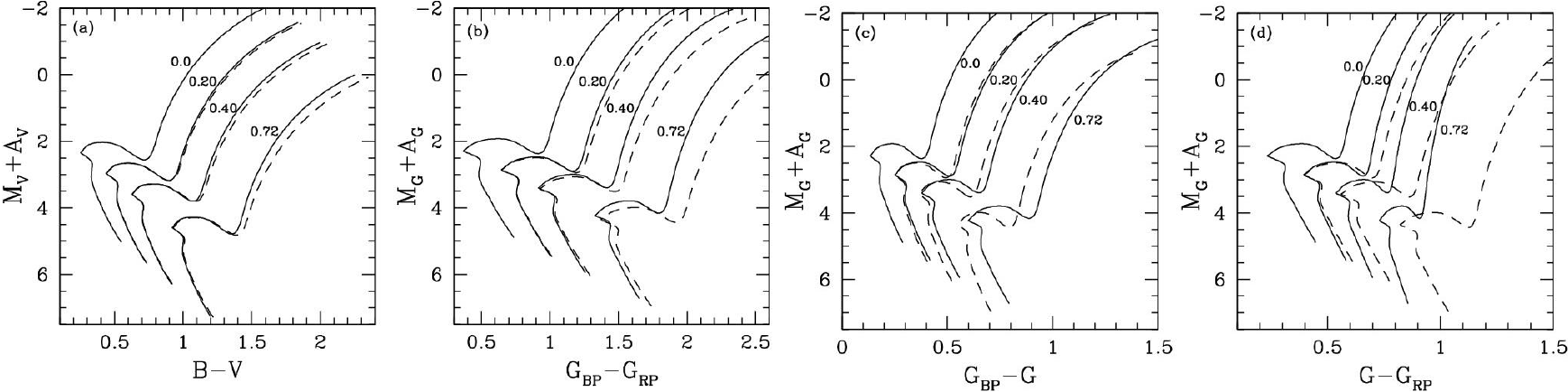}
\caption{Plots on various CMDs of a 2.3 Gyr isochrone for [Fe/H] $= -0.40$, [$\alpha$/Fe] $= 0.0$, and $Y = 0.26$ that employ bolometric corrections which were computed on the assumption of $E(B-V) = 0.0$, 0.2, 0.4, and 0.72. The dashed curves indicate where the same isochrones would be located if it is assumed that the reddening is independent of spectral type.  In these cases, the $R_{\lambda}$ values appropriate to $\approx 7300$~K, nearly unreddened TO stars were assumed to be valid for all values of $E(B-V)$ and applied to all of the stellar models along the isochrone; see the text for details.}
\label{fig:f1}
\end{center}
\end{figure*}

This is illustrated in Figure~\ref{fig:f1}, which plots on several CMDs that have been constructed from Johnson-Cousins $BV$ and \emph{Gaia} photometry, the location of a 2.3 Gyr isochrone for [Fe/H] $= -0.40$, [$\alpha$/Fe] $= 0.0$, and $Y = 0.26$ when it has been transposed to the observed planes using bolometric corrections (BCs) that have been calculated assuming $E(B-V) = 0.0$, 0.2, 0.4, and 0.72.  These BCs, which were obtained by interpolations in the tables provided by \citet{Casagrande14} for the Johnson-Cousins photometric system and by \citet{Casagrande18} for \emph{Gaia} photometry, were generated on the assumption of the standard reddening law (i.e., $R_V = A_V/E(B-V) \approx 3.1$, where $A_V$ is the extinction in the $V$ bandpass; also see \citealt{Cardelli89}).  Note that the values of $E(B-V)$ which were assumed in the Casagrande--VandenBerg computations are so-called {\it nominal} reddenings that apply to early-type stars --- which is the usual convention for many reddening determinations, including those derived by \citet{sfd98} from dust maps.  In this case, for instance, the $B-V$ colors of F-type TO stars are reddened by $\approx 0.92\,E(B-V)$, where $E(B-V)$ is the nominal value of this quantity and the numerical coefficient is equal to the calculated value of $R_B-R_V$ at a temperature of $\approx 7000$~K (\citealt[their Table A1]{Casagrande14}).

\subsection{Application of stellar models to the Tr 5 CMD}
\subsubsection{Scaled solar abundance models}
In this investigation, evolutionary tracks for the main sequence (MS) to the zero-age horizontal branch (ZAHB) phases have been computed using the Victoria code (\citealt{Van24}), while interpolations in the tracks to produce isochrones for ages of interest have been performed using the program described by \citet{Van14}.  Particular care has been taken to treat convective core overshooting in stars that retain convective cores throughout their MS lifetimes given that the morphologies of the TO portions of isochrones relevant to Tr 5 are quite dependent on the extent of core overshooting. Those interested in this aspect of physics are encouraged to refer to previous papers by our group on NGC$\,$6940 (\citealt[hereafter Paper I]{6940IR}) and NGC$\,$752 (\citealt[hereafter Paper II]{752IR}), where the treatment of overshooting and some of the constraints provided by eclipsing binaries and the CMDs of intermediate-age open clusters are discussed at some length (also see \citealt{Van06}).  Little else needs to be said about the stellar models, though it is worthwhile to mention that the stellar evolutionary computations assume the solar mixture of the metals derived by \citet{asplund09}, scaled to the [Fe/H] values of interest, with the possibility of allowing for enhancements in the abundances of the $\alpha$ elements as a group and/or of individual metals (specifically C, N, O, Mg, and Si).  Fully consistent OPAL opacities for stellar interior conditions (\citealt{Iglesias96}) were computed via the OPAL web site\footnote{http://opalopacity.llnl.gov}, along with complementary data for temperatures $\approx 10^4$~K that were generated using the code described by \citet{Ferguson05} for the \citet{Van22} investigation.

The dashed curves in Fig.~\ref{fig:f1} represent the same isochrones as the solid curves except that the effects of reddening are assumed to be independent of spectral type.  To obtain these results, values of $R_{\lambda}$ were calculated using the polynomial fits of $R_{\lambda}$ as a function of \teff\ given by Casagrande \& VandenBerg (see Tables A1 and 2 in their 2014 and 2018 papers, respectively) for the temperature of the hottest stellar model along the isochrone (This model, which is located at the bright end of the blueward ``hook" in the vicinity of the TO, has \teff $= 7385$~K.).  These values of $R_{\lambda}$ were then used to evaluate the shifts in magnitude and color implied by $E(B-V) = 0.2$, 0.4, and 0.72.  This approach preserves the shape of the isochrone while resulting in CMD locations that can be quite different from those obtained when the reddening is properly treated (the solid curves). Of the CMDs considered in Fig.~\ref{fig:f1}, the largest differences are found in the case of $G-G_{RP}$ colors; note, in particular, the large separation between the solid and dashed loci that were generated for $E(B-V) = 0.72$ in panel (d).

\begin{figure*}
\begin{center}
\includegraphics[width=\textwidth]{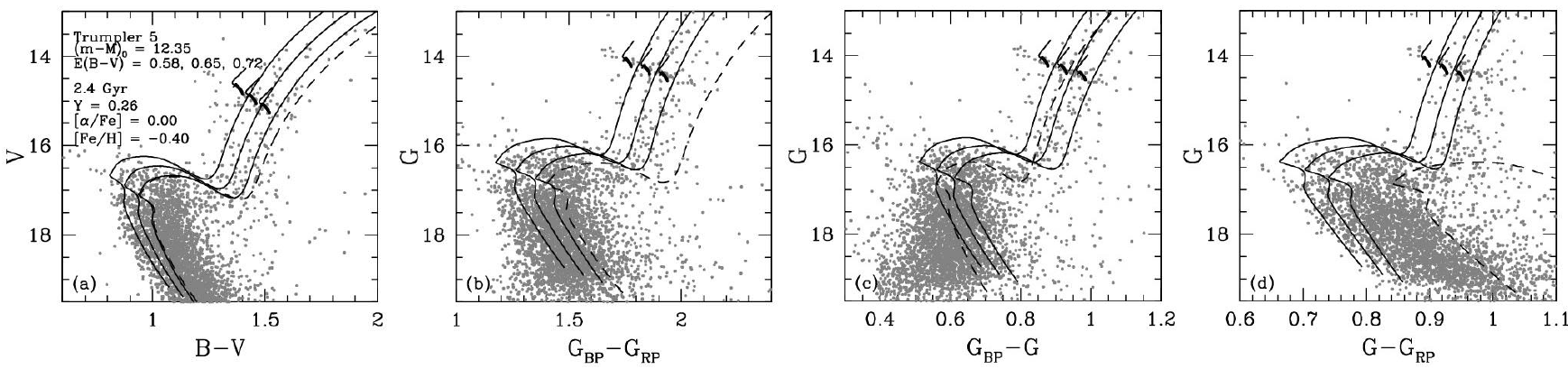}
\caption{Superposition onto the observed Trumpler 5 CMDs of isochrones for the indicated age and chemical abundances, and for E(B-V) values ranging from 0.58 to 0.72 (in the direction from left to right). The dashed loci are identical to the dashed isochrones for $E(B-V) = 0.72$ in the respective panels of Fig.~\ref{fig:f1}. The heavy solid curves indicate the location of ZAHB models with masses ranging from $1.0 {\cal M}_\odot$ to $1.46 {\cal M}_\odot$ in the direction of increasing brightness. The evolutionary track for most of the core He-burning phase of the $1.46 {\cal M}_\odot$ ZAHB model is shown as a thin solid curve originating at the bright end of the ZAHB.}
\label{fig:f2}
\end{center}
\end{figure*}

In Figure~\ref{fig:f2}, 2.4 Gyr isochrones for [Fe/H] $= -0.40$, [$\alpha$/Fe] $= 0.0$, and $Y = 0.26$, assuming $E(B-V) = 0.58$, 0.65, and 0.72, have been superimposed on the CMDs of Tr 5 (Isochrones for the same metal abundances, but with higher O abundances by 0.2 dex, are considered near the end of this section.).  Higher values of $E(B-V)$ are not considered here mainly because the tables of BCs provided by \citet[2018]{Casagrande14} are limited to $E(B-V) \le 0.72$ though extrapolations of the BCs to higher reddenings are carried out later in this analysis.  Note that the ordinate values in the previous figure are readily transformed to apparent $V$ or $G$ magnitudes simply by adding the true distance modulus to the respective values of $M_V+A_V$ or $M_G+A_G$.  In this particular plot, $(m-M)_0$ = 12.35 has been adopted so as to obtain reasonably satisfactory fits of ZAHB models to the observed HB clump stars. These ZAHB models, which were computed following the procedures described by \cite{Van24}, range in mass from $1.0 {\cal M}_\odot$ at the faint end to $1.46 {\cal M}_\odot$ at the bright end.  The most massive model assumes the red-giant tip mass that is predicted by a 2.4 Gyr isochrone if no mass loss is assumed to occur during the ascent of the red giant branch (RGB).  The faintest model would therefore be expected to represent cluster HB stars that have lost as much as $\approx 0.45 {\cal M}_\odot$ prior to reaching the ZAHB.

An evolutionary track for most of the core He-burning phase of the most massive ZAHB model is shown as a thin solid curve.  This was computed using the same version of the MESA code (\citealt{Paxton11}) that was used in our studies of NGC$\,$6940 \citep{6940IR} and NGC$\,$752 \citep{752IR}.  It has already been demonstrated that the Victoria and MESA codes produce nearly indistinguishable evolutionary tracks for all phases of evolution prior to the HB, as well as ZAHB models, when very close to the same physics is adopted (see \citealt{Van16}), and indeed, the respective ZAHB models that were computed for this study for $1.46 {\cal M}_\odot$ differ by only $\Delta\,M_{\rm bol} = 0.003$ mag and $\Delta$\,\teff $= 0.004$ dex, with the Victoria models being slightly brighter and hotter.  These small offsets were applied to the MESA models before transforming them to the observed planes and plotting them in Fig.~\ref{fig:f2}.\footnote{The MESA code has been used to generate core He-burning tracks instead of the Victoria code because the latter has never been developed to deal with the semi-convection process that occurs during the evolution of HB stars.  This choice obviously has no significant impact on our findings; i.e., the very small offsets between the MESA and Victoria ZAHB models are completely inconsequential.}

The vertical spread of the HB stars in Tr 5 appears to be considerably larger on the $B-V,\,V$ diagram than on the CMDs that use $G$ magnitudes for the ordinate parameter.  In fact, given the ranges in $V$ or $G$ that are spanned by the ZAHB and post-ZAHB tracks, one might expect to see more of a scatter in the vertical direction, at a given color, than that shown in the plots of \emph{Gaia} photometry.  The apparent extension of the observed HB to appreciably bluer colors than the model predictions for $E(B-V) = 0.58$ may also be a concern.  Since, in general, higher He abundances tend to produce bluer loops during core He-burning (at least at lower metallicities), similar computations to those shown in Figs.~\ref{fig:f1} and~\ref{fig:f2} were carried out for an initial value of $Y = 0.28$ to explore this possibility.  However, it turns out that the resultant HB tracks (not shown) have CMD locations and morphologies that differ only slightly from those for $Y = 0.26$.  Presumably, the most probable explanation of the bluest HB stars is that they have lower reddenings than the majority of the cluster members.  It is, anyway, much more reasonable that Tr 5 would have an initial He abundance close to $Y = 0.26$ since $Y \approx 0.257$ is favored for 47 Tuc (see, e.g., \citealt{Brogaard17}, \citealt{Denissenkov17}), which has a lower metallicity by only $\sim 0.2$--0.3 dex. Due to Galactic chemical evolution, the He abundances of stellar populations are expected to increase slowly with rising metallicities.

Although the reddening vector implied by the HB clump stars appears to be in quite good agreement with the slope defined by the ZAHBs for different values of $E(B-V)$ (especially in panels b and c), there are clearly zero-point and systematic differences between the isochrones and the observed MS stars. Whereas the isochrones are roughly centered on the distribution of MS stars in panels (b), they are redder than most of the observed stars in panels (c) and offset to the blue in panels (a) and (d).  Moreover, the predicted and observed MS slopes are especially discrepant in panel (c), less so in panel (b). However, the photometric survey carried out by \emph{Gaia} is not very deep and it seems likely that the anomalous slopes are due mostly to photometric errors in the $G_{BP}$ magnitudes at $G \geq 17.5$.  Importantly, no such difficulties are apparent in panel (a), which plots the upper MS stars from the much deeper CMD obtained by \citet{Donati15}, which is well defined down to $V \sim\ 21$.  It is also worth mentioning that stellar models match the MS slopes defined by field Pop. II stars and globular cluster CMDs very well at all metallicities (see \citealt{Van23}); the same study shows that the model \teff\ scale is supported by the best estimates of the temperatures of local subdwarfs.

\begin{figure}
\includegraphics[width=\columnwidth]{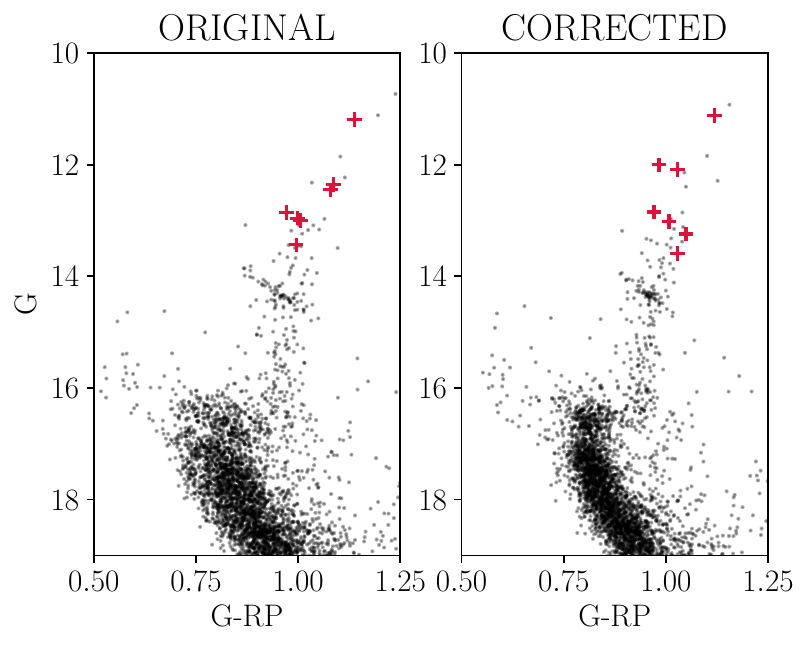}
      \caption{Comparison between the original $G$ vs.\,$G-G_{\rm RP}$ before (left) and after (right) the correction for differential reddening. Spectroscopic targets studied in this work are plotted as red markers.}
      \label{fig:Tr05gaia}
\end{figure}

One cannot help but wonder if some of the aforementioned problems would be reduced if the reddening law that has been assumed in the computation of BCs were altered in some way.  It is interesting that the offsets between the solid and dashed loci for high reddenings in panels (c) and (d) of Fig~\ref{fig:f1} resemble the differences between the isochrones and the observed CMDs in the corresponding panels of Fig.~\ref{fig:f2}.  [Since tables of BCs similar to those provided by \citet{Casagrande14}, but for variations in the assumed reddening law, are not available, this speculation cannot be easily investigated at the present time.  Because the reddening law plays such a critical role in the modeling of the CMDs of highly reddened stellar populations, further studies of this issue should be undertaken.]  On the other hand, the BCs for the $G_{BP}$ passband are likely to be somewhat less trustworthy than those for $G$ and $G_{RP}$, especially in the case of cooler stars, because the former extends well into the ultraviolet where the blanketing is especially strong and therefore considerably more difficult to reproduce with synthetic spectra.  For this reason, the fitting of isochrones to $B-V$ and $G-G_{RP}$ colors should provide the most trustworthy comparisons between theory and observations. Indeed, the apparent consistency of these fits is very encouraging; panels (a) and (d) of Fig.~\ref{fig:f2} both suggest that the majority of the cluster stars are subject to reddenings greater than $E(B-V) = 0.72$.  

All of the CMDs in Fig.~\ref{fig:f2} show clear evidence for a gap in the distribution of stars close to the MS turnoff.  Stars that possess convective cores throughout their MS lifetimes undergo rapid contractions at central H exhaustion in order to ignite H-burning shells around their helium cores. This contraction manifests itself as a blueward hook in both the evolutionary tracks and the isochrones that are derived from them.  Because the contraction is very rapid, one expects to find no more than a few, if any, stars between the red and blue extremities of the hook. The fact that this feature coincides quite well (though not perfectly in some CMDs) with the observed gaps indicates that the age of Tr 5 is close to 2.4 Gyr (if it has the assumed distance and chemical abundances).

It should be possible to improve upon this estimate of the cluster age by correcting the photometry for differential reddening, thereby producing much tighter CMDs for the fitting of isochrones. For this purpose, we employed the procedure outlined by \citep[][see their Section 3]{Milone12}. In essence, we initially generated a map of differential reddening across the field of view of Trumpler\,5 and applied it to correct the photometry. To achieve this, we first defined the fiducial line of cluster stars along the main sequence (MS), the sub giant branch (SGB), and the lower red giant branch (RGB). Subsequently, we selected a sample of cluster members as reference stars and calculated, for each of them, the distance from the fiducial line along the reddening axis. To derive the direction of the reddening vector, we employed the extinction rates of \emph{Gaia} DR2 bands as provided by \cite{Casagrande18}.  Reference stars were exclusively selected from those along the bright MS, the SGB, and the faint RGB. Indeed, the wide angle between the  corresponding fiducial line portion and the reddening vector allows us to disentangle the color and magnitude shift due to differential reddening from the effects of photometric uncertainties.

We adopted the median distance of the nearest 35 reference stars as the assumed differential reddening for each star, while the corresponding uncertainty was determined as the 68.27th percentile of the 35 distances divided by the square root of 34. Notably, in the determination of the differential reddening, each reference star was excluded from the calculation for its own value. The reddening values were then converted into absorption in the G and G$_{RP}$ bands using the relations by \cite{Casagrande18}. To correct for differential reddening in each band, the absorption corresponding to each star was subtracted from its magnitude. The CMD corrected for differential reddening is compared with the original CMD in Figure\,\ref{fig:Tr05gaia}.

\begin{figure*}
\begin{center}
\includegraphics[width=0.75\textwidth]{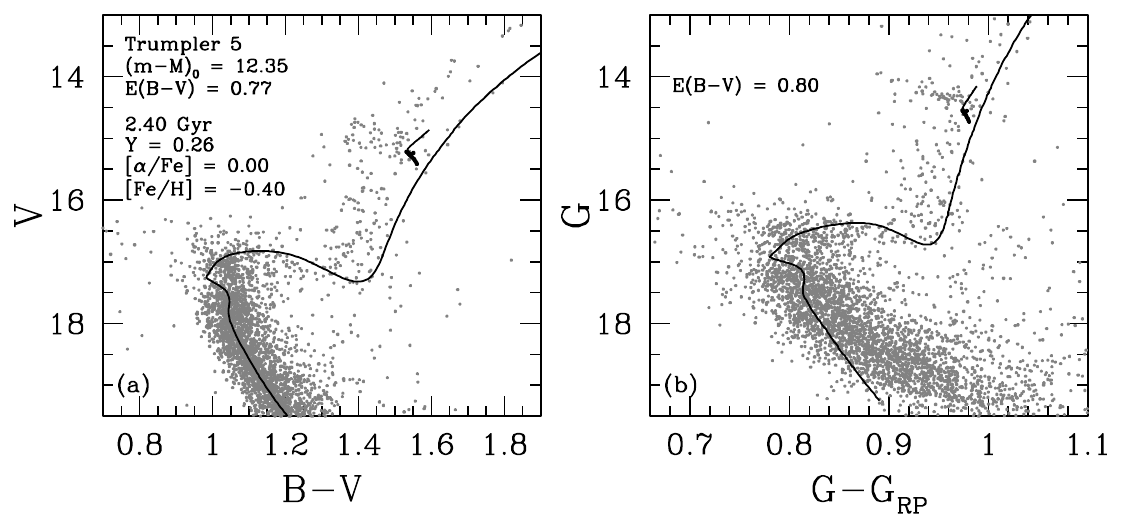}
\caption{Superposition of isochrones for the indicated properties onto the CMDs of Trumpler 5 that have been corrected for differential reddening.  The isochrones were obtained by extrapolating the BCs along the isochrones for $E(B-V) = 0.58$, 0.65, and 0.72 to values consistent with the adopted reddenings.}
\label{fig:f3}
\end{center}
\end{figure*}

The resultant CMDs appear in Figure~\ref{fig:f3}. Though they are significantly tighter than their counterparts in the previous figure, the scatter along the MS is still quite large and the definition of the MS gap has been improved only slightly, if at all.  It is odd that the slope of the MS stars in panel (b) appears to be much shallower than in Fig.~\ref{fig:f2}d, where the MS is bracketed quite well by the solid and dashed isochrones for $E(B-V) = 0.72$.  Although this figure gives the visual impression that the density of MS stars is quite uniform at $V\geq 17$, the median fiducial sequence must have a fairly shallow slope, which would be an indication of photometric errors at faint magnitudes.  The large spreads in the colors of the HB clump stars are also puzzling as the models would suggest that they should span small ranges in color similar to those of the ZAHBs themselves.  At least the luminosity spreads of the core He-burning stars are comparable to the model predictions, though much more so in panel (a) than in panel (b).

To produce isochrones for $E(B-V) > 0.72$, the BCs for $E(B-V) = 0.58$, 0.65, and 0.72 at each point along the isochrone were fitted by quadratics that were subsequently used to extrapolate the BCs to higher reddenings.  [This can be done quite reliably for modest increases in $E(B-V)$ above $E(B-V) = 0.72$ because the BCs are smooth functions of $E(B-V)$.]  Insofar as the resultant isochrones are concerned: the faint end of the hook coincides quite well with the bottom of the observed gap (in both panels), which confirms the indications from Fig.~\ref{fig:f2} that Tr 5 has an age of $\approx 2.4$ Gyr.  In order to match the location of the cluster stars in the vicinity of the TO, it is necessary to adopt $E(B-V) \approx 0.77$ when considering BV photometry, whereas the fits to \emph{Gaia} observations favor a somewhat higher value by $\approx 0.03$ mag.  However, this level of consistency is really quite satisfactory given the possibility of minor zero-point errors in the observed and synthetic photometry and, in particular, the uncertainties in the predicted BCs for different filters at high reddenings.  Errors in the model \teff\ scale could well be mostly responsible for the apparent offsets between the predicted and observed giant branches.  We note, finally, that the isochrones would do a better job of matching the morphology of the gap if the hook feature was more vertical, which could be an indication of deficiencies in the treatment of convective core overshooting. 

\begin{figure*}
\begin{center}
\includegraphics[width=0.75\textwidth]{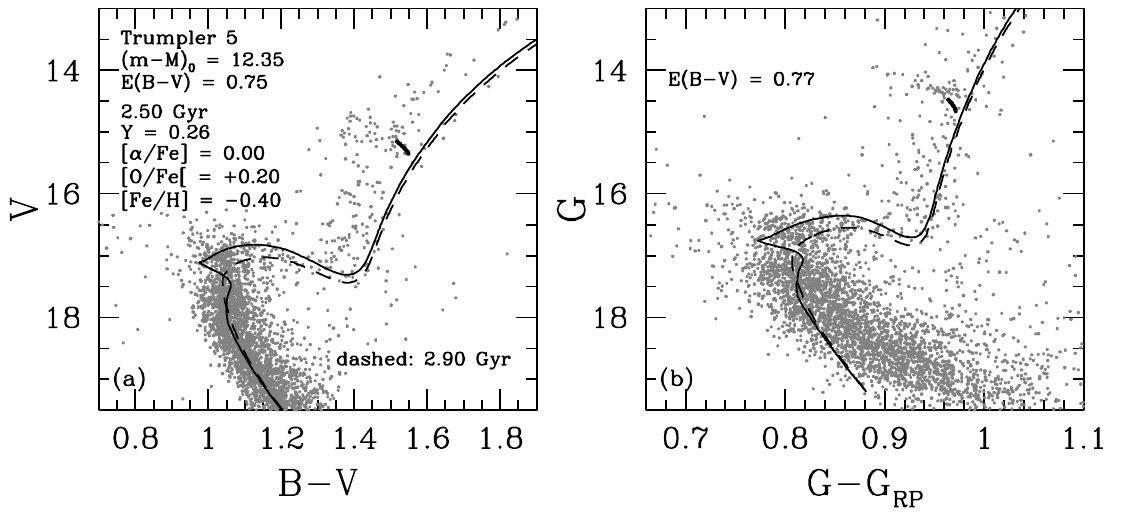}
\caption{As in the previous figure, except that the isochrones assume [O/Fe] $= 0.2$, from which slightly reduced reddening are inferred along with a 0.1 Gyr in the cluster age (to 2.5 Gyr).  The dashed curve shows that MS gaps are not predicted at ages of 2.9 Gyr or higher.}
\label{fig:f4}
\end{center}
\end{figure*}

\subsubsection{Models with enhanced O abundances}
Because the measured O abundance is [O/Fe] $= +0.25$ (recall Section \ref{cno}), a complementary set of stellar models was computed for the same values of [Fe/H] and [$\alpha$/Fe], but with enhanced oxygen by 0.2 dex, given that opacities in the form used by the Victoria code are available for this mixture of the metals.  (These opacities span much smaller ranges in temperature and density than those required by the MESA code; consequently, a core He-burning track was not computed for the [O/Fe] $= +0.2$ case, though there is no reason to expect that it would be very different from the one that was generated for [O/Fe] $= 0.0$.)  

As shown in Figure~\ref{fig:f4}, a 0.2 dex enhancement in the abundance of oxygen has relatively minor consequences for the inferred cluster parameters. The ZAHB models clearly have very similar CMD locations as those for [O/Fe] $= 0.0$ since the adoption of $(m-M)_0 = 12.35$ yields comparable fits of both sets of models to the observed HB stars.  Slightly reduced reddenings are inferred from the fits to the cluster TO because evolutionary tracks for higher O are cooler than those without such enhancements. Interestingly, the fits to the observed TO gap favor a small increase in age to 2.5 Gyr, which is a little surprising because higher oxygen normally has the effect of reducing the age at a given TO luminosity.  However, the transition between stars that retain convective cores until central H-exhaustion, and those which do not, occurs at a somewhat lower mass ($\approx 1.18\ {\cal M}_\odot$) in the models for [O/Fe] $= 0.2$ than in those for [O/Fe] $= 0.0$ (by $\approx 0.03\ {\cal M}_\odot$). Because the amount of core overshooting is assumed to increase sharply with increasing mass just above the transition mass before leveling off (see \citealt{Van06}), the extent of core overshooting is larger in the O-enhanced models that are applicable to the TO stars.  It is well known that overshooting affects the luminosity evolution and the lifetimes of MS stars and these effects give rise to a slightly older age when using the O-enhanced isochrones. These results indicate that Tr 5 has a TO mass that is close to $1.2\ {\cal M}_\odot$.

In view of all of the uncertainties at play, perhaps especially those associated with the BCs at high reddenings (recall Fig.~\ref{fig:f1}), one could hardly expect to obtain much better fits to the CMDs of Tr 5 than those shown in Figs.~\ref{fig:f3} and~\ref{fig:f4}.  Indeed, our results provide quite good support for the reddening-corrected BCs provided by \citet[2018]{Casagrande14}. If anything, a slightly older age may be more accurate because the gap feature would be less pronounced, which would tend to improve the fits to the observed CMDs.  However, the cluster age is unlikely to be much higher than our determination of $\approx 2.5$ Gyr because, as shown by the dashed curve in Fig.~\ref{fig:f4}, isochrones for ages $\geq 2.9$ Gyr (or $\geq 2.7$ Gyr if [O/Fe] $= 0.0$) do not possess blueward hooks near their turnoffs, which means that 2.9 Gyr TO stars are not predicted to have convective cores at central H exhaustion; i.e., they develop radiative cores before reaching this point in their evolution and thus do not undergo a phase of rapid contraction when the central H abundance is depleted.\footnote{The fitting of the 2.9 Gyr isochrone to the Tr 5 CMDs would require the adoption of a smaller distance modulus in order to reproduce the photometry of the SGB stars, as well as a reduced reddening to obtain a satisfactory, simultaneous fit to the upper MS. The main purpose of plotting the older isochrone in Fig.~\ref{fig:f4} is simply to illustrate the differences in the TO morphologies of the 2.5 and 2.9 Gyr isochrones.}
 
A final point: if the dust maps by \citet[hereafter SFD98]{sfd98} and \citet[hereafter SF11]{Schlafly11} are queried\footnote{https://irsa.ipac.caltech.edu/applications/DUST/} for the reddening of Tr 5, one obtains $0.94$ (SDF98) and $0.81$ (SF11)for the mean values of $E(B-V)$, with $1\,\sigma$ uncertainties amounting to about 0.02 mag. In general, the SF11 reddenings are smaller than those given by SDF98 by about 14\%, but differences in the spectral energy distributions of the stars that are used to derive the extinction law appear to be responsible for about half of this reduction (see SF11). As a consequence, the best estimate of the {\it nominal} reddenings should be close to the average of the SDF98 and SF11 determinations, which is $E(B-V) = 0.88$ in the case of Tr 5. As discussed near the beginning of this section, the F-type stars would be reddened by an amount that is approximately 8\% less than this value (i.e., $E(B-V) \approx 0.81$).  Encouragingly, our estimates of the mean reddening from the fits of isochrones to the cluster MS stars are consistent with the reddenings from dust maps to within $\approx 6$\%. Judging from the plot presented in Fig.~\ref{fig:f2}a, the variation in $E(B-V)$ across the face of Tr 5 is $\sim 0.2$ mag (approximately the width of the RGB at $V = 16$).

\section{Discussion}\label{disc}

We used a new approach to derive the model atmospheric parameters of giant stars using only NIR spectra, ideal for stars that are too faint for optical spectroscopy but accessible in the infrared. Our Tr 5 results show strong consistency with the limited literature, validating the reliability of our new approach.

Using the advantages of \emph{Gaia} data, we obtained the detailed kinematic properties of Tr5. We compiled the distances from \emph{Gaia} DR3 and \cite{BailerJ21} that use \emph{Gaia} eDR3. Our CMD analysis derived $(m-M)_{0}$ = 12.35, suggesting that the Tr 5 distance is $\approx 2.5$ Gyr, which is in good agreement with previous studies. We also measured radial velocities from IGRINS spectra of both the \hfilt -band and the \kfilt -band. Our results agree well with the values of \emph{Gaia} and the literature (Figure \ref{fig:RV}). We adopt the membership probabilities from \cite{cantat18}. All of our targets have probabilities >90\%. Kinematic properties have proved that Tr5 lies around at a Galactocentric distance of about 11.5 kpc. This distance shows the importance of Tr5 studies for the chemo-dynamical evolution of the disk and the galaxy since it is located in the region where the radial metallicity gradient changes slope and flattens \citep{Donati15, Magrini2023, Myers22}. We found the metallicity as [Fe/H] = $-$0.42 which is very consistent with the literature.

\begin{figure}
\includegraphics[width=\columnwidth]{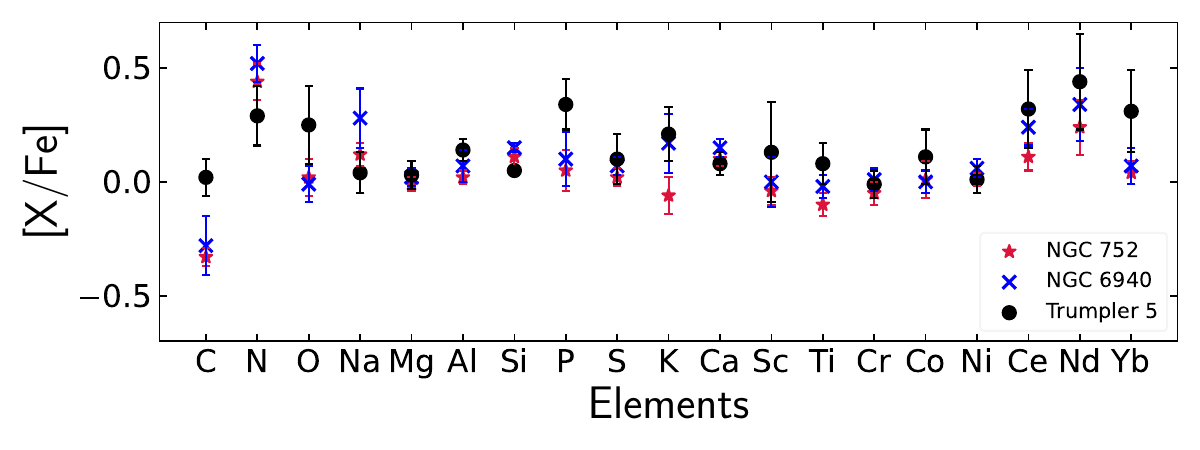}
      \caption{Comparison of Tr5 with other clusters previously studied with IGRINS \citep[NGC 6940 and NGC 752;][respectively]{6940IR, 752IR}.}
      \label{fig:clusters}
\end{figure}

We also compared our results with clusters previously studied by our group using the IGRINS data \citep{6940IR, 752IR}. A comparison of the average cluster abundances of different elements is shown in Figure \ref{fig:clusters}. In general, Tr5 tends to follow the general trend. However, due to lower metallicity of Tr5, some differences arise in some elemental abundances, particularly for C, N, and O. In Tr5, C abundances are around solar, and N anc O abundances are a bit enhanced for our target stars, as expected for stars near this metallicity (e.g. \citealt{Ramirez13,Romano22}).

\begin{figure}
\includegraphics[width=\columnwidth]{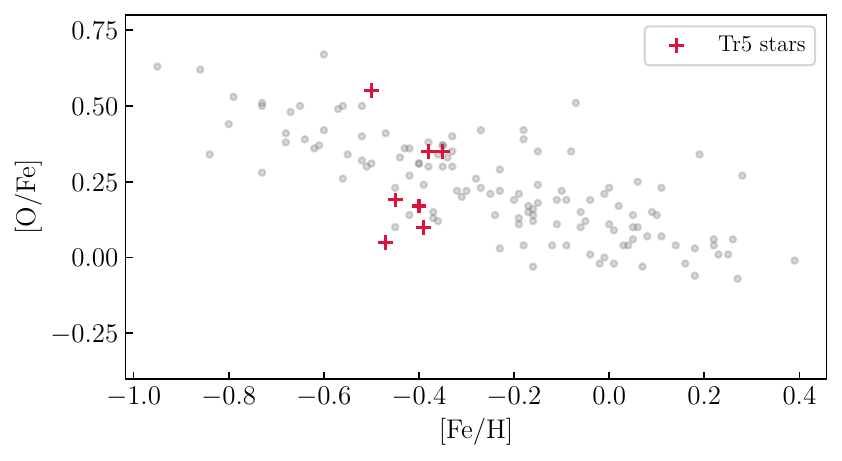}
      \caption{Abundances of O against the metallicity. Spectroscopic targets studied in this work plotted as red markers. Grey points are obtained from \cite{Ramirez13} for local FGK-type disk stars.}
      \label{fig:Oxygen_abundances}
\end{figure}

Our \carbiso\ results confirm that they are giant stars that have an average of \carbiso\ = 19. Comparison of their C and N abundances with similar stars in the same galactocentric distance shows consistent values with the literature \citep{Eilers22}. However, our targets are richer for O abundances compared to \cite{Eilers22}. This can be explained by the lower [Fe/H] of Tr5. Comparison of our abundances of O with \cite{Ramirez13} shows that our stars fall into the spread around the metallicity of Tr5. This can be seen in Figure \ref{fig:Oxygen_abundances}. Our abundances are consistent with the location of Tr5 and its metallicity.

\begin{figure}
\includegraphics[width=\columnwidth]{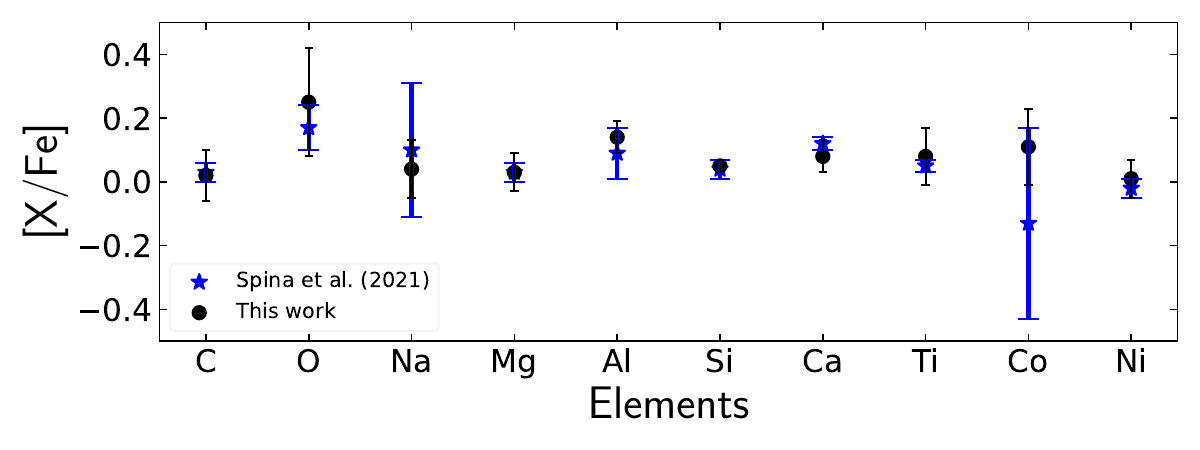}
      \caption{Comparison of average elemental abundances for Tr5 between this work and \cite{Spina2021}.}
      \label{fig:spina_comp}
\end{figure}

\cite{Spina2021} performed membership probabilities and compiled elemental abundances from different surveys. Their sample consists of nine confirmed members from Tr5. They give an average cluster metallicity of [Fe/H] = $-$0.44 ($\sigma$=0.02), which is in quite good agreement with our result: [Fe/H]=$-$0.42, ($\sigma$=0.05). Figure \ref{fig:spina_comp} compares the common elemental abundances of the cluster between the two studies. All elemental abundances are in good agreement except Co, which also shows a very high dispersion in their work.

\begin{figure}
\includegraphics[width=\columnwidth]{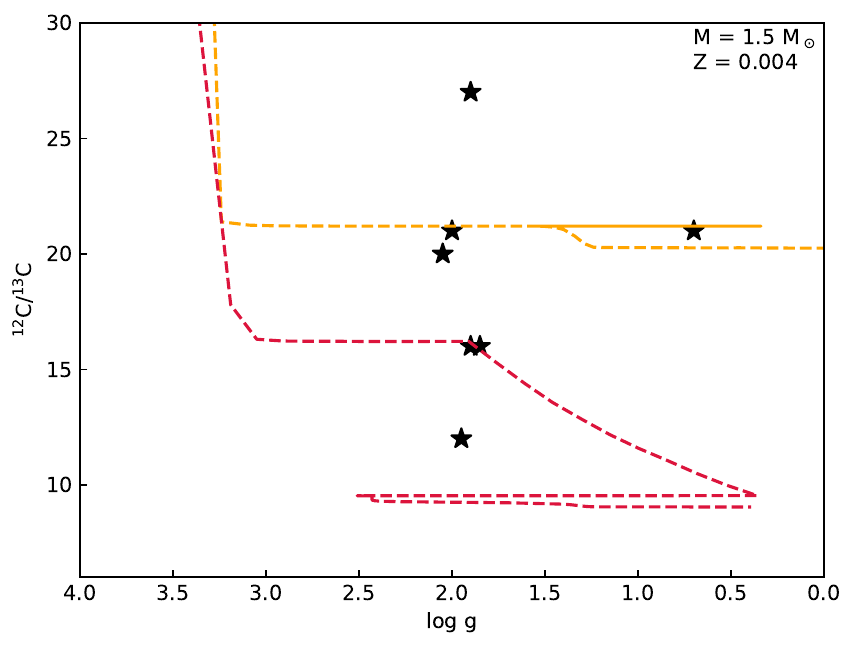}
      \caption{The evolution of the \carbiso\ ratio as a function of log g. Models are taken from \cite{Lagarde2012} for a star with 1.5 \msol\ and z=0.004 metallicity. Orange model shows the evolution in standard prescriptions and red model shows with thermohaline instability and rotation-induced mixing. }
      \label{fig:isotopes}
\end{figure}

We also compared our C and N abundances, along with the \carbiso\ ratios, to 
stellar evolutionary models that incorporate extra-mixing processes in stars. 
The significance of this phenomenon in low-mass stars has been discussed in the 
literature (e.g., by \citealt{Lagarde2012, Lagarde2019, Lagarde24, McCormick23}). 
\cite{Lagarde2019} investigated the impact of thermohaline mixing on C and N 
abundances, and they also suggested that thermohaline mixing could explain the 
observed lower \carbiso\ ratios. In our sample, [C/N] values range from $-$0.10 
to $-$0.43. Given the average metallicity of Tr5 ([Fe/H] = $-$0.42), the models 
from \cite{Lagarde2019} (Figure 8 in \citealt{Lagarde2019}) suggest that the 
[C/N] range of our sample is consistent with both scenarios, with and without the 
effects of thermohaline instability. The same holds also for the \carbiso\ ratios 
(Figure 11 in \citealt{Lagarde2019}). \cite{Lagarde2012} discuss the combined 
impact of rotation-induced mixing and thermohaline instability across a wide 
range of stellar masses and metallicities. We used publicly available 
\cite{Lagarde2012} models to compare our \carbiso\ ratios, which range from 12 to 
27. In Figure \ref{fig:isotopes}, we plot the \carbiso\ ratios from the surface 
abundances provided by \cite{Lagarde2012} for a 1.5 \msol\ star with Z=0.004 
metallicity, including data both with and without the effects of rotation-induced 
mixing and thermohaline instability. Favoring the results presented in Figure 12 
of \cite{Lagarde2019}, the decrease in \carbiso\ ratios can be only explained by 
models that include thermohaline instability. This is also consistent with the 
results found in \cite{Lagarde2019, Lagarde24} for field giants and 
\cite{McCormick23} for open cluster giants.

Abundances of odd-Z and $\alpha$-elements in Tr5 stars are very consistent with other clusters while P abundance shows a significant difference than other clusters with a similar distribution. We used only one line to obtain P abundances which is contaminated with CO molecular line (see Section \ref{alpha}). Value of this enrichment (<[P/Fe]>=0.34) is consistent with field stars \citep{Nandakumar2022}. However, as highlighted in the related section (Section \ref{alpha}), caution is needed when interpreting this result. S abundances obtained from this work agree well with the \cite{Lucertini2023} for Tr5 and Galactic disk stars.

We also measured the abundance of F from HF features, and our analysis of HF lines showed that J06361880 and J06361925 are slightly F-rich. F is produced by the late stages of stellar evolution and mixing helps F reach the surface of the star \citep[][and references therein]{Ryde20}. As expected, four stars from which we inferred the F abundances show a trend similar to that observed for disk stars \cite{Brady24}. Analysis of F gives us a hint to speculate that these four stars have gone through a more complex evolution or evolved more than the rest of the stars.

In this work, we performed the CMD analysis for Tr5 using both BV and \emph{Gaia} magnitudes with and without differential reddening correction. We found that the difference between with and without differential reddening correction is not significant. CMD analysis was performed with solar $\alpha$-element abundance and slightly enhanced O abundance ([O/Fe]=+0.20 dex). The slightly enhanced abundance of O does not have a significant effect on the derived cluster parameters. Reddening value, E(B-V) has been found around 0.76 (See Section \ref{sec:CMD} and Figure \ref{fig:f4}), and the age of the cluster has been suggested as about 2.50 Gyr which is consistent with the literature values and perfectly matches with the age that \cite{Kim03} found (2.4 $\pm$ 0.2 Gyr).

\section{Summary and conclusions}\label{summ}

We conducted a spectroscopic study of the dust-obscured open cluster Tr5, focusing on seven giant stars. While only two of these stars have previously been analyzed with high-resolution spectroscopy, we used the IGRINS instrument to obtain detailed spectra in the H and K bands. To determine the atmospheric parameters of these stars, we studied a new method utilizing infrared spectra. This approach uses the equivalent widths of Ti II lines in the H band to derive initial parameter estimations, and we find the final parameters iteratively. By comparing our results with those obtained from the literature values and LDR temperatures, we found a good agreement, validating the effectiveness of this methodology. Kinematics properties of our targets confirm their memberships. Tr5 has an average distance of $\approx$3 kpc with a mean radial velocity of 50.76 \kmsec\ ($\sigma =$ 0.49 \kmsec ). 

A comprehensive abundance analysis for seven targets chosen from the Tr5 cluster has been performed. To our knowledge, these are the first analysis of these stars from the high-resolution IR spectra. Thanks to the advantages of the IR region and the power of IGRINS, we had a chance to derive abundances of 20 species along with \carbiso\ ratios. IR spectroscopy provides important abundances such as S which is really important for open cluster and the Galactic disk evolution since it is a tracer for Galactic nucleosynthesis. HF abundances and \carbiso\ ratios confirmed that observed stars are RGB stars and indicate a unique star in Tr5, J06361880. Position of this star on the observed CMD and its elemental abundances (enriched in O and F abundances) support that J06361880 is a RGB-tip star.

Isochrones, zero-age horizontal branch (ZAHB) models, and representative core He- burning tracks for [Fe/H]= $-$0.40, [$\alpha$/Fe] = 0.0, and Y = 0.26, with and without enhanced O abundances by 0.2 dex, have been compared with color-magnitude diagrams (CMDs) of Tr 5 that have been constructed from BV and \emph{Gaia} photometry. Analyses of both the observed and differential-reddening-corrected CMDs yield 2.5 Gyr and E(B-V) $\approx$ 0.76 for the best estimates of the cluster age and the mean nominal reddening, respectively, on the assumption of (m-M)$_0$ = 12.35, which is consistent with the predictions of ZAHB models and in good agreement with previous determinations. The importance of using bolometric corrections that have been calculated for the observed values of E(B-V) in highly reddened systems is highlighted.

\begin{acknowledgements}

Our work has been supported by The Scientific and Technological Research Council of Turkey (T\"{U}B\.{I}TAK, project No. 117F497). S. \"Ozdemir acknowledges support from the National Science Centre, Poland, project 2019/34/E/ST9/00133. S. \"Ozdemir would like to express his heartfelt gratitude to his psychologist, Nehir Ulusoy, for her unwavering support and guidance. S. \"Ozdemir also thanks R. Smiljanic and J.~E.~Mart\'{\i}nez Fern\'andez for the fruitful discussions and their support. This research has made use of the SIMBAD database, operated at CDS, Strasbourg, France.  This work used the Immersion Grating Infrared Spectrometer (IGRINS) that was developed under a collaboration between the University of Texas at Austin and the Korea Astronomy and Space Science Institute (KASI) with the financial support of the Mt. Cuba Astronomical Foundation, of the US National Science Foundation under grants AST-1229522 and AST-1702267, of the McDonald Observatory of the University of Texas at Austin, of the Korean GMT Project of KASI, and Gemini Observatory. This paper includes data taken at The McDonald Observatory of The University of Texas at Austin. These results made use of the Lowell Discovery Telescope (LDT) at Lowell Observatory. Lowell is a private, non-profit institution dedicated to astrophysical research and public appreciation of astronomy and operates the LDT in partnership with Boston University, the University of Maryland, the University of Toledo, Northern Arizona University and Yale University. This work has made use of data from the European Space Agency (ESA) mission {\it Gaia} (\url{https://www.cosmos.esa.int/gaia}), processed by the {\it Gaia} Data Processing and Analysis Consortium (DPAC, \url{https://www.cosmos.esa.int/web/gaia/dpac/consortium}). Funding for the DPAC has been provided by national institutions, in particular the institutions participating in the {\it Gaia} Multilateral Agreement.
\end{acknowledgements}

\bibliographystyle{aa}
\bibliography{aa52561-24.bib}

\begin{thebibliography}{123}
\expandafter\ifx\csname natexlab\endcsname\relax\def\natexlab#1{#1}\fi

\bibitem[{{Abdurro'uf} {et~al.}(2022){Abdurro'uf}, {Accetta}, {Aerts}, {Silva Aguirre}, {Ahumada}, {Ajgaonkar}, {Filiz Ak}, {Alam}, {Allende Prieto}, {Almeida}, {Anders}, {Anderson}, {Andrews}, {Anguiano}, {Aquino-Ort{\'\i}z}, {Arag{\'o}n-Salamanca}, {Argudo-Fern{\'a}ndez}, {Ata}, {Aubert}, {Avila-Reese}, {Badenes}, {Barb{\'a}}, {Barger}, {Barrera-Ballesteros}, {Beaton}, {Beers}, {Belfiore}, {Bender}, {Bernardi}, {Bershady}, {Beutler}, {Bidin}, {Bird}, {Bizyaev}, {Blanc}, {Blanton}, {Boardman}, {Bolton}, {Boquien}, {Borissova}, {Bovy}, {Brandt}, {Brown}, {Brownstein}, {Brusa}, {Buchner}, {Bundy}, {Burchett}, {Bureau}, {Burgasser}, {Cabang}, {Campbell}, {Cappellari}, {Carlberg}, {Wanderley}, {Carrera}, {Cash}, {Chen}, {Chen}, {Cherinka}, {Chiappini}, {Choi}, {Chojnowski}, {Chung}, {Clerc}, {Cohen}, {Comerford}, {Comparat}, {da Costa}, {Covey}, {Crane}, {Cruz-Gonzalez}, {Culhane}, {Cunha}, {Dai}, {Damke}, {Darling}, {Davidson}, {Davies}, {Dawson}, {De Lee}, {Diamond-Stanic}, {Cano-D{\'\i}az}, {S{\'a}nchez},
  {Donor}, {Duckworth}, {Dwelly}, {Eisenstein}, {Elsworth}, {Emsellem}, {Eracleous}, {Escoffier}, {Fan}, {Farr}, {Feng}, {Fern{\'a}ndez-Trincado}, {Feuillet}, {Filipp}, {Fillingham}, {Frinchaboy}, {Fromenteau}, {Galbany}, {Garc{\'\i}a}, {Garc{\'\i}a-Hern{\'a}ndez}, {Ge}, {Geisler}, {Gelfand}, {G{\'e}ron}, {Gibson}, {Goddy}, {Godoy-Rivera}, {Grabowski}, {Green}, {Greener}, {Grier}, {Griffith}, {Guo}, {Guy}, {Hadjara}, {Harding}, {Hasselquist}, {Hayes}, {Hearty}, {Hern{\'a}ndez}, {Hill}, {Hogg}, {Holtzman}, {Horta}, {Hsieh}, {Hsu}, {Hsu}, {Huber}, {Huertas-Company}, {Hutchinson}, {Hwang}, {Ibarra-Medel}, {Chitham}, {Ilha}, {Imig}, {Jaekle}, {Jayasinghe}, {Ji}, {Johnson}, {Jones}, {J{\"o}nsson}, {Katkov}, {Khalatyan}, {Kinemuchi}, {Kisku}, {Knapen}, {Kneib}, {Kollmeier}, {Kong}, {Kounkel}, {Kreckel}, {Krishnarao}, {Lacerna}, {Lane}, {Langgin}, {Lavender}, {Law}, {Lazarz}, {Leung}, {Leung}, {Lewis}, {Li}, {Li}, {Lian}, {Liang}, {Lin}, {Lin}, {Lin}, {Lintott}, {Long}, {Longa-Pe{\~n}a}, {L{\'o}pez-Cob{\'a}}, {Lu},
  {Lundgren}, {Luo}, {Mackereth}, {de la Macorra}, {Mahadevan}, {Majewski}, {Manchado}, {Mandeville}, {Maraston}, {Margalef-Bentabol}, {Masseron}, {Masters}, {Mathur}, {McDermid}, {Mckay}, {Merloni}, {Merrifield}, {Meszaros}, {Miglio}, {Di Mille}, {Minniti}, {Minsley}, \& {Monachesi}}]{Apogee2022}
{Abdurro'uf}, {Accetta}, K., {Aerts}, C., {et~al.} 2022, \apjs, 259, 35

\bibitem[{{Abia} {et~al.}(2019){Abia}, {Cristallo}, {Cunha}, {de Laverny}, \& {Smith}}]{Abia19}
{Abia}, C., {Cristallo}, S., {Cunha}, K., {de Laverny}, P., \& {Smith}, V.~V. 2019, \aap, 625, A40

\bibitem[{{Abia} {et~al.}(2010){Abia}, {Cunha}, {Cristallo}, {de Laverny}, {Dom{\'\i}nguez}, {Eriksson}, {Gialanella}, {Hinkle}, {Imbriani}, {Recio-Blanco}, {Smith}, {Straniero}, \& {Wahlin}}]{Abia10}
{Abia}, C., {Cunha}, K., {Cristallo}, S., {et~al.} 2010, \apjl, 715, L94

\bibitem[{{Af{\c s}ar} {et~al.}(2018{\natexlab{a}}){Af{\c s}ar}, {Bozkurt}, {B{\"o}cek Topcu}, {Casetti-Dinescu}, {Sneden}, \& {{\c S}ehito{\v g}lu}}]{afsar18a}
{Af{\c s}ar}, M., {Bozkurt}, Z., {B{\"o}cek Topcu}, G., {et~al.} 2018{\natexlab{a}}, \aj, 155, 240

\bibitem[{{Af{\c s}ar} {et~al.}(2018{\natexlab{b}}){Af{\c s}ar}, {Sneden}, {Wood}, {Lawler}, {Bozkurt}, {B{\"o}cek Topcu}, {Mace}, {Kim}, \& {Jaffe}}]{afsar18b}
{Af{\c s}ar}, M., {Sneden}, C., {Wood}, M.~P., {et~al.} 2018{\natexlab{b}}, ApJ, 865, 44

\bibitem[{{Af{\c{s}}ar} {et~al.}(2023){Af{\c{s}}ar}, {Bozkurt}, {Topcu}, {{\"O}zdemir}, {Sneden}, {Mace}, {Jaffe}, \& {L{\'o}pez-Valdivia}}]{AfsarLDR}
{Af{\c{s}}ar}, M., {Bozkurt}, Z., {Topcu}, G.~B., {et~al.} 2023, \apj, 949, 86

\bibitem[{{Af{\c{s}}ar} {et~al.}(2012){Af{\c{s}}ar}, {Sneden}, \& {For}}]{afsar12}
{Af{\c{s}}ar}, M., {Sneden}, C., \& {For}, B.~Q. 2012, \aj, 144, 20

\bibitem[{{Af\c{s}ar} {et~al.}(2016){Af\c{s}ar}, {Sneden}, {Frebel}, {Kim}, {Mace}, {Kaplan}, {Lee}, {Oh}, {Sok Oh}, {Pak}, {Park}, {Pavel}, {Yuk}, \& {Jaffe}}]{Afsar16}
{Af\c{s}ar}, M., {Sneden}, C., {Frebel}, A., {et~al.} 2016, \apj, 819, 103

\bibitem[{{Aguilera-G{\'o}mez} {et~al.}(2023){Aguilera-G{\'o}mez}, {Jones}, \& {Chanam{\'e}}}]{Aguilera23}
{Aguilera-G{\'o}mez}, C., {Jones}, M.~I., \& {Chanam{\'e}}, J. 2023, \aap, 670, A73

\bibitem[{{Asplund} {et~al.}(2009){Asplund}, {Grevesse}, {Sauval}, \& {Scott}}]{asplund09}
{Asplund}, M., {Grevesse}, N., {Sauval}, A.~J., \& {Scott}, P. 2009, \araa, 47, 481

\bibitem[{{Bailer-Jones} {et~al.}(2021){Bailer-Jones}, {Rybizki}, {Fouesneau}, {Demleitner}, \& {Andrae}}]{BailerJ21}
{Bailer-Jones}, C.~A.~L., {Rybizki}, J., {Fouesneau}, M., {Demleitner}, M., \& {Andrae}, R. 2021, \aj, 161, 147

\bibitem[{{Bertelli Motta} {et~al.}(2017){Bertelli Motta}, {Salaris}, {Pasquali}, \& {Grebel}}]{Bertelli17}
{Bertelli Motta}, C., {Salaris}, M., {Pasquali}, A., \& {Grebel}, E.~K. 2017, \mnras, 466, 2161

\bibitem[{{Bessell} {et~al.}(1998){Bessell}, {Castelli}, \& {Plez}}]{Bessell98}
{Bessell}, M.~S., {Castelli}, F., \& {Plez}, B. 1998, \aap, 333, 231

\bibitem[{{Biazzo} {et~al.}(2007{\natexlab{a}}){Biazzo}, {Frasca}, {Catalano}, \& {Marilli}}]{biazzo07a}
{Biazzo}, K., {Frasca}, A., {Catalano}, S., \& {Marilli}, E. 2007{\natexlab{a}}, Astronomische Nachrichten, 328, 938

\bibitem[{{Biazzo} {et~al.}(2007{\natexlab{b}}){Biazzo}, {Pasquini}, {Girardi}, {Frasca}, {da Silva}, {Setiawan}, {Marilli}, {Hatzes}, \& {Catalano}}]{biazzo07b}
{Biazzo}, K., {Pasquini}, L., {Girardi}, L., {et~al.} 2007{\natexlab{b}}, A\&A, 475, 981

\bibitem[{{B{\"o}cek Topcu} {et~al.}(2015){B{\"o}cek Topcu}, {Af\c{s}ar}, {Schaeuble}, \& {Sneden}}]{752vis}
{B{\"o}cek Topcu}, G., {Af\c{s}ar}, M., {Schaeuble}, M., \& {Sneden}, C. 2015, \mnras, 446, 3562

\bibitem[{{B{\"o}cek Topcu} {et~al.}(2016){B{\"o}cek Topcu}, {Af\c{s}ar}, \& {Sneden}}]{6940vis}
{B{\"o}cek Topcu}, G., {Af\c{s}ar}, M., \& {Sneden}, C. 2016, \mnras, 463, 580

\bibitem[{{B{\"o}cek Topcu} {et~al.}(2020){B{\"o}cek Topcu}, {Af\c{s}ar}, {Sneden}, {Pilachowski}, {Denissenkov}, {Vand enBerg}, {Wright}, {Mace}, {Jaffe}, {Strickland}, {Kim}, \& {Sokal}}]{752IR}
{B{\"o}cek Topcu}, G., {Af\c{s}ar}, M., {Sneden}, C., {et~al.} 2020, \mnras, 491, 544

\bibitem[{{B{\"o}cek Topcu} {et~al.}(2019){B{\"o}cek Topcu}, {Af{\c{s}}ar}, {Sneden}, {Pilachowski}, {Denissenkov}, {VandenBerg}, {Strickland}, {{\"O}zdemir}, {Mace}, {Kim}, \& {Jaffe}}]{6940IR}
{B{\"o}cek Topcu}, G., {Af{\c{s}}ar}, M., {Sneden}, C., {et~al.} 2019, \mnras, 485, 4625

\bibitem[{{Brady} {et~al.}(2024){Brady}, {Pilachowski}, {Grisoni}, {Maas}, \& {Nault}}]{Brady24}
{Brady}, K.~E., {Pilachowski}, C.~A., {Grisoni}, V., {Maas}, Z.~G., \& {Nault}, K.~A. 2024, \aj, 167, 291

\bibitem[{{Bragaglia} {et~al.}(2004){Bragaglia}, {Tosi}, {Marconi}, {Di Fabrizio}, {Andreuzzi}, {Carretta}, {Gratton}, \& {Held}}]{Bragaglia04}
{Bragaglia}, A., {Tosi}, M., {Marconi}, G., {et~al.} 2004, \memsai, 75, 28

\bibitem[{{Brogaard} {et~al.}(2017){Brogaard}, {VandenBerg}, {Bedin}, {Milone}, {Thygesen}, \& {Grundahl}}]{Brogaard17}
{Brogaard}, K., {VandenBerg}, D.~A., {Bedin}, L.~R., {et~al.} 2017, \mnras, 468, 645

\bibitem[{{Cantat-Gaudin} {et~al.}(2020){Cantat-Gaudin}, {Anders}, {Castro-Ginard}, {Jordi}, {Romero-G{\'o}mez}, {Soubiran}, {Casamiquela}, {Tarricq}, {Moitinho}, {Vallenari}, {Bragaglia}, {Krone-Martins}, \& {Kounkel}}]{cantat20}
{Cantat-Gaudin}, T., {Anders}, F., {Castro-Ginard}, A., {et~al.} 2020, \aap, 640, A1

\bibitem[{{Cantat-Gaudin} {et~al.}(2018){Cantat-Gaudin}, {Jordi}, {Vallenari}, {Bragaglia}, {Balaguer-N{\'u}{\~n}ez}, {Soubiran}, {Bossini}, {Moitinho}, {Castro-Ginard}, {Krone-Martins}, {Casamiquela}, {Sordo}, \& {Carrera}}]{cantat18}
{Cantat-Gaudin}, T., {Jordi}, C., {Vallenari}, A., {et~al.} 2018, A\&A, 618, A93

\bibitem[{{Cardelli} {et~al.}(1989){Cardelli}, {Clayton}, \& {Mathis}}]{Cardelli89}
{Cardelli}, J.~A., {Clayton}, G.~C., \& {Mathis}, J.~S. 1989, \apj, 345, 245

\bibitem[{{Carrera} {et~al.}(2019){Carrera}, {Bragaglia}, {Cantat-Gaudin}, {Vallenari}, {Balaguer-N{\'u}{\~n}ez}, {Bossini}, {Casamiquela}, {Jordi}, {Sordo}, \& {Soubiran}}]{Carrera19}
{Carrera}, R., {Bragaglia}, A., {Cantat-Gaudin}, T., {et~al.} 2019, \aap, 623, A80

\bibitem[{{Carrera} {et~al.}(2007){Carrera}, {Gallart}, {Pancino}, \& {Zinn}}]{Carrera07}
{Carrera}, R., {Gallart}, C., {Pancino}, E., \& {Zinn}, R. 2007, \aj, 134, 1298

\bibitem[{{Casagrande} \& {VandenBerg}(2014)}]{Casagrande14}
{Casagrande}, L. \& {VandenBerg}, D.~A. 2014, \mnras, 444, 392

\bibitem[{{Casagrande} \& {VandenBerg}(2018)}]{Casagrande18}
{Casagrande}, L. \& {VandenBerg}, D.~A. 2018, \mnras, 479, L102

\bibitem[{{Charbonnel}(1994)}]{Charbonnel94}
{Charbonnel}, C. 1994, \aap, 282, 811

\bibitem[{{Chun}(2020)}]{Chun2020}
{Chun}, S.-H. 2020, in IAU Symposium, Vol. 351, Star Clusters: From the Milky Way to the Early Universe, ed. A.~{Bragaglia}, M.~{Davies}, A.~{Sills}, \& E.~{Vesperini}, 189--191

\bibitem[{{Cutri} {et~al.}(2003){Cutri}, {Skrutskie}, {van Dyk}, {Beichman}, {Carpenter}, {Chester}, {Cambresy}, {Evans}, {Fowler}, {Gizis}, {Howard}, {Huchra}, {Jarrett}, {Kopan}, {Kirkpatrick}, {Light}, {Marsh}, {McCallon}, {Schneider}, {Stiening}, {Sykes}, {Weinberg}, {Wheaton}, {Wheelock}, \& {Zacarias}}]{2MASS}
{Cutri}, R.~M., {Skrutskie}, M.~F., {van Dyk}, S., {et~al.} 2003, VizieR Online Data Catalog, 2246, 0

\bibitem[{{Denissenkov} {et~al.}(2017){Denissenkov}, {VandenBerg}, {Kopacki}, \& {Ferguson}}]{Denissenkov17}
{Denissenkov}, P.~A., {VandenBerg}, D.~A., {Kopacki}, G., \& {Ferguson}, J.~W. 2017, \apj, 849, 159

\bibitem[{{Donati} {et~al.}(2015){Donati}, {Cocozza}, {Bragaglia}, {Pancino}, {Cantat-Gaudin}, {Carrera}, \& {Tosi}}]{Donati15}
{Donati}, P., {Cocozza}, G., {Bragaglia}, A., {et~al.} 2015, \mnras, 446, 1411

\bibitem[{{Dow} \& {Hawarden}(1970)}]{Dow70}
{Dow}, M.~J. \& {Hawarden}, T.~G. 1970, Monthly Notes of the Astronomical Society of South Africa, 29, 137

\bibitem[{{Eilers} {et~al.}(2022){Eilers}, {Hogg}, {Rix}, {Ness}, {Price-Whelan}, {M{\'e}sz{\'a}ros}, \& {Nitschelm}}]{Eilers22}
{Eilers}, A.-C., {Hogg}, D.~W., {Rix}, H.-W., {et~al.} 2022, \apj, 928, 23

\bibitem[{{Ferguson} {et~al.}(2005){Ferguson}, {Alexander}, {Allard}, {Barman}, {Bodnarik}, {Hauschildt}, {Heffner-Wong}, \& {Tamanai}}]{Ferguson05}
{Ferguson}, J.~W., {Alexander}, D.~R., {Allard}, F., {et~al.} 2005, \apj, 623, 585

\bibitem[{{Ferraro} {et~al.}(2012){Ferraro}, {Lanzoni}, {Dalessandro}, {Beccari}, {Pasquato}, {Miocchi}, {Rood}, {Sigurdsson}, {Sills}, {Vesperini}, {Mapelli}, {Contreras}, {Sanna}, \& {Mucciarelli}}]{Ferraro2012}
{Ferraro}, F.~R., {Lanzoni}, B., {Dalessandro}, E., {et~al.} 2012, \nat, 492, 393

\bibitem[{{Friel}(1995)}]{Friel95}
{Friel}, E.~D. 1995, \araa, 33, 381

\bibitem[{{Fukue} {et~al.}(2015){Fukue}, {Matsunaga}, {Yamamoto}, {Kondo}, {Kobayashi}, {Ikeda}, {Hamano}, {Yasui}, {Arasaki}, {Tsujimoto}, {Bono}, \& {Inno}}]{fukue15}
{Fukue}, K., {Matsunaga}, N., {Yamamoto}, R., {et~al.} 2015, ApJ, 812, 64

\bibitem[{{Gaia Collaboration} {et~al.}(2018){Gaia Collaboration}, {Brown}, {Vallenari}, {Prusti}, {de Bruijne}, {Babusiaux}, \& {Bailer-Jones}}]{GAIA18b}
{Gaia Collaboration}, {Brown}, A.~G.~A., {Vallenari}, A., {et~al.} 2018, A\&A, 616

\bibitem[{{Gaia Collaboration} {et~al.}(2016){Gaia Collaboration}, {Prusti}, {de Bruijne}, {Brown}, {Vallenari}, {Babusiaux}, {Bailer-Jones}, {Bastian}, {Biermann}, {Evans}, {Eyer}, {Jansen}, {Jordi}, {Klioner}, {Lammers}, {Lindegren}, {Luri}, {Mignard}, {Milligan}, {Panem}, {Poinsignon}, {Pourbaix}, {Randich}, {Sarri}, {Sartoretti}, {Siddiqui}, {Soubiran}, {Valette}, {van Leeuwen}, {Walton}, {Aerts}, {Arenou}, {Cropper}, {Drimmel}, {H{\o}g}, {Katz}, {Lattanzi}, {O'Mullane}, {Grebel}, {Holland}, {Huc}, {Passot}, {Bramante}, {Cacciari}, {Casta{\~n}eda}, {Chaoul}, {Cheek}, {De Angeli}, {Fabricius}, {Guerra}, {Hern{\'a}ndez}, {Jean-Antoine-Piccolo}, {Masana}, {Messineo}, {Mowlavi}, {Nienartowicz}, {Ord{\'o}{\~n}ez-Blanco}, {Panuzzo}, {Portell}, {Richards}, {Riello}, {Seabroke}, {Tanga}, {Th{\'e}venin}, {Torra}, {Els}, {Gracia-Abril}, {Comoretto}, {Garcia-Reinaldos}, {Lock}, {Mercier}, {Altmann}, {Andrae}, {Astraatmadja}, {Bellas-Velidis}, {Benson}, {Berthier}, {Blomme}, {Busso}, {Carry}, {Cellino}, {Clementini},
  {Cowell}, {Creevey}, {Cuypers}, {Davidson}, {De Ridder}, {de Torres}, {Delchambre}, {Dell'Oro}, {Ducourant}, {Fr{\'e}mat}, {Garc{\'\i}a-Torres}, {Gosset}, {Halbwachs}, {Hambly}, {Harrison}, {Hauser}, {Hestroffer}, {Hodgkin}, {Huckle}, {Hutton}, {Jasniewicz}, {Jordan}, {Kontizas}, {Korn}, {Lanzafame}, {Manteiga}, {Moitinho}, {Muinonen}, {Osinde}, {Pancino}, {Pauwels}, {Petit}, {Recio-Blanco}, {Robin}, {Sarro}, {Siopis}, {Smith}, {Smith}, {Sozzetti}, {Thuillot}, {van Reeven}, {Viala}, {Abbas}, {Abreu Aramburu}, {Accart}, {Aguado}, {Allan}, {Allasia}, {Altavilla}, {{\'A}lvarez}, {Alves}, {Anderson}, {Andrei}, {Anglada Varela}, {Antiche}, {Antoja}, {Ant{\'o}n}, {Arcay}, {Atzei}, {Ayache}, {Bach}, {Baker}, {Balaguer-N{\'u}{\~n}ez}, {Barache}, {Barata}, {Barbier}, {Barblan}, {Baroni}, {Barrado y Navascu{\'e}s}, {Barros}, {Barstow}, {Becciani}, {Bellazzini}, {Bellei}, {Bello Garc{\'\i}a}, {Belokurov}, {Bendjoya}, {Berihuete}, {Bianchi}, {Bienaym{\'e}}, {Billebaud}, {Blagorodnova}, {Blanco-Cuaresma}, {Boch},
  {Bombrun}, {Borrachero}, {Bouquillon}, {Bourda}, {Bouy}, {Bragaglia}, {Breddels}, {Brouillet}, {Br{\"u}semeister}, {Bucciarelli}, {Budnik}, {Burgess}, {Burgon}, {Burlacu}, {Busonero}, {Buzzi}, {Caffau}, {Cambras}, {Campbell}, {Cancelliere}, {Cantat-Gaudin}, {Carlucci}, {Carrasco}, {Castellani}, {Charlot}, {Charnas}, {Charvet}, {Chassat}, {Chiavassa}, {Clotet}, {Cocozza}, {Collins}, {Collins}, {Costigan}, {Crifo}, {Cross}, {Crosta}, {Crowley}, {Dafonte}, {Damerdji}, {Dapergolas}, {David}, {David}, {De Cat}, {de Felice}, {de Laverny}, {De Luise}, {De March}, {de Martino}, {de Souza}, {Debosscher}, {del Pozo}, {Delbo}, {Delgado}, {Delgado}, {di Marco}, {Di Matteo}, {Diakite}, {Distefano}, {Dolding}, {Dos Anjos}, {Drazinos}, {Dur{\'a}n}, {Dzigan}, {Ecale}, {Edvardsson}, {Enke}, {Erdmann}, {Escolar}, {Espina}, {Evans}, {Eynard Bontemps}, {Fabre}, {Fabrizio}, {Faigler}, {Falc{\~a}o}, {Farr{\`a}s Casas}, {Faye}, {Federici}, {Fedorets}, {Fern{\'a}ndez-Hern{\'a}ndez}, {Fernique}, {Fienga}, {Figueras}, {Filippi},
  {Findeisen}, {Fonti}, {Fouesneau}, {Fraile}, {Fraser}, {Fuchs}, {Furnell}, {Gai}, {Galleti}, {Galluccio}, {Garabato}, {Garc{\'\i}a-Sedano}, {Gar{\'e}}, {Garofalo}, {Garralda}, {Gavras}, {Gerssen}, {Geyer}, {Gilmore}, {Girona}, {Giuffrida}, {Gomes}, {Gonz{\'a}lez-Marcos}, {Gonz{\'a}lez-N{\'u}{\~n}ez}, {Gonz{\'a}lez-Vidal}, {Granvik}, {Guerrier}, {Guillout}, {Guiraud}, {G{\'u}rpide}, {Guti{\'e}rrez-S{\'a}nchez}, {Guy}, {Haigron}, {Hatzidimitriou}, {Haywood}, {Heiter}, {Helmi}, {Hobbs}, {Hofmann}, {Holl}, {Holland}, {Hunt}, {Hypki}, {Icardi}, {Irwin}, {Jevardat de Fombelle}, {Jofr{\'e}}, {Jonker}, {Jorissen}, {Julbe}, {Karampelas}, {Kochoska}, {Kohley}, {Kolenberg}, {Kontizas}, {Koposov}, {Kordopatis}, {Koubsky}, {Kowalczyk}, {Krone-Martins}, {Kudryashova}, {Kull}, {Bachchan}, {Lacoste-Seris}, {Lanza}, {Lavigne}, {Le Poncin-Lafitte}, {Lebreton}, {Lebzelter}, {Leccia}, {Leclerc}, {Lecoeur-Taibi}, {Lemaitre}, {Lenhardt}, {Leroux}, {Liao}, {Licata}, {Lindstr{\o}m}, {Lister}, {Livanou}, {Lobel}, {L{\"o}ffler},
  {L{\'o}pez}, {Lopez-Lozano}, {Lorenz}, {Loureiro}, {MacDonald}, {Magalh{\~a}es Fernandes}, {Managau}, {Mann}, {Mantelet}, {Marchal}, {Marchant}, {Marconi}, {Marie}, {Marinoni}, {Marrese}, {Marschalk{\'o}}, {Marshall}, {Mart{\'\i}n-Fleitas}, {Martino}, {Mary}, {Matijevi{\v{c}}}, {Mazeh}, {McMillan}, {Messina}, {Mestre}, {Michalik}, {Millar}, {Miranda}, {Molina}, {Molinaro}, {Molinaro}, {Moln{\'a}r}, {Moniez}, {Montegriffo}, {Monteiro}, {Mor}, {Mora}, {Morbidelli}, {Morel}, {Morgenthaler}, {Morley}, {Morris}, {Mulone}, {Muraveva}, {Musella}, {Narbonne}, {Nelemans}, {Nicastro}, {Noval}, {Ord{\'e}novic}, {Ordieres-Mer{\'e}}, {Osborne}, {Pagani}, {Pagano}, {Pailler}, {Palacin}, {Palaversa}, {Parsons}, {Paulsen}, {Pecoraro}, {Pedrosa}, {Pentik{\"a}inen}, {Pereira}, {Pichon}, {Piersimoni}, {Pineau}, {Plachy}, {Plum}, {Poujoulet}, {Pr{\v{s}}a}, {Pulone}, {Ragaini}, {Rago}, {Rambaux}, {Ramos-Lerate}, {Ranalli}, {Rauw}, {Read}, {Regibo}, {Renk}, {Reyl{\'e}}, {Ribeiro}, {Rimoldini}, {Ripepi}, {Riva}, {Rixon},
  {Roelens}, {Romero-G{\'o}mez}, {Rowell}, {Royer}, {Rudolph}, {Ruiz-Dern}, {Sadowski}, {Sagrist{\`a} Sell{\'e}s}, {Sahlmann}, {Salgado}, {Salguero}, {Sarasso}, {Savietto}, {Schnorhk}, {Schultheis}, {Sciacca}, {Segol}, {Segovia}, {Segransan}, {Serpell}, {Shih}, {Smareglia}, {Smart}, {Smith}, {Solano}, {Solitro}, {Sordo}, {Soria Nieto}, {Souchay}, {Spagna}, {Spoto}, {Stampa}, {Steele}, {Steidelm{\"u}ller}, {Stephenson}, {Stoev}, {Suess}, {S{\"u}veges}, {Surdej}, {Szabados}, {Szegedi-Elek}, {Tapiador}, {Taris}, {Tauran}, {Taylor}, {Teixeira}, {Terrett}, {Tingley}, {Trager}, {Turon}, {Ulla}, {Utrilla}, {Valentini}, {van Elteren}, {Van Hemelryck}, {van Leeuwen}, {Varadi}, {Vecchiato}, {Veljanoski}, {Via}, {Vicente}, {Vogt}, {Voss}, {Votruba}, {Voutsinas}, {Walmsley}, {Weiler}, {Weingrill}, {Werner}, {Wevers}, {Whitehead}, {Wyrzykowski}, {Yoldas}, {{\v{Z}}erjal}, {Zucker}, {Zurbach}, {Zwitter}, {Alecu}, {Allen}, {Allende Prieto}, {Amorim}, {Anglada-Escud{\'e}}, {Arsenijevic}, {Azaz}, {Balm}, {Beck}, {Bernstein},
  {Bigot}, {Bijaoui}, {Blasco}, {Bonfigli}, {Bono}, {Boudreault}, {Bressan}, {Brown}, {Brunet}, {Bunclark}, {Buonanno}, {Butkevich}, {Carret}, {Carrion}, {Chemin}, {Ch{\'e}reau}, {Corcione}, {Darmigny}, {de Boer}, {de Teodoro}, {de Zeeuw}, {Delle Luche}, {Domingues}, {Dubath}, {Fodor}, {Fr{\'e}zouls}, {Fries}, {Fustes}, {Fyfe}, {Gallardo}, {Gallegos}, {Gardiol}, {Gebran}, {Gomboc}, {G{\'o}mez}, {Grux}, {Gueguen}, {Heyrovsky}, {Hoar}, {Iannicola}, {Isasi Parache}, {Janotto}, {Joliet}, {Jonckheere}, {Keil}, {Kim}, {Klagyivik}, {Klar}, {Knude}, {Kochukhov}, {Kolka}, {Kos}, {Kutka}, {Lainey}, {LeBouquin}, {Liu}, {Loreggia}, {Makarov}, {Marseille}, {Martayan}, {Martinez-Rubi}, {Massart}, {Meynadier}, {Mignot}, {Munari}, {Nguyen}, {Nordlander}, {Ocvirk}, {O'Flaherty}, {Olias Sanz}, {Ortiz}, {Osorio}, {Oszkiewicz}, {Ouzounis}, {Palmer}, {Park}, {Pasquato}, {Peltzer}, {Peralta}, {P{\'e}turaud}, {Pieniluoma}, {Pigozzi}, {Poels}, {Prat}, {Prod'homme}, {Raison}, {Rebordao}, {Risquez}, {Rocca-Volmerange}, {Rosen},
  {Ruiz-Fuertes}, {Russo}, {Sembay}, {Serraller Vizcaino}, {Short}, {Siebert}, {Silva}, {Sinachopoulos}, {Slezak}, {Soffel}, {Sosnowska}, {Strai{\v{z}}ys}, {ter Linden}, {Terrell}, {Theil}, {Tiede}, {Troisi}, {Tsalmantza}, {Tur}, {Vaccari}, {Vachier}, {Valles}, {Van Hamme}, {Veltz}, {Virtanen}, {Wallut}, {Wichmann}, {Wilkinson}, {Ziaeepour}, \& {Zschocke}}]{gaia16}
{Gaia Collaboration}, {Prusti}, T., {de Bruijne}, J.~H.~J., {et~al.} 2016, \aap, 595, A1

\bibitem[{{Gaia Collaboration} {et~al.}(2022){Gaia Collaboration}, {Vallenari}, {Brown}, {Prusti}, {de Bruijne}, {Arenou}, {Babusiaux}, {Biermann}, {Creevey}, {Ducourant}, {Evans}, {Eyer}, {Guerra}, {Hutton}, {Jordi}, {Klioner}, {Lammers}, {Lindegren}, {Luri}, {Mignard}, {Panem}, {Pourbaix}, {Randich}, {Sartoretti}, {Soubiran}, {Tanga}, {Walton}, {Bailer-Jones}, {Bastian}, {Drimmel}, {Jansen}, {Katz}, {Lattanzi}, {van Leeuwen}, {Bakker}, {Cacciari}, {Casta{\~n}eda}, {De Angeli}, {Fabricius}, {Fouesneau}, {Fr{\'e}mat}, {Galluccio}, {Guerrier}, {Heiter}, {Masana}, {Messineo}, {Mowlavi}, {Nicolas}, {Nienartowicz}, {Pailler}, {Panuzzo}, {Riclet}, {Roux}, {Seabroke}, {Sordo{\o}rcit}, {Th{\'e}venin}, {Gracia-Abril}, {Portell}, {Teyssier}, {Altmann}, {Andrae}, {Audard}, {Bellas-Velidis}, {Benson}, {Berthier}, {Blomme}, {Burgess}, {Busonero}, {Busso}, {C{\'a}novas}, {Carry}, {Cellino}, {Cheek}, {Clementini}, {Damerdji}, {Davidson}, {de Teodoro}, {Nu{\~n}ez Campos}, {Delchambre}, {Dell'Oro}, {Esquej},
  {Fern{\'a}ndez-Hern{\'a}ndez}, {Fraile}, {Garabato}, {Garc{\'\i}a-Lario}, {Gosset}, {Haigron}, {Halbwachs}, {Hambly}, {Harrison}, {Hern{\'a}ndez}, {Hestroffer}, {Hodgkin}, {Holl}, {Jan{\ss}en}, {Jevardat de Fombelle}, {Jordan}, {Krone-Martins}, {Lanzafame}, {L{\"o}ffler}, {Marchal}, {Marrese}, {Moitinho}, {Muinonen}, {Osborne}, {Pancino}, {Pauwels}, {Recio-Blanco}, {Reyl{\'e}}, {Riello}, {Rimoldini}, {Roegiers}, {Rybizki}, {Sarro}, {Siopis}, {Smith}, {Sozzetti}, {Utrilla}, {van Leeuwen}, {Abbas}, {{\'A}brah{\'a}m}, {Abreu Aramburu}, {Aerts}, {Aguado}, {Ajaj}, {Aldea-Montero}, {Altavilla}, {{\'A}lvarez}, {Alves}, {Anders}, {Anderson}, {Anglada Varela}, {Antoja}, {Baines}, {Baker}, {Balaguer-N{\'u}{\~n}ez}, {Balbinot}, {Balog}, {Barache}, {Barbato}, {Barros}, {Barstow}, {Bartolom{\'e}}, {Bassilana}, {Bauchet}, {Becciani}, {Bellazzini}, {Berihuete}, {Bernet}, {Bertone}, {Bianchi}, {Binnenfeld}, {Blanco-Cuaresma}, {Blazere}, {Boch}, {Bombrun}, {Bossini}, {Bouquillon}, {Bragaglia}, {Bramante}, {Breedt},
  {Bressan}, {Brouillet}, {Brugaletta}, {Bucciarelli}, {Burlacu}, {Butkevich}, {Buzzi}, {Caffau}, {Cancelliere}, {Cantat-Gaudin}, {Carballo}, {Carlucci}, {Carnerero}, {Carrasco}, {Casamiquela}, {Castellani}, {Castro-Ginard}, {Chaoul}, {Charlot}, {Chemin}, {Chiaramida}, {Chiavassa}, {Chornay}, {Comoretto}, {Contursi}, {Cooper}, {Cornez}, {Cowell}, {Crifo}, {Cropper}, {Crosta}, {Crowley}, {Dafonte}, {Dapergolas}, {David}, {David}, {de Laverny}, {De Luise}, {De March}, {De Ridder}, {de Souza}, {de Torres}, {del Peloso}, {del Pozo}, {Delbo}, {Delgado}, {Delisle}, {Demouchy}, {Dharmawardena}, {Di Matteo}, {Diakite}, {Diener}, {Distefano}, {Dolding}, {Edvardsson}, {Enke}, {Fabre}, {Fabrizio}, {Faigler}, {Fedorets}, {Fernique}, {Fienga}, {Figueras}, {Fournier}, {Fouron}, {Fragkoudi}, {Gai}, {Garcia-Gutierrez}, {Garcia-Reinaldos}, {Garc{\'\i}a-Torres}, {Garofalo}, {Gavel}, {Gavras}, {Gerlach}, {Geyer}, {Giacobbe}, {Gilmore}, {Girona}, {Giuffrida}, {Gomel}, {Gomez}, {Gonz{\'a}lez-N{\'u}{\~n}ez},
  {Gonz{\'a}lez-Santamar{\'\i}a}, {Gonz{\'a}lez-Vidal}, {Granvik}, {Guillout}, {Guiraud}, {Guti{\'e}rrez-S{\'a}nchez}, {Guy}, {Hatzidimitriou}, {Hauser}, {Haywood}, {Helmer}, {Helmi}, {Sarmiento}, {Hidalgo}, {Hilger}, {H{\l}adczuk}, {Hobbs}, {Holland}, {Huckle}, {Jardine}, {Jasniewicz}, {Jean-Antoine Piccolo}, {Jim{\'e}nez-Arranz}, {Jorissen}, {Juaristi Campillo}, {Julbe}, {Karbevska}, {Kervella}, {Khanna}, {Kontizas}, {Kordopatis}, {Korn}, {K{\'o}sp{\'a}l}, {Kostrzewa-Rutkowska}, {Kruszy{\'n}ska}, {Kun}, {Laizeau}, {Lambert}, {Lanza}, {Lasne}, {Le Campion}, {Lebreton}, {Lebzelter}, {Leccia}, {Leclerc}, {Lecoeur-Taibi}, {Liao}, {Licata}, {Lindstr{\o}m}, {Lister}, {Livanou}, {Lobel}, {Lorca}, {Loup}, {Madrero Pardo}, {Magdaleno Romeo}, {Managau}, {Mann}, {Manteiga}, {Marchant}, {Marconi}, {Marcos}, {Marcos Santos}, {Mar{\'\i}n Pina}, {Marinoni}, {Marocco}, {Marshall}, {Polo}, {Mart{\'\i}n-Fleitas}, {Marton}, {Mary}, {Masip}, {Massari}, {Mastrobuono-Battisti}, {Mazeh}, {McMillan}, {Messina}, {Michalik},
  {Millar}, {Mints}, {Molina}, {Molinaro}, {Moln{\'a}r}, {Monari}, {Mongui{\'o}}, {Montegriffo}, {Montero}, {Mor}, {Mora}, {Morbidelli}, {Morel}, {Morris}, {Muraveva}, {Murphy}, {Musella}, {Nagy}, {Noval}, {Oca{\~n}a}, {Ogden}, {Ordenovic}, {Osinde}, {Pagani}, {Pagano}, {Palaversa}, {Palicio}, {Pallas-Quintela}, {Panahi}, {Payne-Wardenaar}, {Pe{\~n}alosa Esteller}, {Penttil{\"a}}, {Pichon}, {Piersimoni}, {Pineau}, {Plachy}, {Plum}, {Poggio}, {Pr{\v{s}}a}, {Pulone}, {Racero}, {Ragaini}, {Rainer}, {Raiteri}, {Rambaux}, {Ramos}, {Ramos-Lerate}, {Re Fiorentin}, {Regibo}, {Richards}, {Rios Diaz}, {Ripepi}, {Riva}, {Rix}, {Rixon}, {Robichon}, {Robin}, {Robin}, {Roelens}, {Rogues}, {Rohrbasser}, {Romero-G{\'o}mez}, {Rowell}, {Royer}, {Ruz Mieres}, {Rybicki}, {Sadowski}, {S{\'a}ez N{\'u}{\~n}ez}, {Sagrist{\`a} Sell{\'e}s}, {Sahlmann}, {Salguero}, {Samaras}, {Sanchez Gimenez}, {Sanna}, {Santove{\~n}a}, {Sarasso}, {Schultheis}, {Sciacca}, {Segol}, {Segovia}, {S{\'e}gransan}, {Semeux}, {Shahaf}, {Siddiqui}, {Siebert},
  {Siltala}, {Silvelo}, {Slezak}, {Slezak}, {Smart}, {Snaith}, {Solano}, {Solitro}, {Souami}, {Souchay}, {Spagna}, {Spina}, {Spoto}, {Steele}, {Steidelm{\"u}ller}, {Stephenson}, {S{\"u}veges}, {Surdej}, {Szabados}, {Szegedi-Elek}, {Taris}, {Taylo}, {Teixeira}, {Tolomei}, {Tonello}, {Torra}, {Torra}, {Torralba Elipe}, {Trabucchi}, {Tsounis}, {Turon}, {Ulla}, {Unger}, {Vaillant}, {van Dillen}, {van Reeven}, {Vanel}, {Vecchiato}, {Viala}, {Vicente}, {Voutsinas}, {Weiler}, {Wevers}, {Wyrzykowski}, {Yoldas}, {Yvard}, {Zhao}, {Zorec}, {Zucker}, \& {Zwitter}}]{gaiadr3}
{Gaia Collaboration}, {Vallenari}, A., {Brown}, A.~G.~A., {et~al.} 2022, arXiv e-prints, arXiv:2208.00211

\bibitem[{{Gilmore} {et~al.}(2012){Gilmore}, {Randich}, {Asplund}, {Binney}, {Bonifacio}, {Drew}, {Feltzing}, {Ferguson}, {Jeffries}, {Micela}, {Negueruela}, {Prusti}, {Rix}, {Vallenari}, {Alfaro}, {Allende-Prieto}, {Babusiaux}, {Bensby}, {Blomme}, {Bragaglia}, {Flaccomio}, {Fran{\c{c}}ois}, {Irwin}, {Koposov}, {Korn}, {Lanzafame}, {Pancino}, {Paunzen}, {Recio-Blanco}, {Sacco}, {Smiljanic}, {Van Eck}, {Walton}, {Aden}, {Aerts}, {Affer}, {Alcala}, {Altavilla}, {Alves}, {Antoja}, {Arenou}, {Argiroffi}, {Asensio Ramos}, {Bailer-Jones}, {Balaguer-Nunez}, {Bayo}, {Barbuy}, {Barisevicius}, {Barrado y Navascues}, {Battistini}, {Bellas Velidis}, {Bellazzini}, {Belokurov}, {Bergemann}, {Bertelli}, {Biazzo}, {Bienayme}, {Bland-Hawthorn}, {Boeche}, {Bonito}, {Boudreault}, {Bouvier}, {Brandao}, {Brown}, {de Bruijne}, {Burleigh}, {Caballero}, {Caffau}, {Calura}, {Capuzzo-Dolcetta}, {Caramazza}, {Carraro}, {Casagrande}, {Casewell}, {Chapman}, {Chiappini}, {Chorniy}, {Christlieb}, {Cignoni}, {Cocozza}, {Colless}, {Collet},
  {Collins}, {Correnti}, {Covino}, {Crnojevic}, {Cropper}, {Cunha}, {Damiani}, {David}, {Delgado}, {Duffau}, {Edvardsson}, {Eldridge}, {Enke}, {Eriksson}, {Evans}, {Eyer}, {Famaey}, {Fellhauer}, {Ferreras}, {Figueras}, {Fiorentino}, {Flynn}, {Folha}, {Franciosini}, {Frasca}, {Freeman}, {Fremat}, {Friel}, {Gaensicke}, {Gameiro}, {Garzon}, {Geier}, {Geisler}, {Gerhard}, {Gibson}, {Gomboc}, {Gomez}, {Gonzalez-Fernandez}, {Gonzalez Hernandez}, {Gosset}, {Grebel}, {Greimel}, {Groenewegen}, {Grundahl}, {Guarcello}, {Gustafsson}, {Hadrava}, {Hatzidimitriou}, {Hambly}, {Hammersley}, {Hansen}, {Haywood}, {Heber}, {Heiter}, {Held}, {Helmi}, {Hensler}, {Herrero}, {Hill}, {Hodgkin}, {Huelamo}, {Huxor}, {Ibata}, {Jackson}, {de Jong}, {Jonker}, {Jordan}, {Jordi}, {Jorissen}, {Katz}, {Kawata}, {Keller}, {Kharchenko}, {Klement}, {Klutsch}, {Knude}, {Koch}, {Kochukhov}, {Kontizas}, {Koubsky}, {Lallement}, {de Laverny}, {van Leeuwen}, {Lemasle}, {Lewis}, {Lind}, {Lindstrom}, {Lobel}, {Lopez Santiago}, {Lucas}, {Ludwig},
  {Lueftinger}, {Magrini}, {Maiz Apellaniz}, {Maldonado}, {Marconi}, {Marino}, {Martayan}, {Martinez-Valpuesta}, {Matijevic}, {McMahon}, {Messina}, {Meyer}, {Miglio}, {Mikolaitis}, {Minchev}, {Minniti}, {Moitinho}, {Momany}, {Monaco}, {Montalto}, {Monteiro}, {Monier}, {Montes}, {Mora}, {Moraux}, {Morel}, {Mowlavi}, {Mucciarelli}, {Munari}, {Napiwotzki}, {Nardetto}, {Naylor}, {Naze}, {Nelemans}, {Okamoto}, {Ortolani}, {Pace}, {Palla}, {Palous}, {Parker}, {Penarrubia}, {Pillitteri}, {Piotto}, {Posbic}, {Prisinzano}, {Puzeras}, {Quirrenbach}, {Ragaini}, {Read}, {Read}, {Reyle}, {De Ridder}, {Robichon}, {Robin}, {Roeser}, {Romano}, {Royer}, {Ruchti}, {Ruzicka}, {Ryan}, {Ryde}, {Santos}, {Sanz Forcada}, {Sarro Baro}, {Sbordone}, {Schilbach}, {Schmeja}, {Schnurr}, {Schoenrich}, {Scholz}, {Seabroke}, {Sharma}, {De Silva}, {Smith}, {Solano}, {Sordo}, {Soubiran}, {Sousa}, {Spagna}, {Steffen}, {Steinmetz}, {Stelzer}, {Stempels}, {Tabernero}, {Tautvaisiene}, {Thevenin}, {Torra}, {Tosi}, {Tolstoy}, {Turon}, {Walker},
  {Wambsganss}, {Worley}, {Venn}, {Vink}, {Wyse}, {Zaggia}, {Zeilinger}, {Zoccali}, {Zorec}, {Zucker}, {Zwitter}, \& {Gaia-ESO Survey Team}}]{GES}
{Gilmore}, G., {Randich}, S., {Asplund}, M., {et~al.} 2012, The Messenger, 147, 25

\bibitem[{{Gratton} {et~al.}(2000){Gratton}, {Sneden}, {Carretta}, \& {Bragaglia}}]{Gratton2000}
{Gratton}, R.~G., {Sneden}, C., {Carretta}, E., \& {Bragaglia}, A. 2000, \aap, 354, 169

\bibitem[{{Gray} \& {Johanson}(1991)}]{gray91}
{Gray}, D.~F. \& {Johanson}, H.~L. 1991, PASP, 103, 439

\bibitem[{{Guer{\c{c}}o} {et~al.}(2019){Guer{\c{c}}o}, {Cunha}, {Smith}, {Hayes}, {Abia}, {Lambert}, {J{\"o}nsson}, \& {Ryde}}]{Guerco19}
{Guer{\c{c}}o}, R., {Cunha}, K., {Smith}, V.~V., {et~al.} 2019, \apj, 885, 139

\bibitem[{{Guer{\c{c}}o} {et~al.}(2022){Guer{\c{c}}o}, {Ram{\'\i}rez}, {Cunha}, {Smith}, {Prantzos}, {Sellgren}, \& {Daflon}}]{Guerco22}
{Guer{\c{c}}o}, R., {Ram{\'\i}rez}, S., {Cunha}, K., {et~al.} 2022, \apj, 929, 24

\bibitem[{{Iglesias} \& {Rogers}(1996)}]{Iglesias96}
{Iglesias}, C.~A. \& {Rogers}, F.~J. 1996, \apj, 464, 943

\bibitem[{{Jian} {et~al.}(2019){Jian}, {Matsunaga}, \& {Fukue}}]{Jian19}
{Jian}, M., {Matsunaga}, N., \& {Fukue}, K. 2019, \mnras, 485, 1310

\bibitem[{{J{\"o}nsson} {et~al.}(2014){J{\"o}nsson}, {Ryde}, {Harper}, {Richter}, \& {Hinkle}}]{Johnsson2014}
{J{\"o}nsson}, H., {Ryde}, N., {Harper}, G.~M., {Richter}, M.~J., \& {Hinkle}, K.~H. 2014, \apjl, 789, L41

\bibitem[{{J{\"o}nsson} {et~al.}(2017){J{\"o}nsson}, {Ryde}, {Spitoni}, {Matteucci}, {Cunha}, {Smith}, {Hinkle}, \& {Schultheis}}]{Johnsson2017}
{J{\"o}nsson}, H., {Ryde}, N., {Spitoni}, E., {et~al.} 2017, \apj, 835, 50

\bibitem[{{Jorissen} {et~al.}(1992){Jorissen}, {Smith}, \& {Lambert}}]{Jorissen92}
{Jorissen}, A., {Smith}, V.~V., \& {Lambert}, D.~L. 1992, \aap, 261, 164

\bibitem[{{Kaluzny}(1998)}]{kaluzny98}
{Kaluzny}, J. 1998, Astron. Astrophys. Suppl. Ser., 133, 25

\bibitem[{{Kharchenko} {et~al.}(2013){Kharchenko}, {Piskunov}, {Schilbach}, {R{\"o}ser}, \& {Scholz}}]{Kharchenko13}
{Kharchenko}, N.~V., {Piskunov}, A.~E., {Schilbach}, E., {R{\"o}ser}, S., \& {Scholz}, R.~D. 2013, \aap, 558, A53

\bibitem[{{Kim} {et~al.}(2009){Kim}, {Jaemann}, \& {Eon-Chang}}]{Kim09}
{Kim}, C.~K., {Jaemann}, K., \& {Eon-Chang}, S. 2009, Journal of The Korean Astronomical Society, 42, 135

\bibitem[{{Kim} \& {Sung}(2003)}]{Kim03}
{Kim}, S.~C. \& {Sung}, H. 2003, Journal of The Korean Astronomical Society, 36, 13

\bibitem[{{Korotin}(2009)}]{Korotin2009}
{Korotin}, S.~A. 2009, Astronomy Reports, 53, 651

\bibitem[{{Kovtyukh} {et~al.}(2006){Kovtyukh}, {Soubiran}, {Bienaym{\'e}}, {Mishenina}, \& {Belik}}]{kovt06}
{Kovtyukh}, V.~V., {Soubiran}, C., {Bienaym{\'e}}, O., {Mishenina}, T.~V., \& {Belik}, S.~I. 2006, \mnras, 371, 879

\bibitem[{{Lagarde} {et~al.}(2012){Lagarde}, {Decressin}, {Charbonnel}, {Eggenberger}, {Ekstr{\"o}m}, \& {Palacios}}]{Lagarde2012}
{Lagarde}, N., {Decressin}, T., {Charbonnel}, C., {et~al.} 2012, \aap, 543, A108

\bibitem[{{Lagarde} {et~al.}(2024){Lagarde}, {Minkevi{\v{c}}i{\={u}}t{\.{e}}}, {Drazdauskas}, {Tautvai{\v{s}}ien{\.{e}}}, {Charbonnel}, {Reyl{\'e}}, {Miglio}, {Kushwahaa}, \& {Bale}}]{Lagarde24}
{Lagarde}, N., {Minkevi{\v{c}}i{\={u}}t{\.{e}}}, R., {Drazdauskas}, A., {et~al.} 2024, \aap, 684, A70

\bibitem[{{Lagarde} {et~al.}(2019){Lagarde}, {Reyl{\'e}}, {Robin}, {Tautvai{\v{s}}ien{\.{e}}}, {Drazdauskas}, {Mikolaitis}, {Minkevi{\v{c}}i{\={u}}t{\.{e}}}, {Stonkut{\.{e}}}, {Chorniy}, {Bagdonas}, {Miglio}, {Nasello}, {Gilmore}, {Randich}, {Bensby}, {Bragaglia}, {Flaccomio}, {Francois}, {Korn}, {Pancino}, {Smiljanic}, {Bayo}, {Carraro}, {Costado}, {Jim{\'e}nez-Esteban}, {Jofr{\'e}}, {Martell}, {Masseron}, {Monaco}, {Morbidelli}, {Sbordone}, {Sousa}, \& {Zaggia}}]{Lagarde2019}
{Lagarde}, N., {Reyl{\'e}}, C., {Robin}, A.~C., {et~al.} 2019, \aap, 621, A24

\bibitem[{{Lambert} {et~al.}(1974){Lambert}, {Dearborn}, \& {Sneden}}]{Lambert1974}
{Lambert}, D.~L., {Dearborn}, D.~S., \& {Sneden}, C. 1974, \apj, 193, 621

\bibitem[{Lee(2015)}]{plp2}
Lee, J.-J. 2015, plp: Version 2.0

\bibitem[{{Lee} {et~al.}(2004){Lee}, {Carney}, \& {Balachandran}}]{Lee2004}
{Lee}, J.-W., {Carney}, B.~W., \& {Balachandran}, S.~C. 2004, \aj, 128, 2388

\bibitem[{{Li} {et~al.}(2013){Li}, {Ludwig}, {Caffau}, {Christlieb}, \& {Zhao}}]{Li2013}
{Li}, H.~N., {Ludwig}, H.~G., {Caffau}, E., {Christlieb}, N., \& {Zhao}, G. 2013, \apj, 765, 51

\bibitem[{{Lucatello} {et~al.}(2011){Lucatello}, {Masseron}, {Johnson}, {Pignatari}, \& {Herwig}}]{Lucatello11}
{Lucatello}, S., {Masseron}, T., {Johnson}, J.~A., {Pignatari}, M., \& {Herwig}, F. 2011, \apj, 729, 40

\bibitem[{{Lucertini} {et~al.}(2023){Lucertini}, {Monaco}, {Caffau}, {Mucciarelli}, {Villanova}, {Bonifacio}, \& {Sbordone}}]{Lucertini2023}
{Lucertini}, F., {Monaco}, L., {Caffau}, E., {et~al.} 2023, \aap, 671, A137

\bibitem[{{Mace} {et~al.}(2016){Mace}, {Kim}, {Jaffe}, {Park}, {Lee}, {Kaplan}, {Yu}, {Yuk}, {Chun}, {Pak}, {Kim}, {Lee}, {Sneden}, {Afsar}, {Pavel}, {Lee}, {Oh}, {Jeong}, {Park}, {Kidder}, {Lee}, {Nguyen Le}, {McLane}, {Gully-Santiago}, {Oh}, {Lee}, {Hwang}, \& {Park}}]{Mace2016}
{Mace}, G., {Kim}, H., {Jaffe}, D.~T., {et~al.} 2016, in Society of Photo-Optical Instrumentation Engineers (SPIE) Conference Series, Vol. 9908, Ground-based and Airborne Instrumentation for Astronomy VI, ed. C.~J. {Evans}, L.~{Simard}, \& H.~{Takami}, 99080C

\bibitem[{{Mace} {et~al.}(2018){Mace}, {Sokal}, {Lee}, {Oh}, {Park}, {Lee}, {Good}, {MacQueen}, {Oh}, {Kaplan}, {Kidder}, {Chun}, {Yuk}, {Jeong}, {Pak}, {Kim}, {Nah}, {Lee}, {Yu}, {Hwang}, {Park}, {Kim}, {Chinn}, {Peck}, {Diaz}, {Rutten}, {Prato}, {Jacoby}, {Cornelius}, {Hardesty}, {DeGroff}, {Dunham}, {Levine}, {Nofi}, {Lopez-Valdivia}, {Weinberger}, \& {Jaffe}}]{Mace2018}
{Mace}, G., {Sokal}, K., {Lee}, J.-J., {et~al.} 2018, in Society of Photo-Optical Instrumentation Engineers (SPIE) Conference Series, Vol. 10702, Ground-based and Airborne Instrumentation for Astronomy VII, ed. C.~J. {Evans}, L.~{Simard}, \& H.~{Takami}, 107020Q

\bibitem[{{Magrini} {et~al.}(2023){Magrini}, {Viscasillas V{\'a}zquez}, {Spina}, {Randich}, {Romano}, {Franciosini}, {Recio-Blanco}, {Nordlander}, {D'Orazi}, {Baratella}, {Smiljanic}, {Dantas}, {Pasquini}, {Spitoni}, {Casali}, {Van der Swaelmen}, {Bensby}, {Stonkute}, {Feltzing}, {Sacco}, {Bragaglia}, {Pancino}, {Heiter}, {Biazzo}, {Gilmore}, {Bergemann}, {Tautvai{\v{s}}ien{\.{e}}}, {Worley}, {Hourihane}, {Gonneau}, \& {Morbidelli}}]{Magrini2023}
{Magrini}, L., {Viscasillas V{\'a}zquez}, C., {Spina}, L., {et~al.} 2023, \aap, 669, A119

\bibitem[{{Matsunaga} {et~al.}(2021){Matsunaga}, {Jian}, {Taniguchi}, \& {Elgueta}}]{Matsunaga21}
{Matsunaga}, N., {Jian}, M., {Taniguchi}, D., \& {Elgueta}, S.~S. 2021, \mnras, 506, 1031

\bibitem[{{McCormick} {et~al.}(2023){McCormick}, {Majewski}, {Smith}, {Hayes}, {Cunha}, {Masseron}, {Weiss}, {Shetrone}, {Almeida}, {Frinchaboy}, {Garc{\'\i}a-Hern{\'a}ndez}, \& {Nitschelm}}]{McCormick23}
{McCormick}, C., {Majewski}, S.~R., {Smith}, V.~V., {et~al.} 2023, \mnras, 524, 4418

\bibitem[{{Milone} {et~al.}(2012){Milone}, {Piotto}, {Bedin}, {Aparicio}, {Anderson}, {Sarajedini}, {Marino}, {Moretti}, {Davies}, {Chaboyer}, {Dotter}, {Hempel}, {Mar{\'\i}n-Franch}, {Majewski}, {Paust}, {Reid}, {Rosenberg}, \& {Siegel}}]{Milone12}
{Milone}, A.~P., {Piotto}, G., {Bedin}, L.~R., {et~al.} 2012, \aap, 540, A16

\bibitem[{{Molaro} {et~al.}(2023){Molaro}, {Aguado}, {Caffau}, {Allende Prieto}, {Bonifacio}, {Gonz{\'a}lez Hern{\'a}ndez}, {Rebolo}, {Zapatero Osorio}, {Cristiani}, {Pepe}, {Santos}, {Alibert}, {Cupani}, {Di Marcantonio}, {D'Odorico}, {Lovis}, {Martins}, {Milakovi{\'c}}, {Murphy}, {Nunes}, {Schmidt}, {Sousa}, {Sozzetti}, \& {Su{\'a}rez Mascare{\~n}o}}]{Molaro23}
{Molaro}, P., {Aguado}, D.~S., {Caffau}, E., {et~al.} 2023, \aap, 679, A72

\bibitem[{{Monaco, L.} {et~al.}(2014){Monaco, L.}, {Boffin, H. M. J.}, {Bonifacio, P.}, {Villanova, S.}, {Carraro, G.}, {Caffau, E.}, {Steffen, M.}, {Ahumada, J. A.}, {Beletsky, Y.}, \& {Beccari, G.}}]{Monaco14}
{Monaco, L.}, {Boffin, H. M. J.}, {Bonifacio, P.}, {et~al.} 2014, A\&A, 564, "L6"

\bibitem[{Myers {et~al.}(2022)Myers, Donor, Spoo, Frinchaboy, Cunha, Price-Whelan, Majewski, Beaton, Zasowski, O’Connell, Ray, Bizyaev, Chiappini, Garcia-Hernandez, Geisler, Jönsson, Lane, Longa-Peña, Minchev, \& Roman-Lopes}]{Myers22}
Myers, N., Donor, J., Spoo, T., {et~al.} 2022, The Astronomical Journal, 164, 85

\bibitem[{{Nandakumar} {et~al.}(2023){Nandakumar}, {Ryde}, \& {Mace}}]{Nandakumar23}
{Nandakumar}, G., {Ryde}, N., \& {Mace}, G. 2023, \aap, 676, A79

\bibitem[{{Nandakumar} {et~al.}(2022){Nandakumar}, {Ryde}, {Montelius}, {Thorsbro}, {J{\"o}nsson}, \& {Mace}}]{Nandakumar2022}
{Nandakumar}, G., {Ryde}, N., {Montelius}, M., {et~al.} 2022, \aap, 668, A88

\bibitem[{{Osorio} {et~al.}(2020){Osorio}, {Allende Prieto}, {Hubeny}, {M{\'e}sz{\'a}ros}, \& {Shetrone}}]{Osorio20}
{Osorio}, Y., {Allende Prieto}, C., {Hubeny}, I., {M{\'e}sz{\'a}ros}, S., \& {Shetrone}, M. 2020, \aap, 637, A80

\bibitem[{{Park} {et~al.}(2014){Park}, {Jaffe}, {Yuk}, {Chun}, {Pak}, {Kim}, {Pavel}, {Lee}, {Oh}, {Jeong}, {Sim}, {Lee}, {Nguyen Le}, {Strubhar}, {Gully-Santiago}, {Oh}, {Cha}, {Moon}, {Park}, {Brooks}, {Ko}, {Han}, {Nah}, {Hill}, {Lee}, {Barnes}, {Yu}, {Kaplan}, {Mace}, {Kim}, {Lee}, {Hwang}, \& {Park}}]{Park14}
{Park}, C., {Jaffe}, D.~T., {Yuk}, I.-S., {et~al.} 2014, Society of Photo-Optical Instrumentation Engineers (SPIE) Conference Series, Vol. 9147, {Design and early performance of IGRINS (Immersion Grating Infrared Spectrometer)}, 91471D

\bibitem[{{Park} {et~al.}(2018){Park}, {Lee}, {Kang}, {Lee}, {Chun}, {Kim}, {Yuk}, {Lee}, {Mace}, {Kim}, {Kaplan}, {Park}, {Sok Oh}, {Lee}, \& {Jaffe}}]{Park18}
{Park}, S., {Lee}, J.-E., {Kang}, W., {et~al.} 2018, \apjs, 238, 29

\bibitem[{{Paxton} {et~al.}(2011){Paxton}, {Bildsten}, {Dotter}, {Herwig}, {Lesaffre}, \& {Timmes}}]{Paxton11}
{Paxton}, B., {Bildsten}, L., {Dotter}, A., {et~al.} 2011, \apjs, 192, 3

\bibitem[{{Piatti} {et~al.}(2004){Piatti}, {Clari{\'a}}, \& {Ahumada}}]{Piatti04}
{Piatti}, A.~E., {Clari{\'a}}, J.~J., \& {Ahumada}, A.~V. 2004, \mnras, 349, 641

\bibitem[{{Pilachowski} \& {Pace}(2015)}]{Pilachowski15}
{Pilachowski}, C.~A. \& {Pace}, C. 2015, \aj, 150, 66

\bibitem[{{Rain} {et~al.}(2021){Rain}, {Carraro}, {Ahumada}, {Villanova}, {Boffin}, \& {Monaco}}]{Rain21}
{Rain}, M.~J., {Carraro}, G., {Ahumada}, J.~A., {et~al.} 2021, \aj, 161, 37

\bibitem[{{Ram{\'\i}rez} \& {Allende Prieto}(2011)}]{ramirez11}
{Ram{\'\i}rez}, I. \& {Allende Prieto}, C. 2011, \apj, 743, 135

\bibitem[{{Ram{\'\i}rez} {et~al.}(2013){Ram{\'\i}rez}, {Allende Prieto}, \& {Lambert}}]{Ramirez13}
{Ram{\'\i}rez}, I., {Allende Prieto}, C., \& {Lambert}, D.~L. 2013, \apj, 764, 78

\bibitem[{{Randich} {et~al.}(2022){Randich}, {Gilmore}, {Magrini}, {Sacco}, {Jackson}, {Jeffries}, {Worley}, {Hourihane}, {Gonneau}, {Viscasillas Vazquez}, {Franciosini}, {Lewis}, {Alfaro}, {Allende Prieto}, {Bensby}, {Blomme}, {Bragaglia}, {Flaccomio}, {Fran{\c{c}}ois}, {Irwin}, {Koposov}, {Korn}, {Lanzafame}, {Pancino}, {Recio-Blanco}, {Smiljanic}, {Van Eck}, {Zwitter}, {Asplund}, {Bonifacio}, {Feltzing}, {Binney}, {Drew}, {Ferguson}, {Micela}, {Negueruela}, {Prusti}, {Rix}, {Vallenari}, {Bayo}, {Bergemann}, {Biazzo}, {Carraro}, {Casey}, {Damiani}, {Frasca}, {Heiter}, {Hill}, {Jofr{\'e}}, {de Laverny}, {Lind}, {Marconi}, {Martayan}, {Masseron}, {Monaco}, {Morbidelli}, {Prisinzano}, {Sbordone}, {Sousa}, {Zaggia}, {Adibekyan}, {Bonito}, {Caffau}, {Daflon}, {Feuillet}, {Gebran}, {Gonzalez Hernandez}, {Guiglion}, {Herrero}, {Lobel}, {Maiz Apellaniz}, {Merle}, {Mikolaitis}, {Montes}, {Morel}, {Soubiran}, {Spina}, {Tabernero}, {Tautvai{\v{s}}iene}, {Traven}, {Valentini}, {Van der Swaelmen}, {Villanova}, {Wright},
  {Abbas}, {Aguirre B{\o}rsen-Koch}, {Alves}, {Balaguer-Nunez}, {Barklem}, {Barrado}, {Berlanas}, {Binks}, {Bressan}, {Capuzzo-Dolcetta}, {Casagrande}, {Casamiquela}, {Collins}, {D'Orazi}, {Dantas}, {Debattista}, {Delgado-Mena}, {Di Marcantonio}, {Drazdauskas}, {Evans}, {Famaey}, {Franchini}, {Fr{\'e}mat}, {Friel}, {Fu}, {Geisler}, {Gerhard}, {Gonzalez Solares}, {Grebel}, {Gutierrez Albarran}, {Hatzidimitriou}, {Held}, {Jim{\'e}nez-Esteban}, {J{\"o}nsson}, {Jordi}, {Khachaturyants}, {Kordopatis}, {Kos}, {Lagarde}, {Mahy}, {Mapelli}, {Marfil}, {Martell}, {Messina}, {Miglio}, {Minchev}, {Moitinho}, {Montalban}, {Monteiro}, {Morossi}, {Mowlavi}, {Mucciarelli}, {Murphy}, {Nardetto}, {Ortolani}, {Paletou}, {Palou{\v{s}}}, {Paunzen}, {Pickering}, {Quirrenbach}, {Re Fiorentin}, {Read}, {Romano}, {Ryde}, {Sanna}, {Santos}, {Seabroke}, {Spagna}, {Steinmetz}, {Stonkut{\'e}}, {Sutorius}, {Th{\'e}venin}, {Tosi}, {Tsantaki}, {Vink}, {Wright}, {Wyse}, {Zoccali}, {Zorec}, {Zucker}, \& {Walton}}]{GESfinal}
{Randich}, S., {Gilmore}, G., {Magrini}, L., {et~al.} 2022, \aap, 666, A121

\bibitem[{{Recio-Blanco} {et~al.}(2023){Recio-Blanco}, {de Laverny}, {Palicio}, {Kordopatis}, {{\'A}lvarez}, {Schultheis}, {Contursi}, {Zhao}, {Torralba Elipe}, {Ordenovic}, {Manteiga}, {Dafonte}, {Oreshina-Slezak}, {Bijaoui}, {Fr{\'e}mat}, {Seabroke}, {Pailler}, {Spitoni}, {Poggio}, {Creevey}, {Abreu Aramburu}, {Accart}, {Andrae}, {Bailer-Jones}, {Bellas-Velidis}, {Brouillet}, {Brugaletta}, {Burlacu}, {Carballo}, {Casamiquela}, {Chiavassa}, {Cooper}, {Dapergolas}, {Delchambre}, {Dharmawardena}, {Drimmel}, {Edvardsson}, {Fouesneau}, {Garabato}, {Garc{\'\i}a-Lario}, {Garc{\'\i}a-Torres}, {Gavel}, {Gomez}, {Gonz{\'a}lez-Santamar{\'\i}a}, {Hatzidimitriou}, {Heiter}, {Jean-Antoine Piccolo}, {Kontizas}, {Korn}, {Lanzafame}, {Lebreton}, {Le Fustec}, {Licata}, {Lindstr{\o}m}, {Livanou}, {Lobel}, {Lorca}, {Magdaleno Romeo}, {Marocco}, {Marshall}, {Mary}, {Nicolas}, {Pallas-Quintela}, {Panem}, {Pichon}, {Riclet}, {Robin}, {Rybizki}, {Santove{\~n}a}, {Silvelo}, {Smart}, {Sarro}, {Sordo}, {Soubiran}, {S{\"u}veges},
  {Ulla}, {Vallenari}, {Zorec}, {Utrilla}, \& {Bakker}}]{RecioBlancoRVS}
{Recio-Blanco}, A., {de Laverny}, P., {Palicio}, P.~A., {et~al.} 2023, \aap, 674, A29

\bibitem[{{Rix} \& {Bovy}(2013)}]{Rix2013}
{Rix}, H.-W. \& {Bovy}, J. 2013, \aapr, 21, 61

\bibitem[{{Romano}(2022)}]{Romano22}
{Romano}, D. 2022, \aapr, 30, 7

\bibitem[{{Ryde} {et~al.}(2020){Ryde}, {J{\"o}nsson}, {Mace}, {Cunha}, {Spitoni}, {Af{\c{s}}ar}, {Jaffe}, {Forsberg}, {Kaplan}, {Kidder}, {Lee}, {Oh}, {Smith}, {Sneden}, {Sokal}, {Strickland}, \& {Thorsbro}}]{Ryde20}
{Ryde}, N., {J{\"o}nsson}, H., {Mace}, G., {et~al.} 2020, \apj, 893, 37

\bibitem[{{Salaris} \& {Cassisi}(2005)}]{Salaris2005}
{Salaris}, M. \& {Cassisi}, S. 2005, {Evolution of Stars and Stellar Populations}

\bibitem[{{Schlafly} \& {Finkbeiner}(2011)}]{Schlafly11}
{Schlafly}, E.~F. \& {Finkbeiner}, D.~P. 2011, \apj, 737, 103

\bibitem[{{Schlegel} {et~al.}(1998){Schlegel}, {Finkbeiner}, \& {Davis}}]{sfd98}
{Schlegel}, D.~J., {Finkbeiner}, D.~P., \& {Davis}, M. 1998, \apj, 500, 525

\bibitem[{{Seshashayana} {et~al.}(2024){Seshashayana}, {J{\"o}nsson}, {D'Orazi}, {Nandakumar}, {Oliva}, {Bragaglia}, {Sanna}, {Romano}, {Spitoni}, {Karakas}, {Lugaro}, \& {Origlia}}]{Bijavara2024}
{Seshashayana}, S.~B., {J{\"o}nsson}, H., {D'Orazi}, V., {et~al.} 2024, \aap, 683, A218

\bibitem[{{Sim} {et~al.}(2014){Sim}, {Le}, {Pak}, {Lee}, {Kang}, {Chun}, {Jeong}, {Yuk}, {Kim}, {Park}, {Pavel}, \& {Jaffe}}]{Kyung14}
{Sim}, C.~K., {Le}, H. A.~N., {Pak}, S., {et~al.} 2014, Advances in Space Research, 53, 1647

\bibitem[{{Smiljanic} {et~al.}(2014){Smiljanic}, {Korn}, {Bergemann}, {Frasca}, {Magrini}, {Masseron}, {Pancino}, {Ruchti}, {San Roman}, {Sbordone}, {Sousa}, {Tabernero}, {Tautvai{\v{s}}ien{\.{e}}}, {Valentini}, {Weber}, {Worley}, {Adibekyan}, {Allende Prieto}, {Barisevi{\v{c}}ius}, {Biazzo}, {Blanco-Cuaresma}, {Bonifacio}, {Bragaglia}, {Caffau}, {Cantat-Gaudin}, {Chorniy}, {de Laverny}, {Delgado-Mena}, {Donati}, {Duffau}, {Franciosini}, {Friel}, {Geisler}, {Gonz{\'a}lez Hern{\'a}ndez}, {Gruyters}, {Guiglion}, {Hansen}, {Heiter}, {Hill}, {Jacobson}, {Jofre}, {J{\"o}nsson}, {Lanzafame}, {Lardo}, {Ludwig}, {Maiorca}, {Mikolaitis}, {Montes}, {Morel}, {Mucciarelli}, {Mu{\~n}oz}, {Nordlander}, {Pasquini}, {Puzeras}, {Recio-Blanco}, {Ryde}, {Sacco}, {Santos}, {Serenelli}, {Sordo}, {Soubiran}, {Spina}, {Steffen}, {Vallenari}, {Van Eck}, {Villanova}, {Gilmore}, {Randich}, {Asplund}, {Binney}, {Drew}, {Feltzing}, {Ferguson}, {Jeffries}, {Micela}, {Negueruela}, {Prusti}, {Rix}, {Alfaro}, {Babusiaux}, {Bensby},
  {Blomme}, {Flaccomio}, {Fran{\c{c}}ois}, {Irwin}, {Koposov}, {Walton}, {Bayo}, {Carraro}, {Costado}, {Damiani}, {Edvardsson}, {Hourihane}, {Jackson}, {Lewis}, {Lind}, {Marconi}, {Martayan}, {Monaco}, {Morbidelli}, {Prisinzano}, \& {Zaggia}}]{Smiljanic14}
{Smiljanic}, R., {Korn}, A.~J., {Bergemann}, M., {et~al.} 2014, \aap, 570, A122

\bibitem[{{Sneden}(1973)}]{sneden73}
{Sneden}, C. 1973, \apj, 184, 839

\bibitem[{{Soubiran} {et~al.}(2018){Soubiran}, {Cantat-Gaudin}, {Romero-G{\'o}mez}, {Casamiquela}, {Jordi}, {Vallenari}, {Antoja}, {Balaguer-N{\'u}{\~n}ez}, {Bossini}, {Bragaglia}, {Carrera}, {Castro-Ginard}, {Figueras}, {Heiter}, {Katz}, {Krone-Martins}, {Le Campion}, {Moitinho}, \& {Sordo}}]{Soubiran18}
{Soubiran}, C., {Cantat-Gaudin}, T., {Romero-G{\'o}mez}, M., {et~al.} 2018, \aap, 619, A155

\bibitem[{{Soubiran} {et~al.}(2016){Soubiran}, {Le Campion}, {Brouillet}, \& {Chemin}}]{Soubiran16}
{Soubiran}, C., {Le Campion}, J.-F., {Brouillet}, N., \& {Chemin}, L. 2016, \aap, 591, A118

\bibitem[{{Spina} {et~al.}(2021){Spina}, {Ting}, {De Silva}, {Frankel}, {Sharma}, {Cantat-Gaudin}, {Joyce}, {Stello}, {Karakas}, {Asplund}, {Nordlander}, {Casagrande}, {D'Orazi}, {Casey}, {Cottrell}, {Tepper-Garc{\'\i}a}, {Baratella}, {Kos}, {{\v{C}}otar}, {Bland-Hawthorn}, {Buder}, {Freeman}, {Hayden}, {Lewis}, {Lin}, {Lind}, {Martell}, {Schlesinger}, {Simpson}, {Zucker}, \& {Zwitter}}]{Spina2021}
{Spina}, L., {Ting}, Y.~S., {De Silva}, G.~M., {et~al.} 2021, \mnras, 503, 3279

\bibitem[{{Stassun} \& {Torres}(2021)}]{Stassun21}
{Stassun}, K.~G. \& {Torres}, G. 2021, \apjl, 907, L33

\bibitem[{Takeda {et~al.}(2016)Takeda, Omiya, Harakawa, \& Sato}]{Taked2016}
Takeda, Y., Omiya, M., Harakawa, H., \& Sato, B. 2016, Publications of the Astronomical Society of Japan, 68, 81

\bibitem[{{Taniguchi} {et~al.}(2021){Taniguchi}, {Matsunaga}, {Jian}, {Kobayashi}, {Fukue}, {Hamano}, {Ikeda}, {Kawakita}, {Kondo}, {Otsubo}, {Sameshima}, {Takenaka}, \& {Yasui}}]{Taniguchi21}
{Taniguchi}, D., {Matsunaga}, N., {Jian}, M., {et~al.} 2021, \mnras, 502, 4210

\bibitem[{{Taniguchi} {et~al.}(2018){Taniguchi}, {Matsunaga}, {Kobayashi}, {Fukue}, {Hamano}, {Ikeda}, {Kawakita}, {Kondo}, {Sameshima}, \& {Yasui}}]{Taniguchi18}
{Taniguchi}, D., {Matsunaga}, N., {Kobayashi}, N., {et~al.} 2018, \mnras, 473, 4993

\bibitem[{{Vaidya} {et~al.}(2020){Vaidya}, {Rao}, {Agarwal}, \& {Bhattacharya}}]{Vaidya20}
{Vaidya}, K., {Rao}, K.~K., {Agarwal}, M., \& {Bhattacharya}, S. 2020, \mnras, 496, 2402

\bibitem[{{VandenBerg}(2023)}]{Van23}
{VandenBerg}, D.~A. 2023, \mnras, 518, 4517

\bibitem[{{VandenBerg}(2024)}]{Van24}
{VandenBerg}, D.~A. 2024, \mnras, 527, 6888

\bibitem[{{VandenBerg} {et~al.}(2006){VandenBerg}, {Bergbusch}, \& {Dowler}}]{Van06}
{VandenBerg}, D.~A., {Bergbusch}, P.~A., \& {Dowler}, P.~D. 2006, \apjs, 162, 375

\bibitem[{{VandenBerg} {et~al.}(2014){VandenBerg}, {Bergbusch}, {Ferguson}, \& {Edvardsson}}]{Van14}
{VandenBerg}, D.~A., {Bergbusch}, P.~A., {Ferguson}, J.~W., \& {Edvardsson}, B. 2014, \apj, 794, 72

\bibitem[{{VandenBerg} {et~al.}(2013){VandenBerg}, {Brogaard}, {Leaman}, \& {Casagrande}}]{Van13}
{VandenBerg}, D.~A., {Brogaard}, K., {Leaman}, R., \& {Casagrande}, L. 2013, \apj, 775, 134

\bibitem[{{VandenBerg} {et~al.}(2016){VandenBerg}, {Denissenkov}, \& {Catelan}}]{Van16}
{VandenBerg}, D.~A., {Denissenkov}, P.~A., \& {Catelan}, M. 2016, \apj, 827, 2

\bibitem[{{VandenBerg} {et~al.}(2022){VandenBerg}, {Edvardsson}, {Casagrande}, \& {Ferguson}}]{Van22}
{VandenBerg}, D.~A., {Edvardsson}, B., {Casagrande}, L., \& {Ferguson}, J.~W. 2022, \mnras, 509, 4189

\bibitem[{{Viscasillas V{\'a}zquez} {et~al.}(2022){Viscasillas V{\'a}zquez}, {Magrini}, {Casali}, {Tautvai{\v{s}}ien{\.{e}}}, {Spina}, {Van der Swaelmen}, {Randich}, {Bensby}, {Bragaglia}, {Friel}, {Feltzing}, {Sacco}, {Turchi}, {Jim{\'e}nez-Esteban}, {D'Orazi}, {Delgado-Mena}, {Mikolaitis}, {Drazdauskas}, {Minkevi{\v{c}}i{\={u}}t{\.{e}}}, {Stonkut{\.{e}}}, {Bagdonas}, {Montes}, {Guiglion}, {Baratella}, {Tabernero}, {Gilmore}, {Alfaro}, {Francois}, {Korn}, {Smiljanic}, {Bergemann}, {Franciosini}, {Gonneau}, {Hourihane}, {Worley}, \& {Zaggia}}]{Viscasillas22}
{Viscasillas V{\'a}zquez}, C., {Magrini}, L., {Casali}, G., {et~al.} 2022, \aap, 660, A135

\bibitem[{{Wallerstein} \& {Helfer}(1959)}]{Wallerstein59}
{Wallerstein}, G. \& {Helfer}, H.~L. 1959, \apj, 129, 720

\bibitem[{{Williams} \& {Viti}(2013)}]{Williams2013}
{Williams}, D.~A. \& {Viti}, S. 2013, {Observational Molecular Astronomy}

\bibitem[{{Wood} {et~al.}(2014){Wood}, {Lawler}, \& {Shetrone}}]{Wood14}
{Wood}, M.~P., {Lawler}, J.~E., \& {Shetrone}, M.~D. 2014, \apjl, 787, L16

\bibitem[{{Woosley} {et~al.}(1990){Woosley}, {Hartmann}, {Hoffman}, \& {Haxton}}]{Woosley90}
{Woosley}, S.~E., {Hartmann}, D.~H., {Hoffman}, R.~D., \& {Haxton}, W.~C. 1990, \apj, 356, 272

\bibitem[{{Yan} {et~al.}(2019){Yan}, {Du}, {Liu}, {Li}, {Shi}, {Chen}, {Ma}, \& {Wu}}]{Yan19}
{Yan}, Y., {Du}, C., {Liu}, S., {et~al.} 2019, \apj, 880, 36

\bibitem[{{Yuk} {et~al.}(2010){Yuk}, {Jaffe}, {Barnes}, {Chun}, {Park}, {Lee}, {Lee}, {Wang}, {Park}, {Pak}, {Strubhar}, {Deen}, {Oh}, {Seo}, {Pyo}, {Park}, {Lacy}, {Goertz}, {Rand}, \& {Gully-Santiago}}]{Yuk10}
{Yuk}, I.-S., {Jaffe}, D.~T., {Barnes}, S., {et~al.} 2010, Society of Photo-Optical Instrumentation Engineers (SPIE) Conference Series, Vol. 7735, {Preliminary design of IGRINS (Immersion GRating INfrared Spectrograph)}, 77351M

\bibitem[{{Zhou} {et~al.}(2023){Zhou}, {Shi}, {Bi}, {Yan}, {Zhang}, {Pan}, \& {Xu}}]{Zhou23}
{Zhou}, Z., {Shi}, J., {Bi}, S., {et~al.} 2023, Universe, 9, 457

\end{thebibliography}

\end{document}